\documentclass[preprint,showpacs,preprintnumbers,amsmath,amssymb,superscriptaddress, nofootinbib]{revtex4}  
\usepackage{graphicx,color}
\usepackage{amsmath,amssymb}
\usepackage{url}
\usepackage{epstopdf}
\usepackage{slashed}
\newcommand{\hs}{\hspace*{0.5cm}}

\newcommand{\be}{\begin{equation}}
\newcommand{\ee}{\end{equation}}
\newcommand{\bea}{\begin{eqnarray}}
\newcommand{\eea}{\end{eqnarray}}

\newcommand{\crn}{\nonumber \\}

\newcommand{\fr}{\frac}

\newcommand{\bc}{\begin{center}}
	\newcommand{\ec}{\end{center}}

\newcommand {\ba}{\begin{array}}
	\newcommand {\ea}{\end{array}}
\newcommand{\ben}{\begin{enumerate}}
	\newcommand{\een}{\end{enumerate}}
	
\usepackage{epsfig,graphicx} 
\usepackage{bm}
\usepackage{dcolumn}
\begin{document}
	\title{$(g-2)_{e,\mu}$ anomalies and decays $h\to e_a e_b$, $Z\to e_ae_b$, and   $e_b\to e_a \gamma$ in a two Higgs doublet model with inverse seesaw neutrinos} 
\author{T.T. Hong} \email{tthong@agu.edu.vn}
\author{Q.Duyet Tran} \email{tqduyet@agu.edu.vn}
\affiliation{An Giang University, Long Xuyen City 880000, Vietnam} 

\affiliation{Vietnam National University, Ho Chi Minh City 700000, Vietnam} 
\author{T.Phong Nguyen}\email{thanhphong@ctu.edu.vn}
\affiliation{Department of Physics, Can Tho University,
	3/2 Street, Can Tho, Vietnam}
\author{L.T. Hue } \email{lethohue@vlu.edu.vn}
\affiliation{Subatomic Physics Research Group, Science and Technology Advanced Institute, Van Lang University, Ho Chi Minh City, Vietnam}
\author{N.H.T. Nha \footnote{Corresponding author}} \email{nguyenhuathanhnha@vlu.edu.vn}
\affiliation{Subatomic Physics Research Group, Science and Technology Advanced Institute, Van Lang University, Ho Chi Minh City, Vietnam}
\affiliation{Faculty of Applied Technology, School of Engineering and Technology,  Van Lang University, Ho Chi Minh City, Vietnam}

\begin{abstract}
The lepton flavor violating decays  $h\to e_b^\pm e_a^\mp $, $Z\to e_b^\pm e_a^\mp$, and   $e_b\to e_a \gamma$ will be discussed in the framework of the  Two Higgs doublet model with presence of new inverse seesaw neutrinos and a singly charged Higgs boson that accommodate both $1\sigma$ experimental data of $(g-2)$ anomalies of the muon and electron.   Numerical results indicate that there exist regions of the parameter space supporting all experimental data of $(g-2)_{e,\mu}$ as well as the promising LFV signals corresponding to the future experimental sensitivities. 
\end{abstract}
	\maketitle
	\allowdisplaybreaks 
\section{Introduction}
In the two Higgs doublet model (2HDM) framework with presence of new seesaw neutrinos, a recent study on lepton flavor violating (LFV) decays of charged leptons $e_b\to e_a \gamma $, (Standard Model-like) SM-like Higgs and neutral gauge bosons $Z,h\to e_b^\pm e_a^\mp$, showed that the $Z\to e_b^\pm e_a^\mp$ decays  are  suppressed even with the future experimental searches, in contrast with the promoting signal of the remaining LFV decays   \cite{Jurciukonis:2021izn}.   On the other hand, the experimental data of charged lepton anomalies $(g-2)_{e,\mu}$ \cite{Parker:2018vye, Fan:2022eto,  Muong-2:2023cdq}  can be accommodated in the 2HDM adding a singly charged Higgs bosons and new heavy leptons \cite{Mondal:2021vou, Hue:2023rks}  such as neutrinos used to explain the neutrino oscillation data through the inverse seesaw (ISS) mechanism.  The new ISS heavy neutrinos give large one-loop contributions, named as "chirally-enhanced" ones,  to both $(g-2)_{e,\mu}$ and LFV decays of charged leptons (cLFV)  \cite{Crivellin:2018qmi}. The same contributions also predict large LFV decays of  the SM-like Higgs boson (LFV$h$) in the 3-3-1 model \cite{Hong:2022xjg}.  In contrast, the LFV decay rates  were predicted to be suppressed in many beyond the SM, including the 3-3-1 models \cite{CortesMaldonado:2011uh}. Therefore, we expect  that there may appear promising signals of LFV decays of the gauge boson $Z$ (LFV$Z$) in the allowed regions accommodating the $(g-2)_{e,\mu}$ data  the mentioned models with ISS neutrinos. This is our main aim in this work. 

 Moreover, our work here will be useful for futher  investigation another class of the beyond the SM (BSM) consisting of both singly charged Higgs bosons and singly charged vector-like (VL) leptons, which also give  "chirally-enhanced" contributions to accommodate the $(g-2)_{\mu}$ data \cite{Crivellin:2018qmi, DeJesus:2020yqx, Cogollo:2020nrc, Dermisek:2020cod, Dermisek:2021ajd,  Dermisek:2022aec, Dermisek:2023tgq, deJesus:2023som}.  Namely, when the future $(g-2)_e$ data is confirmed experimentally, the couplings of these VL particles to both muon and electron can result in interesting consequences on LFV decay rates which should be explored more precisely elsewhere.  

For the 2HDM adding a singly charged Higgs boson $\chi$  and six ISS neutrinos (2HDM$N_{L,R}$) we choose to study LFV decays in  this work, although the one-loop contributions from the $W^\pm$ in the loop do not affect the new deviations of $(g-2)_{e,\mu}$ from the SM predictions, they affect strongly the cLFV decay rates Br$(e_b\to e_a\gamma)$ which are now constrained strictly by experiments. Consequently, the LFV$h$ decay rates are also constrained more strict than the experimental sensitivities, especially in the ISS extension of the SM without new Higg bosons \cite{Herrero:2017myd, Arganda:2016zvc, Arganda:2014dta}.   The same conclusions for the LFV$Z$ decays, where maximal decay rates are orders of $\mathcal{O}(10^{-7})$ for $Z\to \tau^\pm e^\mp, \tau^\pm \mu^\mp$ \cite{Korner:1992an, DeRomeri:2016gum,Hernandez-Tome:2019lkb,  Abada:2022asx}, even for the 2HDM with standard seesaw neutrinos \cite{Jurciukonis:2021izn}. Therefore, new contributions from new singly charged Higgs bosons may give opposite signs to relax the maximal values of these decays rates.

One-loop contributions from diagrams with virtual  $W^\pm$ used in this work will be computed in both unitary and 't Hooft-Feynman gauges, using the same notations introduced in Ref. \cite{Jurciukonis:2021izn}. These formulas can be transformed into the forms  given in Ref. \cite{Abada:2022asx} used to discussed in a simple ISS extension of the SM.   Formulas calculated in the unitary without any need of information of Goldstone boson couplings will be a great advantage applicable for calculating one-loop contributions of new charged gauge boson to the   LFV$Z$ decay amplitudes or new heavy neutral gauge bosons appearing in BSM  being searched for at LHC \cite{ATLAS:2020tre}.  

Experimental data for $(g-2)_{e,\mu}$ anomalies   have been updated recently. In this work we will discuss on the parameter spaces of a 2HDM satisfying the following experimental data:
\begin{itemize}
	\item  $a_{\mu}\equiv (g-2)_{\mu}/2$ data has been  updated from Ref. \cite{Muong-2:2023cdq} showing a deviation from the  SM prediction of  $a^{\mathrm{SM}}_{\mu}= 116591810(43)\times 10^{-11}$ \cite{Aoyama:2020ynm} combined from various different contributions  based on the dispersion approach \cite{Davier:2010nc, Danilkin:2016hnh,  Davier:2017zfy, Keshavarzi:2018mgv, Colangelo:2018mtw, Hoferichter:2019mqg, Davier:2019can, Keshavarzi:2019abf, Kurz:2014wya, Melnikov:2003xd, Masjuan:2017tvw, Colangelo:2017fiz, Hoferichter:2018kwz, Gerardin:2019vio, Bijnens:2019ghy, Colangelo:2019uex, Colangelo:2014qya, Blum:2019ugy, Aoyama:2012wk, Aoyama:2019ryr, Czarnecki:2002nt, Gnendiger:2013pva, Pauk:2014rta, Jegerlehner:2017gek, Knecht:2018sci,Eichmann:2019bqf,Roig:2019reh}. In this work we will use the following deviation  \cite{Ghosh:2023dgk} 
	\begin{equation}\label{eq_damu}
		\Delta a^{\mathrm{NP}}_{\mu}\equiv  a^{\mathrm{exp}}_{\mu} -a^{\mathrm{SM}}_{\mu} =\left(2.49\pm 0.48 \right) \times 10^{-9} (5.1\sigma). 
	\end{equation} 

\item The recent experimental $a_e$ data  was  reported from different groups~\cite{Hanneke:2008tm, Parker:2018vye, Morel:2020dww, Fan:2022eto}, leading to the two inconsistent deviations  between experiments and the SM prediction \cite{Aoyama:2012wj,  Laporta:2017okg, Aoyama:2017uqe,  Terazawa:2018pdc, Volkov:2019phy, Gerardin:2020gpp}. In this work, we accept the following value:  
\begin{equation}\label{eq_dae}
	\Delta a^{\mathrm{NP}}_{e}\equiv  a^{\mathrm{exp}}_{e} -a^{\mathrm{SM}}_{e} = \left( 3.4\pm 1.6\right) \times 10^{-13},
\end{equation}  
where the latest experimental data  for  $  a^{\mathrm{exp}}_{e}$ was given in Ref.  \cite{Fan:2022eto}. There is another $a^{\mathrm{SM}}_{e}$ value derived from  the measurement of the fine-structure constant of Cs-133 atoms \cite{Parker:2018vye},  leading to $\Delta a^{\mathrm{NP}}_{e} = \left(-10.2\pm 2.6\right) \times 10^{-13}$, implying  the $3.9\sigma$ deviation from  the earlier. Although our numerical investigation will use only the $1\sigma$ range given in Eq. \eqref{eq_dae}, the two $(g-2)_e$ data have the same order of magnitudes, therefore the two qualitative results will be the same. 

\item The cLFV rates are constrained experimentally as follows  \cite{BaBar:2009hkt, MEG:2016leq, Belle:2021ysv}: 
\begin{align}
	\label{eq_ebagaex}
	\mathrm{Br}(\tau\rightarrow \mu\gamma)&<4.4\times 10^{-8}, 
	\crn 
	\; \mathrm{Br}(\tau\rightarrow e\gamma)& <3.3\times 10^{-8}, 
	\crn \mathrm{Br}(\mu\rightarrow e\gamma) &< 4.2\times 10^{-13}.
\end{align}
Future sensitivities for these decay will be Br$(\mu \to e\gamma)<6\times 10^{-14}$, Br$(\tau\to e \gamma)< 9.0 \times 10^{-9}$, Br$(\tau\to \mu \gamma)< 6.9 \times 10^{-9}$ \cite{MEGII:2018kmf, Belle-II:2018jsg}.

\item The latest experimental constraints for LFV$h$ decay rates are
\begin{align}
\label{eq_LFVHexp}
 \mathrm{Br}(h\rightarrow \tau \mu) &<1.5\times 10^{-3} \text{\cite{CMS:2021rsq}},
 \crn 
 \mathrm{Br}(h\rightarrow \tau e) &<2.2\times 10^{-3} \text{\cite{CMS:2021rsq}},
 \crn 
 \mathrm{Br}(h\rightarrow \mu e) &<6.1\times 10^{-5}  \text{\cite{ATLAS:2019xlq}}.
\end{align}
The future sensitivities at the HL-LHC and $e^+e^-$ colliders may be  orders of  $\mathcal{O}(10^{-4}) $ \cite{Qin:2017aju, Barman:2022iwj, Aoki:2023wfb}, $\mathcal{O}(10^{-4}) $, and $\mathcal{O}(10^{-5}) $ \cite{Qin:2017aju} for the three above LFV$h$ decays, respectively .

\item The latest experimental constraints for LFV$Z$ decay rates  are
\begin{align}
	\label{eq_LFVZexp}
	\mathrm{Br}(Z\rightarrow \tau^\pm \mu^\mp) &<6.5\times 10^{-6}  \text{\cite{ATLAS:2021bdj}},
	\crn 
	\mathrm{Br}(Z\rightarrow \tau^\pm e^\mp) &<5.0\times 10^{-6} \text{\cite{ATLAS:2021bdj}},
	\crn 
	\mathrm{Br}(Z\rightarrow \mu^\pm e^\mp) &<2.62\times 10^{-7}\text{\cite{ATLAS:2022uhq}},
\end{align}
The future sensitivities will be $10^{-6}$, $10^{-6}$, and $7\times 10^{-8}$ at HL-LHC \cite{Dam:2018rfz};  and $10^{-9}$, $10^{-9}$, and $10^{-10}$ at FCC-ee \cite{Dam:2018rfz, FCC:2018byv}, respectively.
\end{itemize} 

Our work is arranged as follows. In section \ref{eq_oneloopLFVZ},  we discuss on the one-loop contributions of the $W$ mediation to the decay amplitudes  $Z\to e_b^\pm e_a^\mp$, using the notations introduced in Ref. \cite{Jurciukonis:2021izn}.  In section \ref{eq_2HDMNR}, we will  investigate the three LFV decay classes, namely  $e_b\to e_a\gamma$,  $Z\to e^\pm_be^\mp_a$, and $h\to e^\pm_be^\mp_a$ in the 2HDM$N_{L,R}$ framework, concentrating on the regions of the parameter space accommodating the $1\sigma$ range of the $(g-2)_{e,\mu}$ experimental data.  

\section{ \label{eq_oneloopLFVZ} One-loop contributions of $W$ mediation to the decay amplitude  $Z\to e^+_b e^-_a$}
In this section we will  determine   analytic formulas of all  diagrams relevant to one-loop contributions of the gauge boson $W^{\pm}$ to the decay amplitude $Z\to e_b^+e_a^-$  in the  unitary gauge, using the notations introduced in Ref. \cite{Jurciukonis:2021izn}. Although the calculation is limited in the two Higgs doublet models, in which the results in the  't Hooft-Feynman gauge were  introduced \cite{Jurciukonis:2021izn}, the calculations  in the unitary gauge can be generalized  for  many BSM   predicting new neutral and charged gauge bosons.  This is very convenient because the relevant couplings of new goldstone bosons can be ignored. 

In the 2HDM framework, the one-loop contributions of the $W^\pm$  to the decay amplitude $Z\to e^+_a e^-_b$ will be  calculated in the unitary gauge,  based on the  well-known  Lagrangian parts constructed previously \cite{Jurciukonis:2021izn,Grimus:2002ux}. We summarize here the necessary ingredients.  
\begin{itemize}
	\item We consider the 2HDM model consists of  $K$ exotic right-handed neutrinos $N_{IR}$ ($I=1,2,\dots, K$) as $SU(2)_L$ singlets,  in stead of 3 discussed in Ref. \cite{Jurciukonis:2021izn}. Because all exotic $N_R$ are $SU(2)_L$ singlets, they do not couple to gauge bosons $Z$ and $W^\pm$.   Lagrangian  for the  charged current is the same form given in Ref.  \cite{Jurciukonis:2021izn}, namely   
	\begin{align}
	\label{eq_Lcc}
	\mathcal{L}_{cc}&= \frac{e}{\sqrt{2} s_W}\sum_{a =1}^3\sum_{i=1}^{K+3} \left( U_{a i} \overline{e_a} \gamma^{\mu} P_L n_i W^-_{\mu} + U^{*}_{a i} \bar{n}_i\gamma^{\mu} P_L e_a W^+_{\mu}\right),
	\end{align}
	where $U_{a i}$ is the $3\times (K+3)$ mixing matrix of  three active neutrinos and new heavy ones, $UU^{\dagger}=I$,  which is defined as a submatrix  of the following total $(K+3)\times (K+3)$ neutrino mixing matrix:
	\begin{equation}
		\label{eq_Unu}
		U^{\nu}:= 
		\begin{pmatrix}
			U\\
			X^*
		\end{pmatrix}. 
	\end{equation}
	Here we used the general from of neutrino mixing matrix introduced in Ref. \cite{Jurciukonis:2021izn}, in which the total neutrino mass matrix is a $(K+3)\times (K+3)$ symmetric one denoted as $\mathcal{M}^{\nu}$. The original basis is $\nu'_L: =(\nu_L, (N_R)^c)^T$ and $\nu'_R:= (\nu_L)^c=((\nu_L)^c, N_R)^T$. The respective physical basis of neutrinos is $n_{L,R}:=(n_{1L,R},n_{2L,R},\dots , n_{(K+3)L,R})^T$, which consist of  left and right components of the physical Majorana states $n_i=n_i^c\equiv (n_{iL}, n_{iR})^T $ with $n_{iR}=(n_{iL})^c$. Useful relations are:
	\begin{align}
	\label{eq_Mnu}
\mathcal{L}^{\nu}_{\mathrm{mass}}&=  -\frac{1}{2} \overline{\nu'_R} \mathcal{M}^{\nu} \nu_L +\mathrm{H.c.}, 
\crn U^{\nu T}\mathcal{M}^{\nu}U^{\nu} &= \hat{\mathcal{M}}^{\nu} =\mathrm{diag} \left( \hat{m}_{\nu}, \hat{M}_N \right), 
\crn \nu'_{L}&=U^{\nu} n_L,\;  \nu'_{R}=U^{\nu*} n_R.
\end{align}
We note here that $\mathcal{M}^{\nu}$ considered here is more general than the standard seesaw (ss) form.  Three physical active neutrinos have masses being included in the matrix $\hat{m}_{\nu}=\mathrm{diag}(m_{n_1}, m_{n_2},m_{n_3})$, which must guarantee the neutrino oscillation data. As the charged lepton states are assumed to be physical from the beginning,  $\equiv U_{ab}\; \forall a,b=1,2,3$ can be derived in terms of the experimental parameters relating to the neutrino oscillation data. The sub-matrices have the following properties.
\begin{align}
\label{eq_UX}
	\left( UU^{\dagger}\right)_{ab} =& \left( U^{\nu}U^{\nu}\right)_{ab}=\delta_{ab}\; \forall a,b=1,2,3;
\crn  \left( X^*X^{T}\right)_{IJ} =& \left( U^{\nu}U^{\nu}\right)_{(I+a)(J+b)}=\delta_{(I+a)(J+b)}\; \forall a,b=1,2,3; I,J=1,2,\dots, K.  
\end{align}
They are originated from the unitary property and dependent to the particular form of $\mathcal{M}^{\nu}$. 
	
	\item  Lagrangian  for the neutral current in this model is:
	\begin{align}
	\label{eq_Lnn}
	\mathcal{L}_{nc}&= e Z_{\mu} \sum_{a=1}^3  \bar{e}_a \gamma^{\mu}\left[t_L P_L +t_R  P_R\right] e_a 
	+\frac{e Z_{\mu}}{4s_W c_W} \sum_{i,j=1}^{K+3}\bar{n}_i \gamma^{\mu}\left[q_{ij} P_L -q_{ji} P_R\right] n_j , 
	\end{align}
	where 
	\begin{equation}\label{eq_qịj}
	t_R=\frac{s_W}{c_W},\; t_L=\frac{s^2_W-c^2_W}{2s_Wc_W},\;	q_{ij}=\left(U^\dagger U\right)_{ij}, \; -\frac{1}{t_R}=2t_L -t_R. 
	\end{equation}
	\item The triple-gauge-boson couplings  $ZWW$, which is changed into the momentum notations, is written as follows: 
	\begin{align}
	\label{eq_LZWW}
	\mathcal{L}_{ZWW} =& -\frac{e}{t_r} Z_{\mu}W^+_{\nu}W^-_{\alpha} \Gamma^{\mu\nu \alpha}(p_0,p_+,p_-),
	\crn \Gamma^{\mu\nu \alpha}(p_0,p_+,p_-) &\equiv g^{\nu \alpha}\left( p_{+} -p_-\right)^{\mu} + g^{ \alpha \mu}\left( p_{-} -p_0\right)^{\nu} + g^{  \mu \nu}\left( p_{0} -p_+\right)^{\alpha}. 
	\end{align} 
	
\end{itemize}
To calculate in the 't Hooft-Feynman gauge, one needs more particular information of the goldstone boson couplings. Namely,  Lagrangian for couplings of the goldstone boson $G^\pm_W$ corresponding to $W^\pm$ is   
\begin{align}
\label{eq_LGY}
\mathcal{L}^{G}&= \sum_{a=1}^3 \sum_{i=1}^{6} \frac{g}{\sqrt{2}m_W}\left[ G^-_W  \bar{e}_a   U^{\nu}_{a i}\left( m_i P_R - m_{e_a} P_L \right) n_i +G^+_W  \bar{n}_i U^{\nu*}_{a i} \left( m_i P_L -m_{e_a} P_R \right) e_a \right]
\crn &-   e  t_R m_W Z_{\mu}  \left( W^{+\mu} G^-_W + \mathrm{h.c.}\right)    -iet_L Z_{\mu}  \left[ \left( \partial^{\mu}G^+_W\right)G^-_W  -G^+_W\partial^{\mu}G^-_W\right].
\end{align}
The Feynman rules for $Z \overline{e_a}e_a$ and $Z \overline{n_i}n_j$ couplings are   shown in Table \ref{t_Zcouplings}. 
\begin{table}[h]
	\begin{tabular}{|c|c|c|c|}
		\hline
		Vertex & factor & Vertex &factor\\
		\hline
		$Z_{\mu} \overline{e_a}e_a$ & $i\gamma^{\mu} \left(t_L P_L +t_RP_R\right)$ & $Z\overline{n_i} n_j$ & $ \frac{ie}{2s_W c_W} \gamma^{\mu} \left[ q_{ij} P_L -q_{ji} P_R\right]$\\
		\hline 
		\end{tabular}
	\caption{ Feynman rules for one-loop contributions to $ Z \to e_b^+e_a$.    \label{t_Zcouplings}}
\end{table}
It is emphasized that the factor for $Z \overline{n_i}n_j$ vertex for two Majorana neutrinos  in Table \ref{t_Zcouplings} is different from  that appears in Lagrangian  \eqref{eq_Lnn}  by a factor two, which consistent with Refs. \cite{Jurciukonis:2021izn, Dreiner:2008tw}  \footnote{This factor does not appear for the Feynman rule in Ref. \cite{Abada:2022asx}}. 

In the unitary gauge, the propagator of Goldstone boson $\Delta^{(u)}_{G_W}=0$, therefore all contributions from diagrams consisting of goldstone propagators  vanish.  In this work, we will give all relevant one-loop contributions in the unitary gauge, then we compare them with the  previous results calculated in the  't Hoof-Feynman gauge.

\subsection{Decays $Z\to e^+_b e^-_a $}
\begin{figure}[ht]
	\centering 
	\includegraphics[width=16cm]{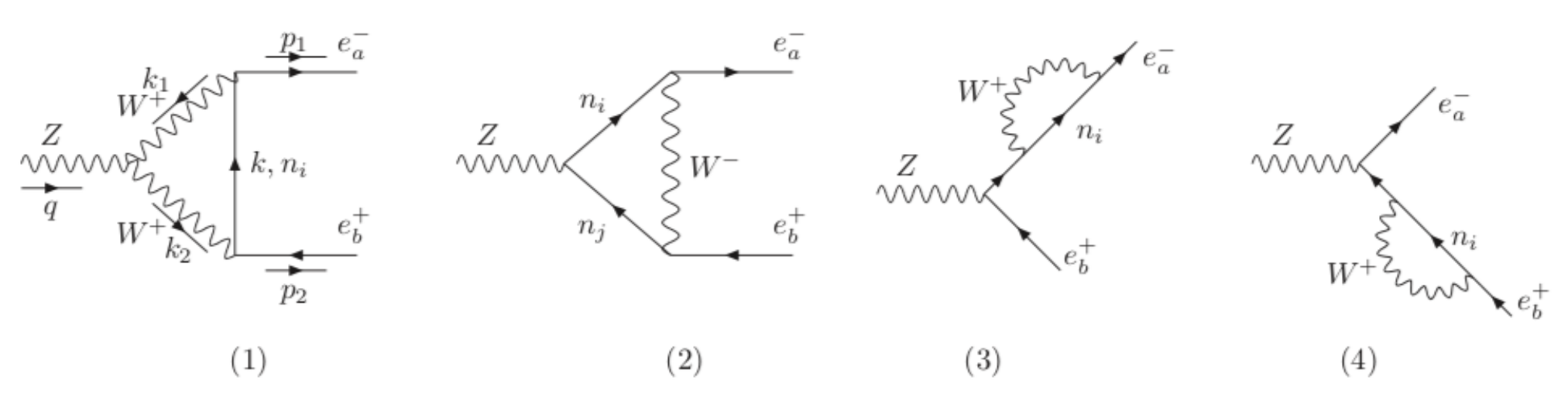}
\caption{ One-loop Feynman diagrams in the unitary gauge. }
 \label{fig_2HDMU}
\end{figure}

The effective amplitude for the decays  $Z\to e^\pm_b (p_2) e^{\mp}_a (p_1)$ is written following the notations introduced in  Refs. \cite{Jurciukonis:2021izn, DeRomeri:2016gum}, namely:
\begin{align} \label{eq_Mzeab}
i\mathcal{M}(Z\to e^+_be^-_a)
	%
	= & \frac{ie}{16\pi^2} \overline{u}_{a}\left[ \slashed{\varepsilon} \left( \bar{a}_l P_L + \bar{a}_r P_R\right) +  (p_1.\varepsilon) \left( \bar{b}_l P_L + \bar{b}_r P_R\right)  \right]v_{b},
\end{align}
where $\varepsilon_{\alpha}(q)$ is  the polarization of $Z$ and $u_a(p_1)$, and $v_b(p_2)$ Dirac spinors of $e_a^-$ and $e^+_b$.  We also used the relation $q.\varepsilon=0$ to derive that $p_2.\varepsilon=-p_1.\varepsilon$, hence simplify the relevant expression introduced in Ref. \cite{Jurciukonis:2021izn}. In this work, the form factors $\bar{a}_{l,r}$ and $\bar{b}_{l,r}$  get contributions from one-loop corrections. The on-shell conditions of the final leptons and $Z$ are $p_1^2= m_1^2 =m_a^2$, $p_2^2=m_2^2 =m_b^2$, and $q^2=m_Z^2$. The respective partial decay width is 
\begin{align} 
\label{eq_GAZeba}
\Gamma (Z\to e^+_b e^-_a)= 	\frac{\sqrt{\lambda}}{16\pi m_Z^3}\times \left(\frac{e}{16\pi^2}\right)^2 \left( \frac{\lambda M_0}{12 m^2_Z} +M_1 +\frac{ M_2}{3 m^2_Z}\right),
\end{align}
where $\lambda= m^4_Z +m^4_{b} +m^4_{a} -2(m^2_Zm^2_{a} +m^2_Zm^2_{b} +m^2_{a}m^2_{b})$, 
and 
\begin{align}
\label{eq_Mi}
M_0= & (m^2_Z -m_{a}^2 -m_{b}^2)\left(|\bar{b}_l|^2 +|\bar{b}_r|^2\right)  -4 m_{a} m_{b} \mathrm{Re}\left[ \bar{b}_l  \bar{b}^*_r\right]
\crn&
 - 4m_{b} \mathrm{Re}\left[ \bar{a}^*_r \bar{b}_l   + \bar{a}^*_l \bar{b}_r  \right] -  4m_{a}\mathrm{Re}\left[ \bar{a}^*_l \bar{b}_l   + \bar{a}^*_r  \bar{b}_r  \right] , 
\crn M_1 = & 4 m_{a}m_{b} \mathrm{Re}\left[\bar{a}_l\bar{a}_r^* \right],
\crn  M_2 = &  \left[ 2 m^4_Z - m_Z^2\left( m_{a}^2 + m_{b}^2\right) - \left( m_{a}^2 - m_{b}^2\right)^2  \right] \left( |\bar{a}_l|^2 +|\bar{a}_r|^2\right).
\end{align}
In the unitary gauge, the relevant 4 one-loop diagrams relating to the $W^\pm$ mediation are shown in Fig. \ref{fig_2HDMU}.
 Detailed steps to derive the form factors  by hand  are given in Appendix \ref{app_Zeab}. Consequently,  the one-loop contributions to the form factors  $\bar{a}^{nWW}_{l,r}$ and  $\bar{b}^{nWW}_{l,r}$ corresponding to the diagram (1) in Fig. \ref{fig_2HDMU} are written in terms of the Passarino-Veltman (PV) functions \cite{tHooft:1978jhc} defined in appendix \ref{app_PVLT}, namely: 
\begin{align}
 \bar{a}^{nWW}_l	=&\dfrac{e^2 }{2s^2_Wt_R} \sum_{i=1}^{K+3}U_{ai}^\nu U_{bi}^{\nu*}
  \left\{ \left[ -4+  \dfrac{(m^2_Z -2m_W^2)m^2_{n_i}}{m^4_W} \right] C_{00}  +2\left(m^2_Z- m^2_a -m^2_b\right)X_3 
\right.\crn &\left. \hspace{3.7cm}
	 -\frac{1}{m_W^2}  \left[\frac{}{} m^2_Z  \left(2m^2_{n_i}C_0 +m^2_aC_1 +m^2_bC_2\right)
\right.\right.\crn	& \left.\left. 
 	\hspace{5cm} -m^2_{n_i}\left(B_0^{(1)} +B_0^{(2)}\right) -m^2_aB_1^{(1)} -m^2_bB_1^{(2)}   \right]  \right\} , \label{eq_al1u} 
	%
	%
\\	\bar{a}^{nWW}_ r=&  \dfrac{e^2 m_a m_b}{2s^2_Wt_R}  \sum_{i=1}^{K+3}U_{ai}^\nu U_{bi}^{\nu*} \left[  \left(-4 + \frac{ m^2_Z}{m_W^2} \right) X_3 
 + \dfrac{m^2_Z -2 m_W^2}{m^4_W}C_{00} \right], \label{eq_ar1u}
	%
\\	\bar{b}^{nWW}_l= & \dfrac{e^2 m_a}{2s^2_Wt_R}   \sum_{i=1}^{K+3}U_{ai}^\nu U_{bi}^{\nu*} \left[ 4 \left( X_3 -X_1\right) 	+\dfrac{m^2_Z -2m_W^2}{m^4_W}   \left( m^2_{n_i}X_{01} +m^2_bX_2 \right)   -\frac{2m_Z^2}{m_W^2} C_2  
  \right], \label{eq_bl1u} 
	%
\\	\bar{b}^{nWW}_r =&\dfrac{e^2 m_b}{2s^2_Wt_R}  \sum_{i=1}^{K+3}U_{ai}^\nu U_{bi}^{\nu*} \left[4\left(X_3   - X_2\right)  +\dfrac{m^2_Z -2 m_W^2}{m^4_W}\left( m^2_{n_i}X_{02} +m^2_aX_1\right) -\frac{2m^2_Z}{m_W^2} C_1 \right], \label{eq_br1u}
\end{align}
where $B^{(k)}_{0,1}=B^{(k)}_{0,1}(p_k^2; m_{n_i}^2,m_W^2)$, $C_{00,0,k,kl}=C_{00,0,k,kl}(m_a^2,m_Z^2,m_b^2; m_{n_i}^2, m_W^2,m_W^2)$, and  $X_{0,k,kl}$  are defined in terms of the PV-functions in Eq. \eqref{eq_Xidef} for all $k,l=1,2$. 

Similarly, the one-loop contributions from diagram (2) in Fig. \ref{fig_2HDMU} are: 
\begin{align}
	\bar{a}^{Wnn}_l =& \dfrac{e^2}{4m^2_Ws^3_Wc_W} 
	\crn&\times  \sum_{i,j=1}^{K+3}U_{ai}^\nu U_{bj}^{\nu*} \left\lbrace q_{ij} \left[  m^2_W \Big(4C_{00}  +2 m^2_aX_{01} +2m^2_bX_{02} -2m^2_Z \left( C_{12}+X_0\right)\Big) \right.\right.\crn
	&\left.\left. \hspace{3.4cm} -\left(m^2_{n_i}-m^2_a\right)B_0^{(1)} -\left(m^2_{n_j}-m^2_b\right)B_0^{(2)}  +m^2_aB_1^{(1)} +m^2_bB_1^{(2)}
	\right.\right.\crn
	&\left.\left. \hspace{3.4cm}+ \left(m_{n_j}^2m_{a}^2 +m_{n_i}^2m_{b}^2-m_a^2m_b^2 \right) X_0  -m_{n_i}^2m_{n_j}^2C_0 
	\right.\right.\crn
	&\left.\left. \hspace{3.4cm} -m_{n_i}^2m_{b}^2 C_1 -m_{n_j}^2m_{a}^2 C_2  \frac{}{}\right] \right.\crn
	&\left. \hspace{2.6cm} +q_{ji}m_{n_i}m_{n_j}\bigg[2m^2_WC_0 -2C_{00} -m^2_aC_{11} -m^2_bC_{22} 
\right.\crn &\left. \hspace{5cm}	
+\left(m^2_Z -m^2_a -m^2_b\right)C_{12}\bigg]\right\rbrace,  \label{eq_al2u} 
	%
	\\ \bar{a}^{Wnn}_r  =&  
	\dfrac{e^2 m_am_b}{4m^2_Ws^3_Wc_W} 
	 \sum_{i,j=1}^{K+3}U_{ai}^\nu U_{bj}^{\nu*} q_{ij}\left[\frac{}{} 2C_{00}+ 2m_W^2X_0 +m^2_aX_1 +m^2_bX_2 
\right.\crn&\left. \hspace{4.7cm}	 -m_Z^2 C_{12}  -m_{n_i}^2C_1 -m_{n_j}^2C_2\right], \label{eq_ar2u}
	\\ \bar{b}^{Wnn}_l  =&  \dfrac{2e^2 m_a}{4m^2_Ws^3_Wc_W} \sum_{i,j=1}^{K+3}U_{ai}^\nu U_{bj}^{\nu*} \left[ q_{ij} \left(-2m_W^2 X_{01}   -m^2_bX_2 +m_{n_j}^2C_2\right) +q_{ji}m_{n_i}m_{n_j}(X_1 -C_1) \right],   \label{eq_bl2u} 
	%
	\\ \bar{b}^{Wnn}_r =& \dfrac{2 e^2 m_b}{4m^2_Ws^3_Wc_W} \sum_{i,j=1}^{K+3}U_{ai}^\nu U_{bj}^{\nu*} \Big[ q_{ij} \left( -2m_W^2X_{02}  -m^2_aX_1  +m_{n_i}^2C_1\right) +q_{ji}m_{n_i}m_{n_j}(X_2 -C_2)
	\Big], \label{eq_br2u}
\end{align} 
where  $B^{(1)}_{0,1}=B^{(1)}_{0,1}(p_1^2; m_W^2, m_{n_i}^2)$, $B^{(2)}_{0,1}=B^{(2)}_{0,1}(p_2^2; m_W^2, m_{n_j}^2)$, and  $C_{00,0,k,kl}=C_{00,0,k,kl}(m_a^2,m_Z^2,m_b^2;  m_W^2,m_{n_i}^2,m_{n_j}^2)$ for all $k,l=1,2$. 

The   form factors for sum  contributions  from two diagrams (3) and (4) in Fig. \ref{fig_2HDMU} are  
\begin{align}
	\bar{a}^{nW}_l =& \dfrac{e^2   t_L}{2m^2_Ws^2_W(m^2_a -m^2_b)}  \sum_{i=1}^{K+3}U_{ai}^\nu U_{bi}^{\nu*}
\left\lbrace 2m^2_{n_i}(m^2_a B_0^{(1)}- m^2_b B_0^{(2)})  + m^4_a B_1^{(1)}  -m^4_b B_1^{(2)}
	\right.\crn&\left.\hspace{5.5cm}+  \left[2 m^2_W +m^2_{n_i} \right] \left(m^2_aB_1^{(1)} -m^2_bB_1^{(2)} \right)  \right\rbrace. \label{eq_al3u} 
	%
	\\ \bar{a}^{nW}_ r =& \dfrac{e^2   m_a m_b t_R}{2m^2_Ws^2_W(m^2_a -m_b^2)}  \sum_{i=1}^{K+3}U_{ai}^\nu U_{bi}^{\nu*}   \left\lbrace 2m^2_{n_i}(B_0^{(1)}- B_0^{(2)})  + m^2_aB_1^{(1)}  -m^2_b B_1^{(2)}
	\right.\crn&\left. \hspace{5.5cm} +  \left(2m^2_W +m^2_{n_i}\right) \left(B_1^{(1)} -B_1^{(2)} \right)  \right\rbrace , \label{eq_ar3u}
	\\ \bar{b}^{nW}_{l} =&    \bar{b}^{nW}_{r} = 0, \label{eq_blr3u}
\end{align}
where $B^{(k)}_{0,1}=B^{(1)}_{0,1}(p_k^2; m_{n_i}^2,m_W^2)$ with $k=1,2$. 

We have used $d=4$ for all finite parts, and the unitary property of  $U^{\nu}$: $\sum_{i=1}^{K+3} U^{\nu}_{ai}U^{\nu*}_{bi}=\delta_{ab}$, and $\sum_{i,j=1}^{K+3} U^{\nu}_{ai}U^{\nu*}_{bi}q_{ij}=\delta_{ab}$,  based on brief explanations given in Appendix \ref{app_Zeab}.  The results of  all form factors listed above were also crosschecked using FORM package \cite{Vermaseren:2000nd, Kuipers:2012rf}. They are  also consistent with those introduced in  Ref. \cite{Hue:2023rks} for decay amplitude $e_b \to e_a \gamma$ in the limit $t_R=t_L=1$ and $g^R=0$.   For completeness, we list in  appendix \ref{app_Sloop}  the one-loop contributions from singly charged Higgs bosons to the LFV$Z$ amplitude, which is totally  consistent with previous results \cite{Jurciukonis:2021izn}. We note that the  results calculated  in the unitary gauge  presented in our work and ref. \cite{Abada:2022asx} are consistent in general, except two parts expressed more detailed in appendix \ref{eq_ZeabHF}.

\begin{figure}[ht]
	\centering 
	\includegraphics[width=12cm]{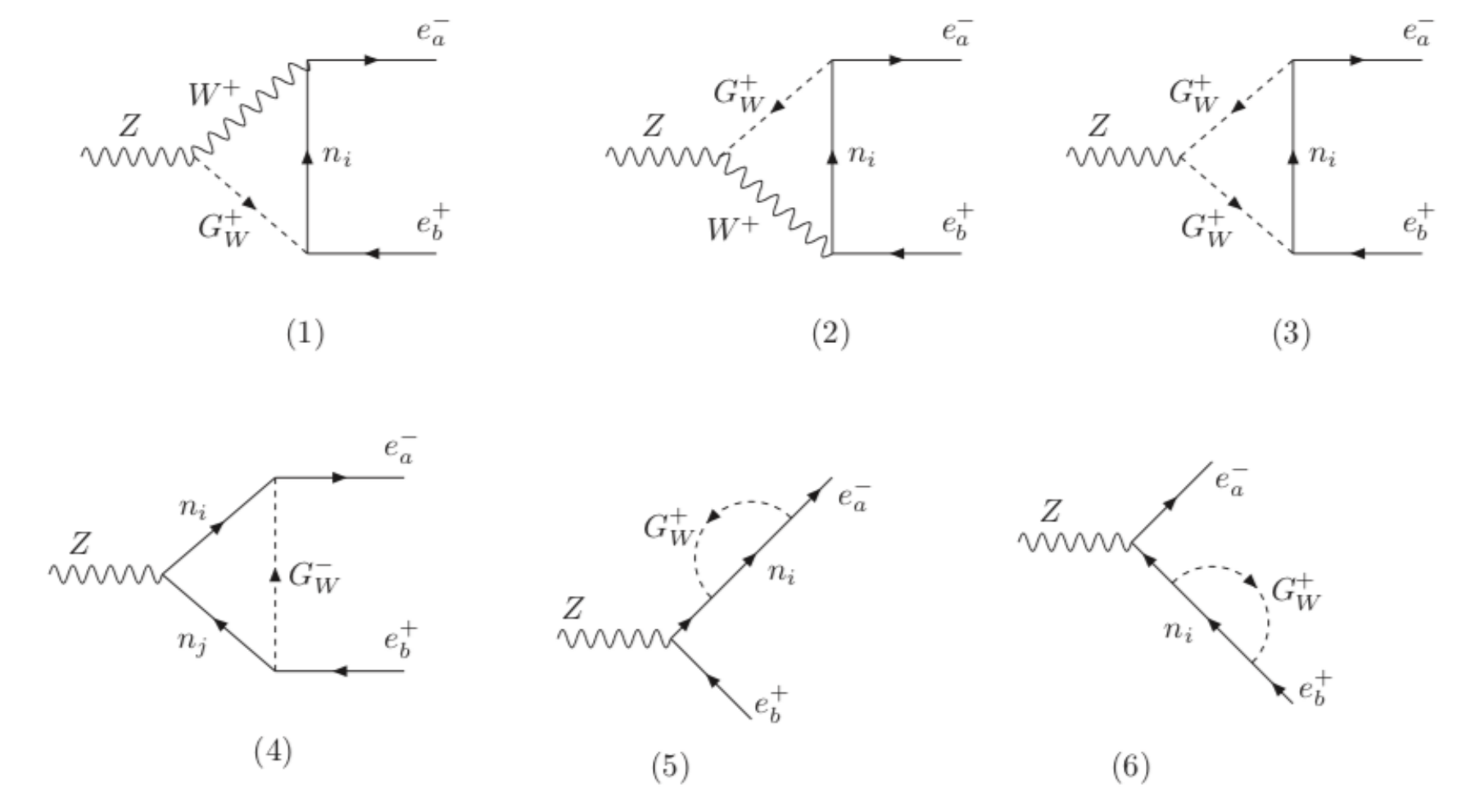}
	\caption{One-loop Feynman diagrams consisting of goldstone boson exchanges}
	\label{fig_2HDMHF}
\end{figure}

In the 't Hooft-Feynman gauge, apart from 4 diagrams listed in Fig. \ref{fig_2HDMU}, the relevant diagrams consisting the goldstone boson exchanges are shown in Fig. \ref{fig_2HDMHF}. 
The one-loop form factors are contributions from 10 diagrams discussed in Ref. \cite{Jurciukonis:2021izn}, which were checked carefully by us to confirm the complete consistency  with the results derived from our calculation, see the detailed discussion in Appendix  \ref{app_Zeab}.   We also show that the two results calculated in both  unitary and 't Hooft-Feynman gauges are the same.  The proof is summarized as follows.  The first diagram in Fig. \ref{fig_2HDMU} relates to the class of  four relevant diagrams in the 't Hooft Feynman gauge, in which the $W$ propagators may be replaced with  those of the respective Goldstone boson $G^\pm_W$. Consequently, we denote the following deviations between the two gauges:    
\begin{align}
	\label{eq_dnWW}
	\delta \bar{a}^{nWW}_l \equiv &  	\bar{a}^{nWW}_l  -\bar{a}'_{nWW,l}- \bar{a}'_{nGW,l}- \bar{a}'_{nWG,l}- \bar{a}'_{nGG,l}
	\crn =& \frac{e^2}{2m_W^2 s_W^2t_R}\sum_{i=1}^{K+3} U_{ai}^\nu U_{bi}^{\nu*}
\left\{  m_{n_i}^2 \left( B^{(1)}_0 + B^{(2)}_0\right) +m_a^2 B^{(1)}_1 +m_b^2 B^{(2)}_1 \right\}
	\crn =& \frac{e^2}{4m_W^2 s_W^2t_R}
\crn &\times  \sum_{i=1}^{K+3} U_{ai}^\nu U_{bi}^{\nu*}
\left\{ 2A_0(m_{n_i}^2) - \left( m_a^2 B^{(1)}_0 + m_b^2 B^{(2)}_0\right) + \left(3m_{n_i}^2-m_W^2 \right) \left( B^{(1)}_0 + B^{(2)}_0\right)  \right\},
	\crn \delta \bar{a}^{nWW}_r =&  	\bar{a}^{nWW}_r -\bar{a}'_{nWW,r}- \bar{a}'_{nGW,r}- \bar{a}'_{nWG,r}- \bar{a}'_{nGG,r}=0,
	%
	%
	\crn \delta \bar{b}^{nWW}_l =&  	\bar{b}^{nWW}_l -\bar{b}'_{nWW,l}- \bar{b}'_{nGW,l}- \bar{b}'_{nWG,l}- \bar{b}'_{nGG,l}
	=0,
	\crn \delta \bar{b}^{nWW}_r =&  	\bar{b}^{nWW}_r -\bar{b}'_{nWW,r}- \bar{b}'_{nGW,r}- \bar{b}'_{nWG,r}- \bar{b}'_{nGG,r} =0,
\end{align}
where  $B^{(k)}_{0}=B^{(k)}_{0}\left(p_k^2;m_{n_i}^2, m_W^2\right) $ with $k=1,2$;  $t_{L,R}$ given in Eqs. \eqref{eq_qịj},  and $m_Z=m_W/c_W$ were used to derive the zero values of $\delta \bar{b}_{nWW,l}$ and $\delta \bar{b}_{nWW,r}$.  The functions $B^{(k)}_1$ is replaced with Eqs. \eqref{eq_Bi} to obtain the final results of $\delta \bar{a}_{nWW,l}$.

Similarly, we consider the second class of the diagrams containing  two neutrino propagators as follows:
\begin{align}
	\label{eq_dWnn}
	\delta \bar{a}^{Wnn}_l =&  	\bar{a}^{Wnn}_l -\bar{a}'_{Wnn,l} - \bar{a}'_{Gnn,l}
	\crn =& \frac{e^2}{4 m_W^2s_W^3c_W}\sum_{i,j=1}^{K+3}q_{ij} \left[ -m_{n_i}^2 B^{(1)}_0 - m_{n_j}^2 B^{(2)}_0 + m^2_a B^{(1)}_1 + m^2_b B^{(2)}_1 + m^2_a B^{(1)}_0 + m^2_b B^{(2)}_0\right] , 
\crn =&\frac{e^2}{4 m_W^2s_W^3c_W}\sum_{i=1}^{K+3}  \left[ -m_{n_i}^2 \left( B^{(1)}_0 + B^{(2)}_0\right) + m^2_a B^{(1)}_1 + m^2_b B^{(2)}_1 + m^2_a B^{(1)}_0 + m^2_b B^{(2)}_0\right] 
\crn =&-\frac{e^2}{8 m_W^2s_W^3c_W}\sum_{i=1}^{K+3} \left\{ 2A_0(m_{n_i}^2) - \left( m_a^2 B^{(1)}_0 + m_b^2 B^{(2)}_0\right) + \left(3m_{n_i}^2-m_W^2 \right) \left( B^{(1)}_0 + B^{(2)}_0\right)  \right\},
\crn \delta \bar{a}^{Wnn}_r =&  \delta \bar{b}^{Wnn}_l = \delta \bar{b}^{Wnn}_r  =0,
\end{align}
where $$B^{(k)}_{0}= B^{(k)}_{0}\left(p_k^2;m_W^2, m_{n_i}^2\right) =B^{(k)}_{0}\left(p_k^2;m_{n_i}^2, m_W^2\right),$$ 
and $k=1,2$. 
We note that the relation \eqref{eq_PVrelation0} relating to $C_{00}$ results in $\delta \bar{a}_{nWW,r}=0$. 

Finally, the class of the two-point diagrams give the following deviations:
\begin{align}
	\label{eq_dnW}
	\delta \bar{a}^{nW}_l =&  	\bar{a}^{nW}_l -\bar{a}'_{nW,l} -\bar{a}'_{nG,l}
	\crn =& \frac{e^2 t_L}{2 m_W^2 s_W^2}\sum_{i=1}^{K+3}U_{ai}^\nu U_{bi}^{\nu*} \left[ m_{n_i}^2 \left( B^{(1)}_0 + B^{(2)}_0\right) + m_a^2 B^{(1)}_1 + m_b^2 B^{(2)}_1 \right] 
	\crn =& \frac{e^2 t_L}{4 m_W^2 s_W^2}
\crn &\times  \sum_{i=1}^{K+3} U_{ai}^\nu U_{bi}^{\nu*}
\left\{ 2A_0(m_{n_i}^2) - \left( m_a^2 B^{(1)}_0 + m_b^2 B^{(2)}_0\right) + \left(3m_{n_i}^2-m_W^2 \right) \left( B^{(1)}_0 + B^{(2)}_0\right)  \right\},
	\crn \delta \bar{a}^{nW}_r =&  	\bar{a}^{nW}_r -\bar{a}'_{nW,r} -\bar{a}'_{nG,r}= 0,
\end{align}
where $B^{(k)}_{0} =B^{(k)}_{0}\left(  p_k^2;m_{n_i}^2, m_W^2\right)$ with $k=1,2$. 
As a result, it is easy to derive that 
\begin{align}
	\label{eq_dal}
\delta \bar{a}^{nWW}_l +\delta \bar{a}^{Wnn}_l +\delta \bar{a}^{nW}_l \varpropto   \frac{e^2}{4m_W^2s_W^2}\times \left( \frac{1}{t_R} - \frac{1}{2s_Wc_W} + t_L\right)=0,
\end{align}
i.e, the two results calculated in the two unitary and 't Hooft-Feynman gauges coincide with each other. 
\section{ \label{eq_2HDMNR} The 2HDM with inverse seesaw neutrinos}
\subsection{Particle content and couplings}
In this work, we will study a model discussed recently to explain experimental data of   $(g-2)_{e,\mu}$ anomalies, where all LFV processes mentioned above will be  discussed, namely the particle content is of of the leptons and Higgs sector is listed in Table \ref{t_particle}, which is a  particular model (2HDM$N_{L,R}$) mentioned in Ref. \cite{Hue:2023rks}. 
\begin{table}[ht]
	\centering 
	\begin{tabular}{|c||c|c|c|c||c|c|c|c|}
		\hline
		Symmetry	&  $L_L$ &$e_R$  & $N_L$  & $N_R$ & $H_1$ & $H_2$ & $\varphi$ & $\chi^-$ \\
		\hline
		$SU(3)_C$	 & $\mathbf{1}$  &  $\mathbf{1}$&   $\mathbf{1}$ &  $\mathbf{1}$&  $\mathbf{1}$ &  $\mathbf{1}$ &  $\mathbf{1}$ & $\mathbf{1}$   \\
		\hline
		$SU(2)_L$  & $\mathbf{2}$ &  $\mathbf{1}$&  $\mathbf{1}$ &  $\mathbf{1}$&  $\mathbf{2}$ & $\mathbf{2}$  & $\mathbf{1}$  &  $\mathbf{1}$ \\
		\hline
		$U(1)_Y$  & $ -\frac{1}{2}$  & $-1$ & $0$ & $0$ &   $ \frac{1}{2}$ &    $ \frac{1}{2}$& $0$ & $-1$ \\
				\hline
		$\mathbb{Z}_2$  & $-$ & $-$ &  $+$&  $+$& $-$ & $+$ &$+$  & $-$ \\
		\hline
	\end{tabular}	
	\caption{Particle content of the 2HDM$N_{L,R}$ \label{t_particle}}
\end{table}
This model is also  a simple version without the gauge symmetry $U(1)_{B-L}$ mentioned  in Ref. \cite{Mondal:2021vou}. We do not mention the quark sector because it is is irrelevant with our discussions and can be found in many well-known works, see  reviews in Refs.  \cite{Mondal:2021vou, Branco:2011iw}. 

Accordingly, the electric charge operator and  covariant derivative corresponding to the electroweak gauge group $SU(2)_L \times U(1)_Y$ are:
\begin{align}
	\label{eq_QDmu}
	Q &=T^3 +Y,
	\\	D_{\mu} & = \partial_{\mu}-ig_2 T^aW^a_{\mu} -g_1 YB_{\mu},
\end{align}
where  $a=\overline{1,3}$, $g_2$, and $g_1$ are respectively the gauge couplings of the gauge fields $G^a_{\mu}$, $W^a_{\mu}$, $B_{\mu}$, and $B'_{\mu}$. 
The Higgs doublets are expanded as follows:
\begin{align}
	H_i&= \begin{pmatrix}
		H^+_i\\
		H^0_i
	\end{pmatrix}, \; H^0_i=\frac{v_i +r_i +i z_i}{\sqrt{2}}, \quad \varphi= \frac{v_{\varphi} +r' +i z'}{\sqrt{2}}; i=1,2. 
   \label{eq_Higgs1}
\end{align}

The Yukawa Lagrangian of leptons is \cite{Mondal:2021vou}
\begin{align}
	\label{eq_LellY}
	-\mathcal{L}^{\ell}_Y =  \overline{L_L} y_{\ell} H_1 e_R + \overline{L_L} f\tilde{H}_2 N_R + \overline{N_L}y^{\chi}  e_R \chi^+ + \overline{N_L}y_N N_R \varphi +  \overline{(N_L)^C} \frac{\lambda_{L}}{\Lambda}N_L \varphi^2  +\mathrm{h.c.},  
\end{align} 
where $\tilde{H}_2= i\sigma_2 H_2^*$ , $y_{\ell}$, $f$, $Y_N$, $y^{\chi}$, and  $\lambda_L$  are $3\times 3$ matrices,  with repective entries  $y_{\ell, ab}$, $f_{ab}$, $g_{ab}$,  and  $\lambda_{L,ab}$ with  $a,b=1,2,3$.  The five-dimension effective matrix   $\mu_L$ generates small Majorana values consistent with the ISS form. We note that to forbid unnecessary Yukawa couplings appearing in Eq. \eqref{eq_LellY},   a gauge symmetry $U(1)_{B-L}$  or a discrete symmetry like $Z_3$  introduced respectively in Ref. \cite{Mondal:2021vou} or \cite{Hue:2023rks} must be imposed. These new symmetries will not affect our results in the limit of large $v_{\varphi}$ hence we will not discuss details here. \footnote{We thank the referee for pointing out this point.}

\begin{figure}[ht]
	\centering 
	\includegraphics[width=6cm]{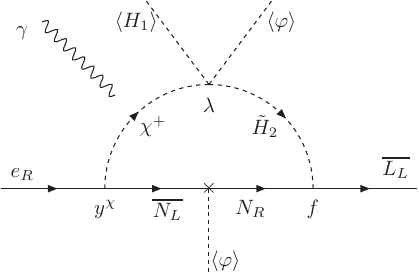}
	\caption{One-loop Feynman diagram for chirally-enhanced contribution  for $(g-2)_{e_a}$ and cLFV decays $e_b\to e_a \gamma$. }
	\label{fig:chirally-enhancement}
\end{figure}

As we will show details below, the Yukawa part  in Eq. \eqref{eq_LellY} generates one-loop contributions containing  chirally-enhancement corresponding to the Feynman diagram shown in Fig. \ref{fig:chirally-enhancement},  
see a similar diagram discussed in Ref. \cite{Cherchiglia:2023utd}. The quartic Higgs-self coupling $\lambda$ comes from the Higgs potential listed in appendix \ref{app_2HDMnu}.

The mass of leptons are derived from Eq. \eqref{eq_LellY}, keeping the VEV terms as follows 
\begin{align}
	\label{eq_LYellmass}
	-\mathcal{L}^Y_{\ell, \mathrm{mass}} =& m_{e_a} \bar{e}_a e_a + \left[ \frac{1}{2} \left( \overline{(\nu_L)^C},\; \overline{N_R},\; \overline{(N_L)^C}\right) \mathcal{M}^{\nu} \left( \nu_L,\; (N_R)^C,\; N_L\right)^T +\mathrm{H.c.}\right], 
\end{align}
where 
\begin{align}
	\mathcal{M}^{\nu}&= 
	\left(
	\begin{array}{cc}
		\mathcal{O}_{3\times3} & M_D^T \\
		M_D & M_{N} \\
	\end{array}
	\right), \;  M_D= \begin{pmatrix}
		m_D	\\
		\mathcal{O}_{3\times3}
	\end{pmatrix}, 
	\;  M_N=\left(
	\begin{array}{cc}
		\mathcal{O}_{3\times3}& M_R \\
		M^T_R & \mu_L \\
	\end{array}
	\right), 
	\crn  m_D &\equiv \frac{f^{\dagger}v_2}{\sqrt{2}}, \; M_R \equiv y_N^{\dagger} \frac{v_{\varphi}}{\sqrt{2}}, \; \mu_L \equiv \lambda_L v^2_{\varphi}/\Lambda,
	\label{eq_L0ISSnumass}	
\end{align}
and  $\mathcal{O}_{3\times3}$ is a zero matrix. The total mass matrix $\mathcal{M}^{\nu}$ will be identified with that given in Eq. \eqref{eq_Mnu}, where $K=6$. The analytic form of the Dirac mass matrix $m_D$ was chosen generally following Ref. \cite{Casas:2001sr}.

The first term in Eq. \eqref{eq_LellY} generate charged lepton masses, i.e., Eq. \eqref{eq_LYellmass}, where we choose the diagonal form to avoid  tree level cLFV decay: 
\begin{equation}
	\label{eq_mea}
	m_{e_a}= \frac{ \left( y_{\ell}\right)_{aa}v_1}{\sqrt{2}} \to 	\left( y_{\ell}\right)_{aa}=  \frac{g m_{a}}{ \sqrt{2}m_W c_{\beta}},
\end{equation}
where 
\begin{align}
\label{eq_deftb}	
	t_{\beta} \equiv\tan\beta= v_2/v_1, \; s_{\beta}=\sin\beta,\; c_{\beta}=\cos\beta. 
\end{align}
The neutrino mass matrix is diagonalized through the following mixing matrix \cite{ParticleDataGroup:2022pth}:
\begin{equation}
	\label{eq_hatMN}
	U^{\nu T}\mathcal{M}^{\nu}  U^\nu = \widehat{\mathcal{M}}^{\nu} = \mathrm{diag}(m_{n_1},m_{n_2},...,m_{n_{9}} ) =\mathrm{diag}(\widehat{m}_\nu,\widehat{M}_N),
\end{equation}
where $\widehat{m}_\nu=\mathrm{diag}(m_{n_1},m_{n_2}, m_{n_3})$ and 	$\widehat{M}_N$ consist of active and new heavy neutrinos, respectively. The relations between the flavor and mass base are 
\begin{equation} 
	\label{eq_nupL} \nu'_L =U^{\nu}n_L, \; \text{and} \;  (\nu'_L)^c =U^{\nu*}(n_L)^c =U^{\nu*}n_R, 
\end{equation}
where four-component spinor for Majorana neutrinos $ n_i=(n_{iL},n_{iR})^T$, and 
\begin{align}
	\label{eq_niL}
	\nu'_L &\equiv \left( \nu_L,\; (N_R)^C,\; N_L\right)^T \leftrightarrow (\nu'_L)^C \equiv \left( (\nu_L)^C,\; N_R,\; (N_L)^C \right)^T, 
	\crn n_L & \equiv \left( n_{1L},\; n_{2L},\dots n_{9L}\right)^T \leftrightarrow  n_R\equiv (n_L)^C = \left( (n_{1L})^C,\; (n_{2L})^C, \dots (n_{9L})^C\right)^T. 
\end{align}

The total neutrino mixing matrix are written in the popular ISS form 
\begin{align}	
	U^\nu= \begin{pmatrix}(I_3 - \dfrac{1}{2} RR^\dagger)U_{\mathrm{PMNS}} & RV\\ -R^\dagger U_{\mathrm{PMNS}} & (I_K - \dfrac{1}{2}R^\dagger  R)V \end{pmatrix} +\mathcal{O}(R^3).
\end{align}
This matrix is also identified with the general form given in Eq. \eqref{eq_Mnu} to determine the relevant couplings relating to the gauge boson $Z$.

Defining   $M'=M_R\mu_L^{-1}M_R^T \sim M^2\mu_L^{-1}\gg m_D$, 
The ISS relations are:  
\begin{align}
	R&= M^{\dagger}_D{M'^*_N}^{-1}  = \left(-m_D^{\dagger}M'^{*-1},\hs m^\dagger_D\left(M^\dagger\right)^{-1} \right),
	\crn m_{\nu}&=-M^T_DM_N'^{-1}M_D =  m_D^T \left( M^T\right)^{-1}\mu_LM^{-1}m_D,
	\crn V^*\hat{M}_NV^{\dagger} &\simeq M_{N} + \frac{1}{2}R^TR^*M_N +\frac{1}{2}M_N R^{\dagger}R.  
	\label{eq_RISS}
\end{align}
We will apply the following simple ISS framework in numerical investigation \cite{Thao:2017qtn}:
\begin{align}
	\label{eq_mDRISS}
 M_R\equiv &  M_0 I_3,
\; V\simeq  \frac{1}{\sqrt{2}}  \begin{pmatrix}
	-iI_3& I_3 \\
	iI_3& I_3
\end{pmatrix}, 
\; 	m_D= M_0\hat{x}_\nu^{1/2}  U^\dagger_{\mathrm{PMNS}}, 
	\crn R =& \left( -U_{\mathrm{PMNS}}\frac{\sqrt{\mu_L \hat{m}_{\nu}}}{M_0},\;  U_{\mathrm{PMNS}}\hat{x}_\nu^{1/2} \right) \simeq \left(\mathcal{O}_{3\times3},\;  U_{\mathrm{PMNS}}\hat{x}_\nu^{1/2} \right),
\end{align}
where  $\; \hat{x}_\nu\equiv \frac{\hat{m}_\nu}{\mu_0}$ satisfying the ISS condition max$[\left( |\hat{x}_\nu|\right)_{ab}]\ll1$ with all $a,b=1,2,3$. The precise form of $U^{\nu}$ in terms of $\hat{x}_{\nu}$  used here was given in Ref. \cite{Hue:2021zyw}.  Consequently, all six neutrino masses are nearly degenerate, $m_{n_i}\simeq M_0\;\forall i= \overline{4,9}$.  Also, the total neutrino mixing matrix $U^{\nu}$ will be determined from the formulas given in Eq. \eqref{eq_mDRISS}.

A detailed calculation was shown in appendix \ref{app_2HDMnu} to derive all physical Higgs states including masses and mixing parameters.  The Yukawa couplings of Higgs and two leptons are derived from Eq. \eqref{eq_LellY}.  Defining  Higgs boson couplings  with  leptons  that
\begin{align}
	\label{eq_lahij}
	\lambda^{h}_{ij}=& \sum_{c=1}^3 \left(  U^{\nu}_{cj}U^{\nu*}_{ci} m_{n_i} + U^{\nu}_{ci}U^{\nu*}_{cj} m_{n_j} \right),
	\crn 	\lambda^{L,G}_{ai}=& - \sum_{b=1}^3U^{\nu}_{(b+3)i}(m_D)_{ba},
	%
	\quad 	\lambda^{R,G}_{ai}= U^{\nu*}_{ai}m_a,
	\crn 	\lambda^{L,1}_{ai}=& -c_{\phi} t_{\beta}^{-1} \sum_{b=1}^3U^{\nu}_{(b+3)i} (m_D)_{ba},
	%
	\quad \lambda^{R,1}_{ai}=  \left( -U^{\nu*}_{ai}m_a t_{\beta} c_{\phi} + \frac{\sqrt{2} m_Ws_{\phi}}{g} \sum_{b=1}^{3}y^{\chi}_{ba}U^{\nu*}_{(b+6)i}\right) ,
	\crn 	\lambda^{L,2}_{ai}=& s_{\phi} t_{\beta}^{-1} \sum_{b=1}^3U^{\nu}_{(b+3)i} (m_D)_{ba},
	%
	\quad  	\lambda^{R,2}_{ai}=   U^{\nu*}_{ai}m_a t_{\beta} s_{\phi} + \frac{\sqrt{2} m_Wc_{\phi}}{g} \sum_{b=1}^{3}y^{\chi}_{ba}U^{\nu*}_{(b+6)i}.
\end{align}
leading to the following coupling of  scalars and leptons: 
\begin{align}
	\label{eq_lakc}
	\mathcal{L}^{hll}_Y=& h\sum_{a=1}^3 \frac{s_{\alpha}}{c_{\beta}}\times \frac{gm_{a}}{\sqrt{2} m_W}  -  \frac{gc_{\alpha}}{ 4m_Ws_{\beta}} \times h\sum_{i,j} \overline{ n_i} \left[ \lambda^h_{ij} P_L + \lambda^{h*}_{ij} P_R\right] n_j 
	\crn&- \frac{g}{\sqrt{2} m_W} \sum_{k=0}^2\sum_{a=1}^3\sum_{i} \left[ c_k\overline{ n_i} \left(\lambda^{L,k}_{ai} P_L + \lambda^{R,k}_{ai} P_R\right) e_a+\mathrm{h.c}.\right]  ,
\end{align}
where $c^\pm_0=G^\pm_W$ is the Goldstone boson absorbed by $W^\pm$. We note that $\lambda^h_{ij}=\lambda^h_{ji}$, i.e., the coupling $h\overline{n_i}n_j$ is symmetric following the rules defined in Ref. \cite{Dreiner:2008tw} and consistent with previous works \cite{Pilaftsis:1992st, Arganda:2014dta}.

In the numerical investigation, we use the parameter
\begin{equation}
	\label{eq_deltadef}
	\delta \equiv \frac{\pi}{2} +\alpha -\beta
\end{equation} 
 so that in the limit $\delta \to0$ leads to the consistency with the SM for  couplings of the SM-like Higgs boson $h$ appearing in the model under consideration, namely $s_{\beta -\alpha}=c_{\delta} \to 1$, $c_{\beta -\alpha}=s_{\delta} \to 0$,  $s_{\alpha}/s_{\beta}= -\left( c_{\delta} t_{\beta}^{-1}  -s_{\delta}\right) \to -t_{\beta}^{-1}$, and $s_{\alpha}/c_{\beta}= -\left( c_{\delta} -t_{\beta} s_{\delta}\right) \to -1$.  The Feynman rules for couplings of  the SM-like Higgs boson relating to the decay $h\to e_b^+e_a^-$ considered in this work are listed in Table \ref{t_hxy},
\begin{table}[h]
	\begin{tabular}{|c|c|c|c|}
		\hline
		Vertex & Coupling  & Vertex & Coupling\\
		\hline
		$h\overline{n_i} n_j$& $ -\frac{ig c_{\alpha}}{ 2m_Ws_{\beta}}\left[ \lambda^h_{ij} P_L + \lambda^{h*}_{ij} P_R\right] $&  $h\overline{e_a} e_a$& $\frac{i gm_{a}s_{\alpha}}{2  m_Wc_{\beta}}$\\
		\hline
		$c_k\overline{n_i} e_a$& $ \frac{-i g}{\sqrt{2} m_W}\left[ \lambda^{L, k}_{ai} P_L + \lambda^{R,k}_{ai} P_R\right] $&   $c^-_k\overline{e_a}n_i$&  $ \frac{-i g}{\sqrt{2} m_W} \left[ \lambda^{L, k*}_{ai} P_R + \lambda^{R, k*}_{ai} P_L\right] $ \\
		\hline
		$hW^+_{\mu}W^-_{\nu}$ & $ig^{\mu\nu}m_Wc_{\delta}$ & $hG^{+}G^{-}$& $ -i \frac{c_{\delta } m_h^2}{v_H^2}$\\
		\hline 
		$hG^+W^-_{\mu}$& $-\frac{ig}{2}c_{\delta} (p_0 -p_+)^{\mu}$ &$hG^-W^-_{\mu}$&  $\frac{ig}{2}c_{\delta} (p_0 -p_-)^{\mu}$\\
		\hline 
		$hc^+_1W^-_{\mu}$&  $ -\frac{ig}{2} s_{\delta} c_{\phi} (p_0 -p_+)^{\mu}$ &$hc^-_1W^+_{\mu}$&  $\frac{ig}{2}s_{\delta} c_{\phi} (p_0 -p_-)^{\mu}$\\
		\hline 
		$hc^+_2W^-_{\mu}$& $ \frac{ig}{2} s_{\delta} s_{\phi} (p_0 -p_+)^{\mu}$ &$hc^-_2W^+_{\mu}$& $-\frac{ig}{2} s_{\delta} s_{\phi} (p_0 -p_-)^{\mu}$ \\
		\hline 
		$hc^+_1c^-_1$& $i \lambda^{hcc}_{11}$ & $hc^+_2c^-_2$ & $i \lambda^{hcc}_{22}$\\
		\hline 
		$hc^+_1c^-_2$& $i \lambda^{hcc}_{12}$& $hc^+_2c^-_1$ & $i \lambda^{hcc}_{21}=i \lambda^{hcc}_{12}$\\
		\hline 
	\end{tabular}
	\caption{ Feynman rules for SM-like Higgs couplings giving  one-loop contributions to $h \to e_ae_b$ in the 2HDM$N_{L,R}$,  where $p_{0,\pm \mu}$ denotes the incoming momenta of  neutral and charged scalars.       \label{t_hxy}}
\end{table}
where
\begin{align}
	\label{eq_lahkl}
	\lambda^{hcc}_{11} =& \left[ \frac{s_{2\beta}}{2} c_{\phi}^2  ( s_{\alpha} s_{\beta} \lambda_1  -c_{\alpha} c_{\beta} \lambda_2)  -c_{\delta } c_{\phi }^2 \lambda_3 +\frac{s_{2\beta}}{2}  c_{\phi}^2 c_{(\beta +\alpha)}\ \lambda_{345}
	%
	+s_{\phi}^2 (c_{\beta}  s_{\alpha} \lambda_{1\chi} -c_{\alpha}  s_{\beta} \lambda_{2\chi})\right]v_H  
 \crn& + \frac{c_{\delta} s^2_{2\phi}  \left(m_{c_2}^2 - m_{c_1}^2 \right)}{2v_H},
	\crn \lambda^{hcc}_{22} =& \left[\frac{s_{2\beta}}{2}s_{\phi}^2 ( s_{\alpha} s_{\beta}\lambda_1  -c_{\alpha} c_{\beta} \lambda_2)   -c_{\delta } s_{\phi }^2 \lambda_3 
	%
%
	+ \frac{s_{2\beta}}{2} s_{\phi}^2  c_{(\beta +\alpha)} \lambda_{345}  +c_{\phi}^2 (c_{\beta} s_{\alpha} \lambda_{1\chi}   -c_{\alpha} s_{\beta}  \lambda_{2\chi}) \right]v_H 
\crn&  - \frac{c_{\delta} s^2_{2\phi}  \left(m_{c_2}^2 - m_{c_1}^2 \right)}{2v_H},
	\crn \lambda^{hcc}_{12} =&  \frac{s_{2\phi}}{2}   \left[  \frac{s_{2\beta}}{2} \left(- s_{\alpha} s_{\beta} \lambda_1  +c_{\alpha} c_{\beta} \lambda_2 \right) +  c_{\delta }  \lambda_3 - \frac{s_{2\beta}}{2} c_{(\beta +\alpha)}\lambda_{345} +  (c_{\beta}  s_{\alpha} \lambda_{1\chi} -c_{\alpha}  s_{\beta} \lambda_{2\chi})  \right]v_H  
	\crn&+ \frac{c_{\delta} s_{4\phi}  \left(m_{c_2}^2 - m_{c_1}^2 \right)}{4v_H},
\end{align}
where the coupling $\lambda$ appearing in  $\lambda^{hcc}_{ij}$ was replaced with the expression given in Eq. \eqref{eq_fparameter}.

The Feynman rules for couplings of  the gauge boson $Z$ relating to the decay $Z\to e_b^+e_a^-$  are listed in Table \ref{t_zxy}, 
\begin{table}[h]
	\begin{tabular}{|c|c|c|c|}
		\hline
		Vertex & Coupling  & Vertex & Coupling\\
		\hline
		$Z_{\mu}\overline{n_i} n_j$& $ \frac{ie}{2s_W c_W}   \gamma^{\mu}\left[q_{ij} P_L -q_{ji} P_R\right] $ &  $Z_{\mu}\overline{e_a} e_a$& $ie \gamma^{\mu}\left( t_L P_L +t_R  P_R\right)$\\
		\hline
		$Z_{\mu}W^+_{\nu}W^-_{\alpha}$ & $ - \frac{ie}{t_r} \Gamma^{\mu\nu \alpha}(p_0,p_+,p_-)$  & $Z_{\mu}G^{+}G^{-}$& $-iet_L(p_+ -p_-)^{\mu}$\\
		\hline 
		$Z_{\mu}G^+W^-_{\nu}$&$-iem_Wt_R g^{\mu\nu}$ &$Z_{\mu}G^-W^+_{\nu}$& $-iem_Wt_R g^{\mu\nu}$\\
		\hline 
		$Zc^+_1c^-_1$&$\frac{ie \left(c_{\phi }^2-2 s_W^2\right)}{2 c_W s_W} \left(p_+ -p_-\right)^{\mu}$& $Zc^+_2c^-_2$ & $\frac{ie \left(s_{\phi }^2-2 s_W^2\right)}{2 c_W s_W} \left(p_+ -p_-\right)^{\mu}$\\
		\hline 
		$Zc^+_1c^-_2$& $-\frac{ie s_{2\phi }}{4 c_W s_W} \left(p_+ -p_-\right)^{\mu}$ & $Zc^+_2c^-_1$ & $-\frac{ie s_{2\phi }}{4 c_W s_W} \left(p_+ -p_-\right)^{\mu}$\\
		\hline 
	\end{tabular}
	\caption{ Feynman rules for  $Z$ couplings giving  one-loop contributions to $Z \to e_b^+e_a^-$ in the 2HDM$N_{L,R}$, where $p_{0,\pm \mu}$ denotes the incoming momenta of  neutral and charged scalars.    \label{t_zxy}}
\end{table}
where $W^3_{\mu} =s_W A_{\mu} +c_W Z_{\mu}$ and $B_{\mu} = c_W A_{\mu} -s_W Z_{\mu}$, resulting the   couplings of $Z$ being consistent with those given in Eqs. \eqref{eq_Lnn}, \eqref{eq_LZWW},  and \eqref{eq_LGY} for the 2HDM without the singlet scalar $ \chi^\pm$. 

\subsection{Decays $h\to e_a e_b$}
\begin{figure}[ht]
	\centering 
	\includegraphics[width=8 cm]{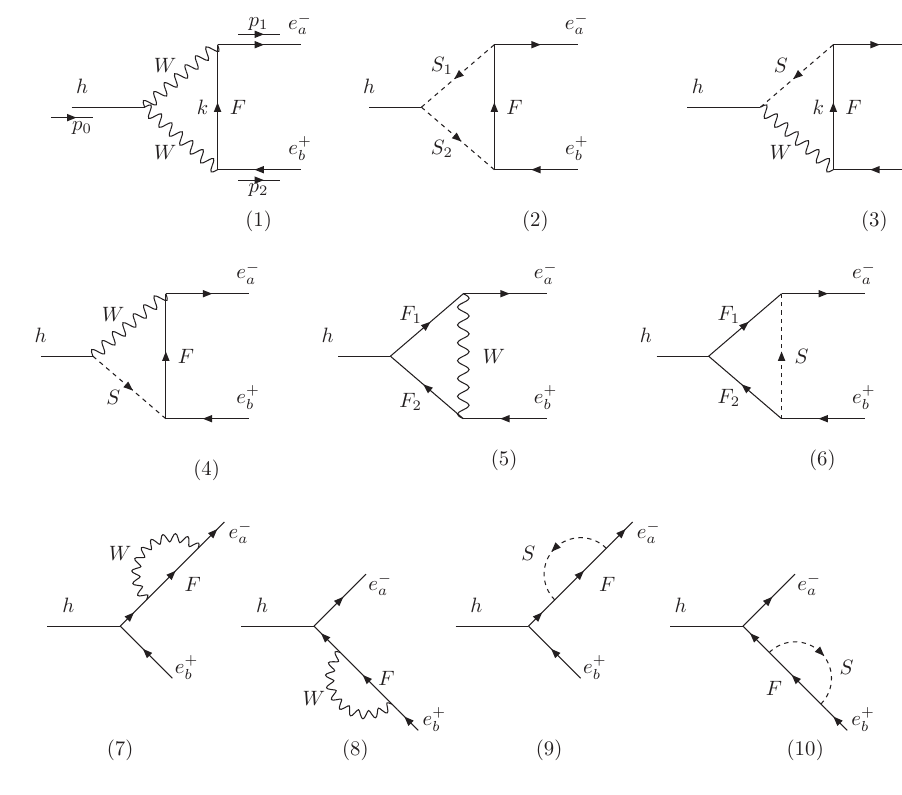}
	\caption{One-loop Feynman diagrams of the decays $h\to e_b^+e_a^- $ in the 2HDM$N_{L,R}$}
	\label{fig_heab2HDM}
\end{figure}

The effective Lagrangian of the LFV$h$ decay  $h \rightarrow e_a^{\pm}e_b^{\mp}$ is
$$ \mathcal{L}^{\mathrm{LFV}h}= h \left(\Delta^{(ab)}_{L} \overline{e_a}P_L e_b +\Delta^{(ab)}_{R} \overline{e_a}P_R e_b\right) + \mathrm{H.c.},$$
where scalar factors $\Delta_{(ab)L,R}$  arise from the loop contributions.  The one-loop diagrams of the decays $h\to e^-_ae^+_b$ in the unitary gauge  are shown in Fig. \ref{fig_heab2HDM}. 
The partial width of  the decay is   \cite{Arganda:2004bz}
\begin{equation}
	\Gamma (h \rightarrow e_ae_b)\equiv\Gamma (h\rightarrow e_a^{-} e_b^{+})+\Gamma (h \rightarrow e_a^{+} e_b^{-})
	\simeq   \fr{ m_{h}}{8\pi}\left(\vert \Delta^{(ab)}_L\vert^2+\vert \Delta^{(ab)}_R\vert^2\right), \label{eq_LFVwidth}
\end{equation}
with the condition  $m_{h}\gg m_{a,b}$  being  masses of  charged leptons. The on-shell conditions for external particles are $p^2_{1,2}=m_{e_a,e_b}^2$  and $ q^2 \equiv( p_1+p_2)^2=m^2_{h}$. The corresponding branching ratio is  Br$(h\rightarrow e_ae_b)= \Gamma (h\rightarrow e_ae_b)/\Gamma^{\mathrm{total}}_{h}$ where $\Gamma^{\mathrm{total}}_{h}\simeq 4.1\times 10^{-3}$ GeV \cite{LHCHiggsCrossSectionWorkingGroup:2016ypw}.    Formulas of  $\Delta_{(ab)L,R}$ are given as follows \begin{align}
	\label{eq_DeLR}
	\Delta^{(ab)}_{L,R} &=\Delta^{(ab)W}_{L,R} +\Delta^{(ab)c}_{L,R} +\Delta^{(ab)Wc}_{L,R}, 
\crn \Delta^{(ab) W}_{L,R} &=\Delta^{(ab) nWW}_{L,R} +\Delta^{(ab) Wnn}_{L,R} +\Delta^{nW}_{(ab)L,R},
\crn \Delta^{(ab)c}_{L,R} &= \sum_{k,l=1,2}\Delta^{(ab)nc_kc_l}_{L,R} +\sum_{k=1}^2 \left( \Delta^{(ab) c_knn}_{L,R} +\Delta^{(ab) nc_k}_{L,R}\right),
\crn \Delta^{(ab) Wc}_{L,R} &= \sum_{k=1}^2 \left( \Delta^{ (ab) c_kWn}_{L,R} +\Delta^{(ab) nWc_k}_{L,R}\right),
\end{align} 
where detailed analytic forms are given in appendix \ref{app_heab}. 

\subsection{Decays $Z\to e^+_be^-_a$}
In this model, the one-lopp contributions to the LFVZ decays $Z\to e^+_be^-_a$ consists of  the two parts originated from exchanges of $W^\pm$ and singly charged Higgs bosons $c^\pm_k$ with Feynman rules listed in Table \ref{t_zxy}. Analytic formulas of one-loop contributions from these Higgs were listed in appendix \ref{app_Zeab}, consistent with previous results \cite{Grimus:2002ux}. The partial decay widths are given in Eq. \eqref{eq_GAZeba}, where the form factors are 
\begin{align}
\label{eq_abZeba}
	\bar{a}_{L,R}&= a^W_{l,r} +a^{c^\pm}_{l,r},  \; \bar{b}_{L,R}= b^W_{l,r} +b^{c^\pm}_{l,r}, 
\crn 	\bar{a}^{W}_{l,r} & =\bar{a}^{nWW}_{l,r}+\bar{a}^{Wnn}_{l,r} +\bar{a}^{nW}_{l,r},
\crn 	\bar{b}^{W}_{l,r} & =\bar{b}^{nWW}_{l,r}+\bar{b}^{Wnn}_{l,r} +\bar{b}^{nW}_{l,r},
\crn 	\bar{a}^{c^\pm}_{l,r} & = \sum_{p,q=1}^2 \bar{a}^{nc_pc_q}_{l,r} + \sum_{k}^2 \left( \bar{a}^{c_knn}_{l,r} +\bar{a}^{nc_k}_{l,r}\right),
\crn 	\bar{b}^{c^\pm}_{l,r} & = \sum_{p,q=1}^2 \bar{b}^{nc_pc_q}_{l,r}  + \sum_{k}^2 \left( \bar{b}^{c_knn}_{l,r} +\bar{b}^{nc_k}_{l,r}\right).
\end{align}

\subsection{$(g-2)_{e,\mu}$ and decays $e_b\to e_a\gamma$}

The branching ratios of the cLFV decays are formulated as follows \cite{Lavoura:2003xp, Hue:2017lak, Crivellin:2018qmi}:
\begin{align}
	\label{eq_brebaga}
	\mathrm{Br}(e_b\to e_a\gamma)= \frac{48\pi^2}{G_F^2 m_b^2}\left( \left| c_{(ab)R}\right|^2 + \left| c_{(ba)R}\right|^2\right) \mathrm{Br}(e_b\to e_a \overline{\nu_a}\nu_b),
\end{align}
where $G_F=g^2/(4\sqrt{2}m_W^2)$, Br$(\mu\to e \overline{\nu_e}\nu_{\mu})\simeq 1$,  Br$(\tau\to e \overline{\nu_e} \nu_{\tau}) \simeq 0.1782$, Br$(\tau\to \mu \overline{\nu_\mu}\nu_{\tau})\simeq 0.1739$ \cite{ParticleDataGroup:2022pth},  and 
\begin{align}
	\label{eq_cabR}	
	c_{(ab)R}&=  c^{c^\pm}_{(ab)R} \left(h^\pm\right) + c^W_{(ab)R}, \quad   c_{(ba)R}= c_{(ab)R}[a \to b,\; b\to a], 
	\crn c^{c^\pm}_{(ab)R} &= \sum_{k=1}^2 c_{(ab)R} \left(c^\pm_k\right),
	\crn 	c_{(ab)R} \left(c^\pm_k\right)& = \frac{g^2e\;}{32 \pi^2 m^2_Wm^2_{c_k} }  
	\crn&\times\sum_{i=1}^{9}  \left[ \lambda^{L,k*}_{ia } \lambda^{R,k}_{ib }m_{n_i} f_{\Phi}(x_{i,k}) + \left( m_{b} \lambda^{L,k*}_{ia } \lambda^{L,k}_{ib } + m_{a} \lambda^{R,k*}_{ia } \lambda^{R,k}_{ib }\right)  \tilde{f}_{\Phi}(x_{i,k}) \right],
\crn 	c_{(ab)R}(W)& \simeq\frac{e G_Fm_{e_b}}{4\sqrt{2} \pi^2}   \left[  -\frac{5\delta_{ab}}{12} +  \left(U_{\mathrm{PMNS}}\hat{x}_{\nu}U_{\mathrm{PMNS}}^\dagger \right)_{ab} \times \left(\tilde{f}_V \left(x_W\right)+ \frac{5}{12}\right) \right]
\end{align}
with $x_{i,k}\equiv m^2_{n_i}/m^2_{c_k}$,   $x_W=M_0^2/m_W^2$, and \cite{Crivellin:2018qmi}
\begin{align}
	\label{eq_fphiX}	
	f_\Phi (x)&= 2\tilde{g}_\Phi(x)=\frac{x^2-1 -2x\ln x}{4(x-1)^3},\crn 
	g_\Phi&=\frac{x-1 -\ln x}{2(x-1)^2}, \crn 
	\tilde{f}_\Phi(x)&= \frac{2x^3 +3x^2 -6x +1 -6x^2 \ln x}{24(x-1)^4},
\crn 	\tilde{f}_V(x) &= \frac{-4x^4 +49x^3 -78 x^2 +43x -10 -18x^3\ln x}{24(x-1)^4}.
\end{align}
The one-loop contributions from the singly charged Higgs boson $c^\pm_{1,2}$ and $W^\pm$ exchanges to $a_{e_a}$ are: 
\begin{align}
	\label{eq_Hpm1}
	a^{c^\pm}_{e_a}&=-\frac{4m_{a}}{e} \left( \sum_{k=1}^2 \mathrm{Re}[c_{(aa)R}(c^\pm_k)] +\Delta 	c_{(aa)R}(W)\right),
\end{align}
  where $\Delta 	c_{(aa)R}(W)=\left(U_{\mathrm{PMNS}}\hat{x}_{\nu}U_{\mathrm{PMNS}}^\dagger \right)_{aa} \times \left(\tilde{f}_V \left(x_W\right)+ \frac{5}{12}\right)$ is the deviation between the 2HDM$N_{L,R}$ and the SM.

Up to the order $\mathcal{O}(R^2)$ of the neutrino mixing matrix, the non-zero one-loop contributions relating to $c^\pm_{1,2}$ were determined precisely, see Refs. \cite{Hue:2021xzl, Hue:2021zyw}.  Accordingly, the main contribution to $a_{e_a}(c^\pm)$  is  
\begin{equation}\label{eq_aea0}
	a_{e_a,0}(c^\pm)=	\frac{G_Fm^2_{a}}{\sqrt{2}\pi^2}
	\times \mathrm{Re}\left\{  \left[ \frac{vt_{\beta}^{-1}c_{\alpha}s_{\alpha}}{\sqrt{2}m_{a}}U_{\mathrm{PMNS}} \hat{x}_{\nu}^{1/2}  y^{\chi}\right]_{aa} \left[ x_1f_{\Phi}(x_1) - x_2f_{\Phi}(x_2)\right]
	\right\},
\end{equation}
where $x_k=M_0^2/m^2_{c_k}$.  In numerical calculation, the following diagonal form of  $c_{(ab)R,0}$   will be chosen for  discussing the Yukawa coupling matrix $ y^{\chi}$ at the beginning 
\begin{equation}\label{eq_cab0}
	c_{(ab)R,0}\varpropto\;   \left[ U_{\mathrm{PMNS}} \hat{x}_{\nu}^{1/2}  y^{\chi} \right]_{ab} \varpropto\;  \delta_{ab}.
\end{equation}
The non-zero values of  $c_{(ab)R}$ and $c_{(ba)R}$ with $b\neq a$   may  give large contributions  to the  cLFV rates, see a detailed discussion in Ref. \cite{Thao:2023gvs}. 
Correspondingly,   the formula of  	$a_{e_a,0}$ is proportional to a diagonal matrix $y^d$ satisfying: 
\begin{align} \label{eq_defYd}
	&U_{\mathrm{PMNS}} \times \mathrm{diag}\left( \frac{m_{n_1}}{m_{n_3}},\; \frac{m_{n_2}}{m_{n_3}},\;1\right)^{1/2}  y^{\chi}=	 y^d  \equiv \mathrm{diag} \left(y^d_{11}, \; y^d_{22}, y^d_{33}\right), 
\end{align}
where $m_{n_3}>m_{n_2}>m_{n_1}$ corresponding to the normal order of the neutrino oscillation data will be chosen in the numerical investigation. Then,   the  main contributions from charged Higgs bosons to $a_{e_a}$ was shown to be \cite{Hue:2021xzl} 
\begin{equation}\label{eq_aea01}
	a_{e_a,0}=	\frac{G_Fm^2_{a} \sqrt{x_0}}{\sqrt{2}\pi^2}
	\times \mathrm{Re}\left[ \frac{vt_{\beta}^{-1}c_{\alpha}s_{\alpha}}{\sqrt{2}m_{a}}y^d\right]_{aa} \left[ x_1f_{\Phi}(x_1) - x_2f_{\Phi}(x_2)\right], 
\end{equation}
where $x_k\equiv M_0^2/m^2_{c_k}$ ($k=1,2$), and 
\begin{equation}\label{eq_x0}
	x_0\equiv \frac{m_{n_3}}{\mu_0}.
\end{equation}
Note that $	c_{(ab)R,0}$ vanishes with $a\neq b$, therefore do not affect the Br$(e_b\to e_a \gamma)$.  The values of entries $y^{\chi}$ will be scanned around the diagonal forms of $y^d$ to guarantee the cLFV constraints of experiments.

\subsection{Numerical discussion}
We will use the best-fit values  of the neutrino osculation data \cite{ParticleDataGroup:2022pth} corresponding to  the normal order (NO) scheme  with $m_{n_1}<m_{n_2}<m_{n_3}$, namely 
\begin{align}
	\label{eq_d2mijNO}
	&s^2_{12}=0.318,\;   s^2_{23}= 0.574,\; s^2_{13}= 0.022 ,\;  \delta= 194 \;[\mathrm{Deg}] , 
	\crn &\Delta m^2_{21}=7.5 \times 10^{-5} [\mathrm{eV}^2], \;
	\Delta m^2_{32}=2.47\times 10^{-3} [\mathrm{eV}^2].
\end{align}
The active mixing matrix and neutrino masses are determined  as follows 
\begin{align}
	\label{eq_NO1}
	\hat{m}_{\nu}&= \left( \hat{m}^2_{\nu}\right)^{1/2}= \mathrm{diag} \left( m_{n_1}, \; \sqrt{m^2_{n_1} +\Delta m^2_{21}},\; \sqrt{m^2_{n_1} +\Delta m^2_{21} +\Delta m^2_{32}} \right),
	\crn U_{\mathrm{PMNS}}&=\left(
	\begin{array}{ccc}
		c_{12} c_{13} & c_{13} s_{12} & s_{13} e^{-i \delta } \\
		-c_{23} s_{12}-c_{12} s_{13} s_{23} e^{i \delta } & c_{12}c_{23}-s_{12} s_{13} s_{23} e^{i \delta } & c_{13} s_{23} \\
		s_{12} s_{23}-c_{12} c_{23} s_{13} e^{i \delta } & -c_{23} s_{12} e^{i \delta } s_{13}-c_{12} s_{23} & c_{13} c_{23} \\
	\end{array}
	\right).
\end{align}
We fix $m_{n_1}=0.01$ eV for simplicity in numerical investigation. This choice of active neutrino masses  satisfies the constraint  from Plank2018 \cite{Planck:2018vyg}, $	\sum_{i=a}^{3}m_{n_a}\leq 0.12\; \mathrm{eV}$.

The non-unitary of the active neutrino mixing matrix $\left(	I_3-\frac{1}{2}RR^{\dagger} \right) U_{\mathrm{PMNS}}$ is constrained by other phenomenology such as electroweak precision~\cite{ Agostinho:2017wfs, Coutinho:2019aiy}, leading to a very strict constraint of $\eta\equiv	\frac{1}{2}\left| RR^{\dagger}\right| \varpropto\;   \hat{x}_\nu \varpropto x_0$ in the ISS framework \cite{Mondal:2021vou, Biggio:2019eeo, Escribano:2021css}.  

The  well-known numerical parameters  are \cite{ParticleDataGroup:2022pth} 
\begin{align}
	\label{eq_ex}
	g &=0.652,\; G_F=1.1664\times 10^{-5}\; \mathrm{GeV},\; s^2_{W}=0.231,\; \alpha_e=1/137,\; e =\sqrt{4\pi \alpha_e},
\crn m_W&=80.377 \; \mathrm{GeV}, m_Z= 91.1876\; \mathrm{GeV},\; m_h=125.25  \; \mathrm{GeV},
\crn \Gamma_h&=4.07\times 10^{-3}\;\mathrm{GeV},\;  \Gamma_Z= 2.4955\; \mathrm{GeV},\
 \crn
	m_e&=5\times 10^{-4} \;\mathrm{GeV},\; m_{\mu}=0.105 \;\mathrm{GeV}, \; m_{\tau}=1.776  \; \mathrm{GeV}. 
\end{align}
For the free parameters of the 2HD$N_{L,R}$ model, the numerical scanning ranges are 
\begin{align} \label{eq_scanP0}
	&M_0,  m_{c_{1,2}} \in  \left[ 1,\;10\right] \; \; \mathrm{TeV};\;\lambda_1, |\lambda_4|, |\lambda_5|\in \left[0,\;4\pi\right] ; 
	\crn& t_{\beta} \in \left[ 5,30\right] ; \; x_0 \in \left[ 10^{-6},5\times 10^{-4}\right]; \;\phi \in  \left[0,\pi\right]; \;|y^d_{ab}|\leq 3.5 \; \forall a,b=1,2,3.  
\end{align}
 The matching condition  with  SM requires small $|s_\delta|$. Therefore, we fix $s_{\delta}=0$ in the numerical investigation.  
The Higgs self-couplings and Higgs masses appearing in Eq. \eqref{eq_lahkl} are $\lambda_1$, $\lambda_4$ , $\lambda_5$,   $\lambda_{1\chi}$, and $\lambda_{2\chi}$, in which some of them are  given in Eq. \eqref{eq_fparameter}. The related independent parameters   are chosen as   $\lambda_{1\chi}$, and $\lambda_{2\chi}$, apart from those given in Eq. \eqref{eq_scanP0}.  For simplicity, we will fix $\lambda_{1\chi}=\lambda_{2\chi}=0$.  In the numerical investigation we take lower bounds that $m_H,m_A\geq 500$ GeV.  The couplings  $hc^+_kc^-_l$ given in \eqref{eq_lahkl} are determined after confirming numerically that   all Higgs couplings must satisfy the two conditions of bounded from below and unitarity  limits mentioned in appendix \ref{app_2HDMnu}. All Yukawa couplings of the matrices $y^\chi$ and $f$ must satisfy the pertubative limits, therefore we choose the safe upped bounds that $|y^\chi_{ab}|,|f_{ab}| \leq3<\sqrt{4\pi}$. 

In the following discussion on numerical results, we just collect allowed points in the scanning ranges given in Eq. \eqref{eq_scanP0},  in which they satisfy all experimental LFV constraints listed in Eqs. \eqref{eq_ebagaex},  \eqref{eq_LFVHexp}, and \eqref{eq_LFVZexp}. In addition the $(g-2)_{e,\mu}$ data is chosen at $1\sigma$ deviations derived from two Eqs. \eqref{eq_damu} and \eqref{eq_dae}. 

\begin{figure}[ht]
	\centering
 \begin{tabular}{ccc}
		\includegraphics[width=5.5cm]{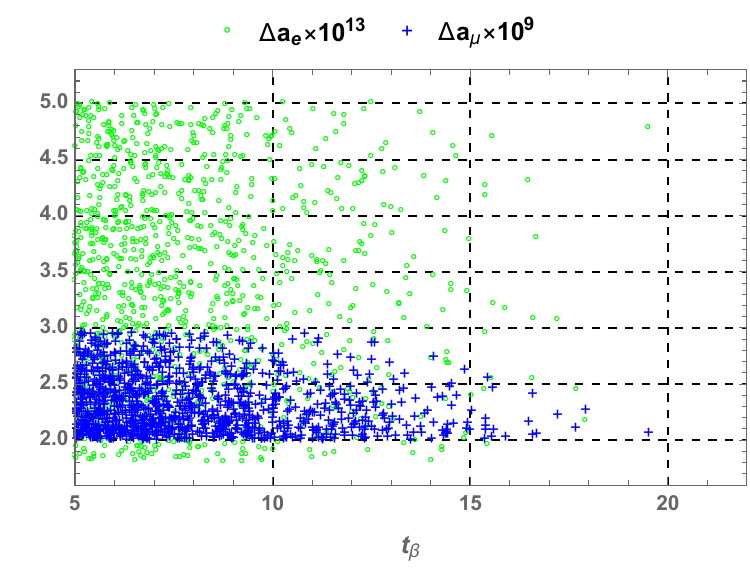}
		&
\includegraphics[width=5.5cm]{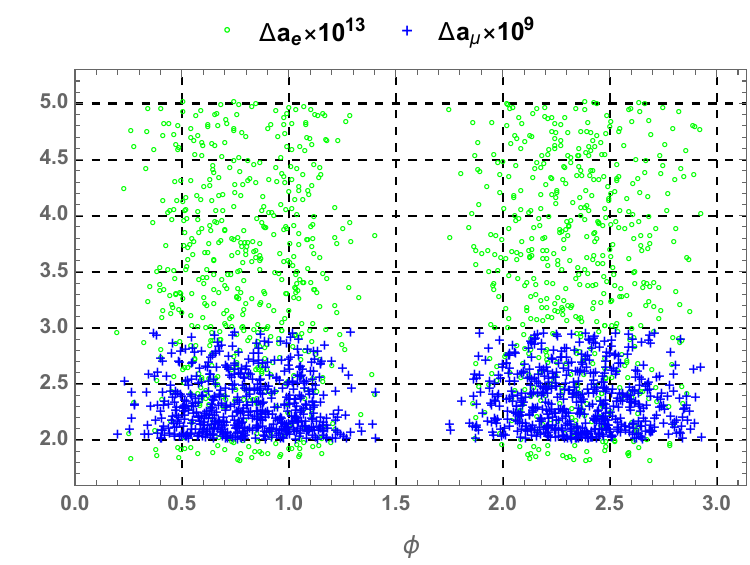} 
&
	\includegraphics[width=5.5cm]{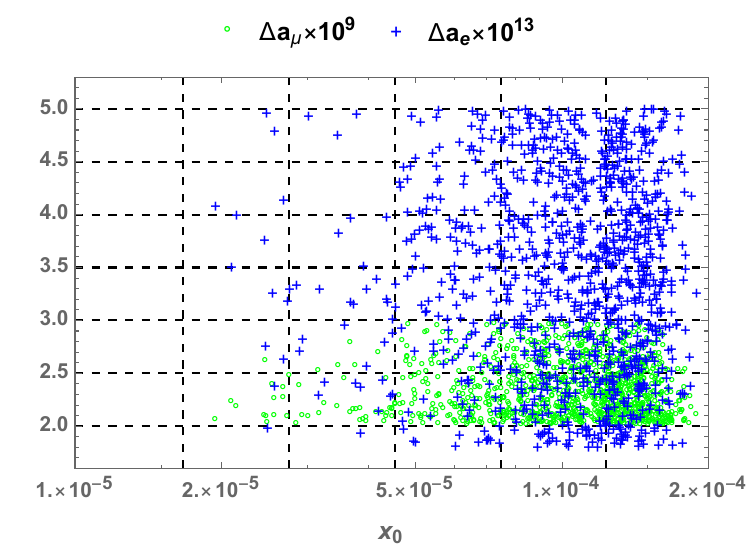}\\ 
	\end{tabular}
	\caption{ The  correlations between $\Delta a_{e,\mu}$  and  $t_{\beta}$ (left), $\phi$ (center), and $x_0$ (right) in the limits of  $\delta=0$ and  $y^d_{11},y^d_{22}\neq0$.}\label{fig_Xaemu}
\end{figure}

Firstly, we focus on the simplest case of $\delta=0$, and only $y^d_{11}, y^d_{22} \neq 0$, while $y^d_{ab}=0$ for $(ab)=(33)$ and $a\neq b$. The correlations between $\Delta a_{e,\mu}$ with $t_\beta$, $\phi$, and $x_0$ are shown in Fig. \ref{fig_Xaemu}.
The allowed ranges of these three parameters are: $5\leq t_{\beta}<20$, $0.2<\phi<2.93 $ and $ 10^{-5}<x_0<4\times 10^{-4}$.  The imprort property is that $s_{\phi}c_{\phi}$ always non-zero. 

\begin{figure}[ht]
	\centering
	\begin{tabular}{cc}
		\includegraphics[width=7.5cm]{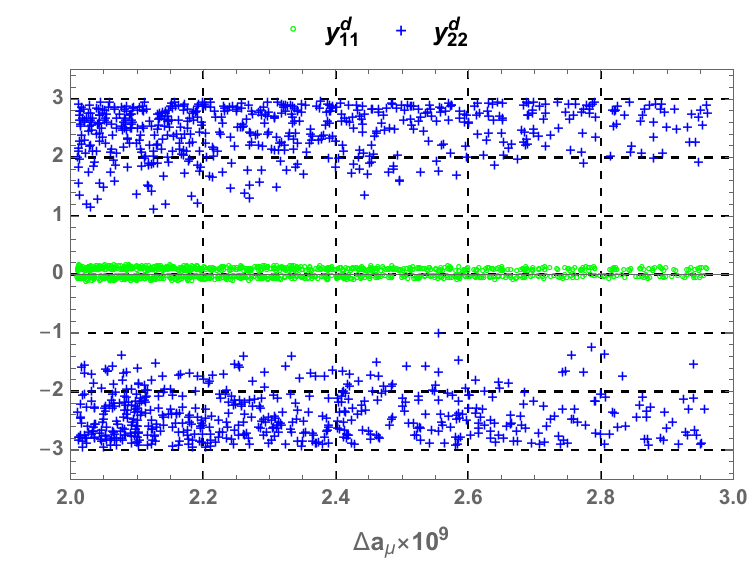}
		&
		\includegraphics[width=7.5cm]{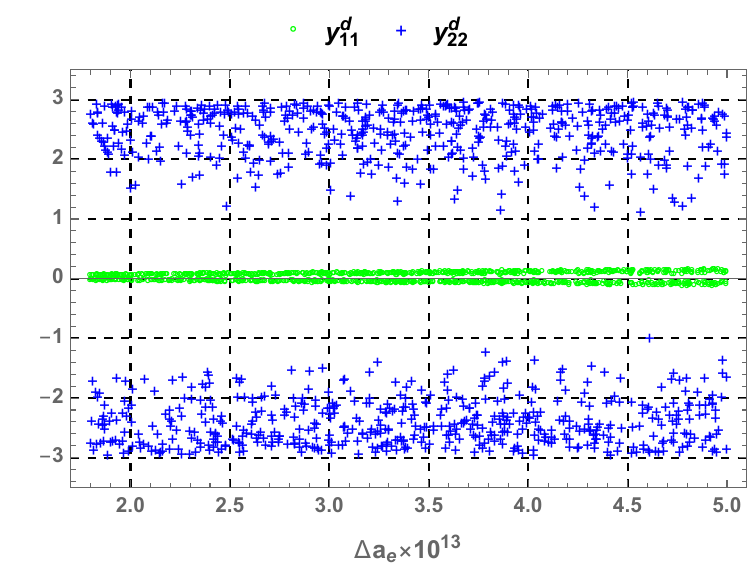} 
	\\ 
	\end{tabular}
	\caption{ The  correlations between  $\Delta a_{e,\mu}$  vs  $y^d_{11}$  and $y^d_{22}$  in the limits of  $\delta=0$ and  $y^d_{11},y^d_{22}\neq0$.}\label{fig_aemuydii}
\end{figure}

The correlations between $\Delta a_{e,\mu}$ with $y^d_{11}$, and $y^d_{22}$ are shown in Fig. \ref{fig_aemuydii}.
The allowed regions are $0.03\leq |y^d_{11}|\leq 0.15$ and $1\leq |y^d_{22}|\leq3$. 

\begin{figure}[ht]
	\centering
	\begin{tabular}{cc}
		\includegraphics[width=7.5cm]{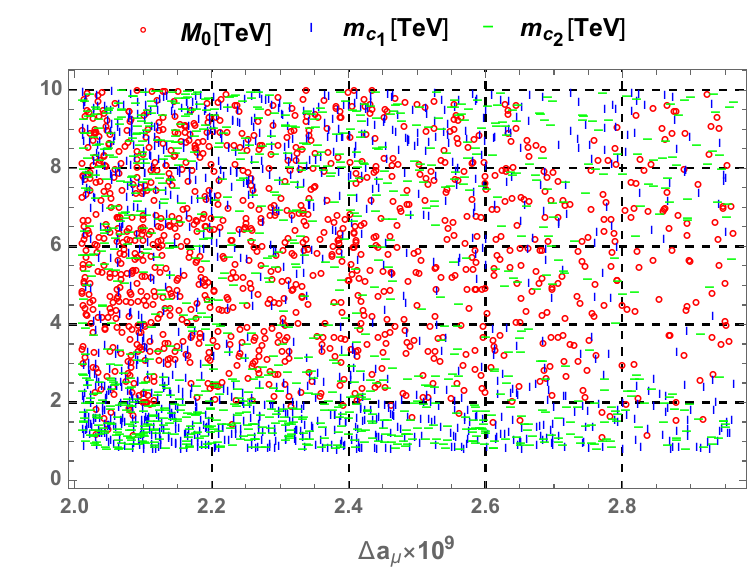}
		&
		\includegraphics[width=7.5cm]{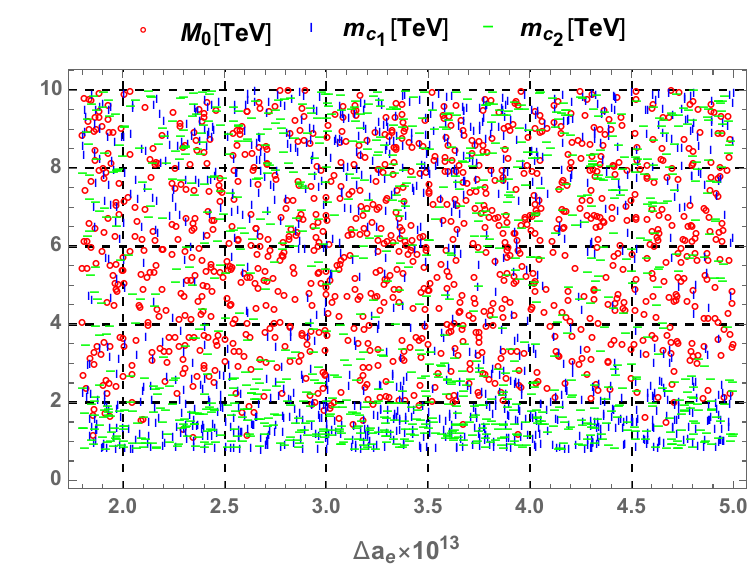} 
		\\ 
	\end{tabular}
	\caption{ The  correlations between   $\Delta a_{e,\mu}$ vs neutrinos and charged Higgs masses in the limits of  $\delta=0$ and  $y^d_{11},y^d_{22}\neq0$.}\label{fig_aemumX}
\end{figure}

The correlations between $\Delta a_{e,\mu}$ with neutrinos and charged Higgs masses  are shown in Fig. \ref{fig_aemumX}.
There are not constaints of  $M_0$ and charged Higgs boson masses in the scanning regions given in Eq. \eqref{eq_scanP0}.  

\begin{figure}[ht]
	\centering
	\begin{tabular}{cc}
		\includegraphics[width=7.5cm]{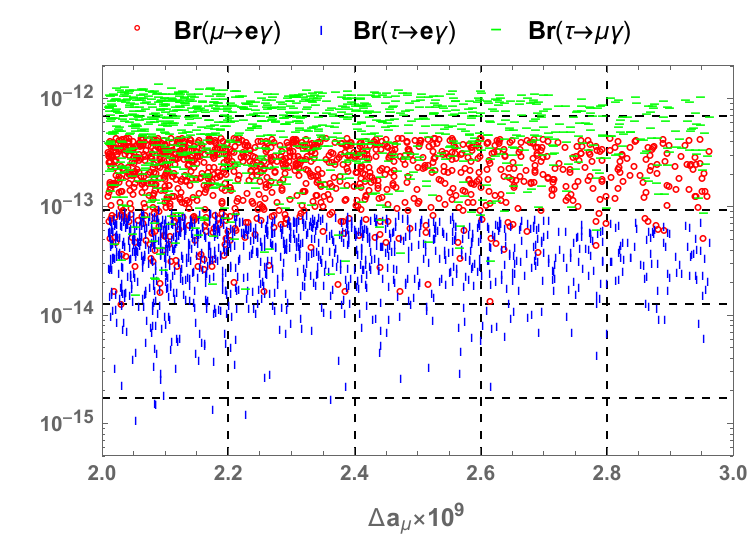}
		&
		\includegraphics[width=7.5cm]{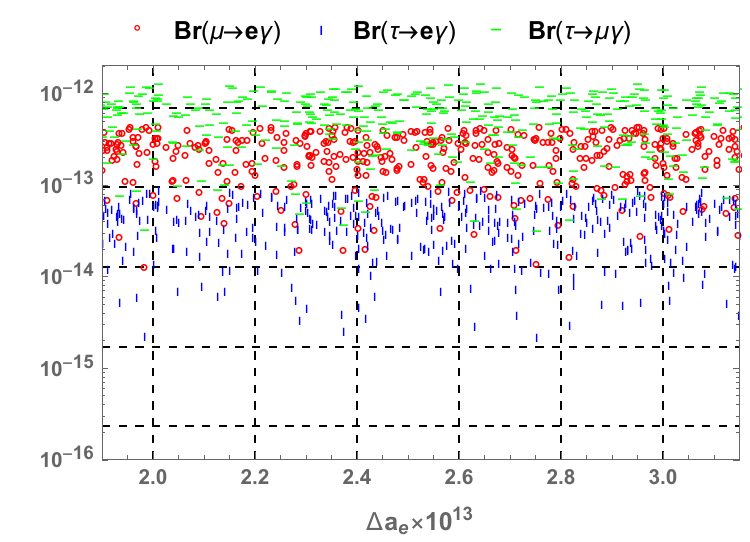} 
		\\ 
	\end{tabular}
	\caption{ The  correlations between   $\Delta a_{e,\mu}$ vs  cLFV decays Br$(e_b\to e_a \gamma)$  in the limits of  $\delta=0$ and  $y^d_{11},y^d_{22}\neq0$.}\label{fig_aemuebaga}
\end{figure}

The correlations between $\Delta a_{e,\mu}$ with cLFV decays Br$(e_b\to e_a \gamma)$ are shown in Fig. \ref{fig_aemuebaga}.
The decay $\mu \to e\gamma$ can reach the experimental bound, but two  decay modes  are much smaller than the near future experimental sensitivities, namely  Fig. \ref{fig_aemuebaga} shows  two upper values of Br$(\tau \to e \gamma) < 10^{-13}$ and  Br$(\tau \to \mu \gamma) <1.5\times 10^{-12}$.

\begin{figure}[ht] 
	\centering
	\begin{tabular}{cc}
		\includegraphics[width=7.5cm]{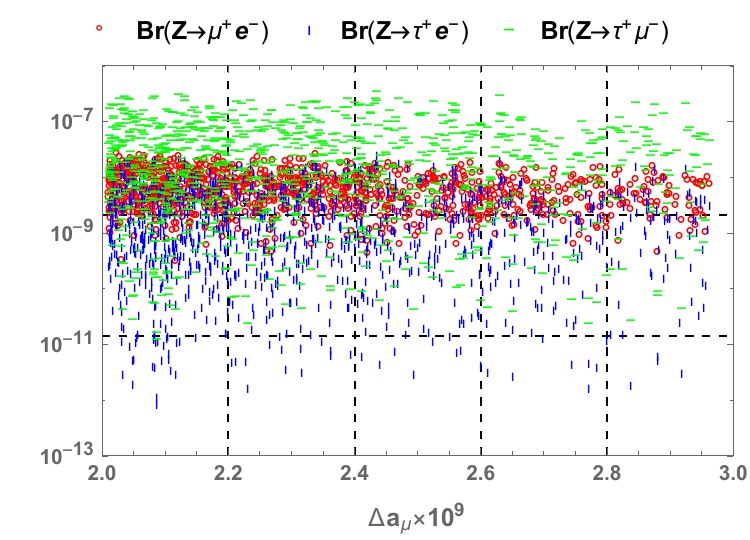}
		&
		\includegraphics[width=7.5cm]{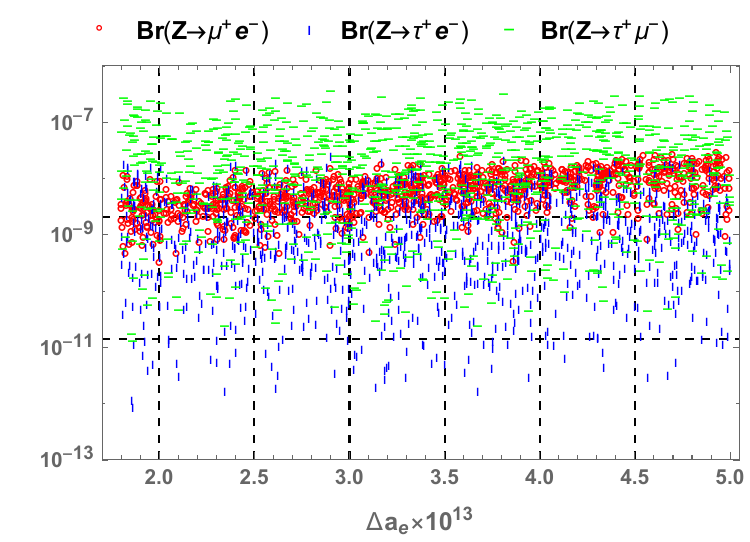} 
		\\ 
	\end{tabular}
	\caption{ The  correlations between   $\Delta a_{e,\mu}$ vs Br$(Z\to e_b^+ e_a^-)$ in the limits of  $\delta=0$ and  $y^d_{11},y^d_{22}\neq0$.}\label{fig_aemZba}
\end{figure}

The correlations between $\Delta a_{e,\mu}$ with LFV decays Br$(Z\to e_b^+ e_a^-)$ are shown in Fig. \ref{fig_aemZba}.
These three decays can reach the future sensitivities. The upper bound are found as follows: max[Br$(Z\to \mu^+e^-)]\simeq  2.75\times 10^{-8}$, max[Br$(Z\to \tau^+ e^-)]\simeq  2.43\times 10^{-8}$, and max[Br$(Z\to \tau^+ \mu^-)] \simeq  3.53 \times 10^{-7}$. Although these three decays rates are smaller than the recent experimental upper bounds, they all reach the near future sensitivities of experiments.  This property in the 2HDM$N_{L,R}$ is different from the model discussed in ref. \cite{Jurciukonis:2021izn}, where all LFV$Z$ decay rates are predicted to be suppressed. Also,  Br$(Z\to \mu^\pm e^\mp)<10^{-9}$ was confirmed in Ref. \cite{Abada:2022asx}.

\begin{figure}[ht]
	\centering
	\begin{tabular}{cc}
		\includegraphics[width=7.5cm]{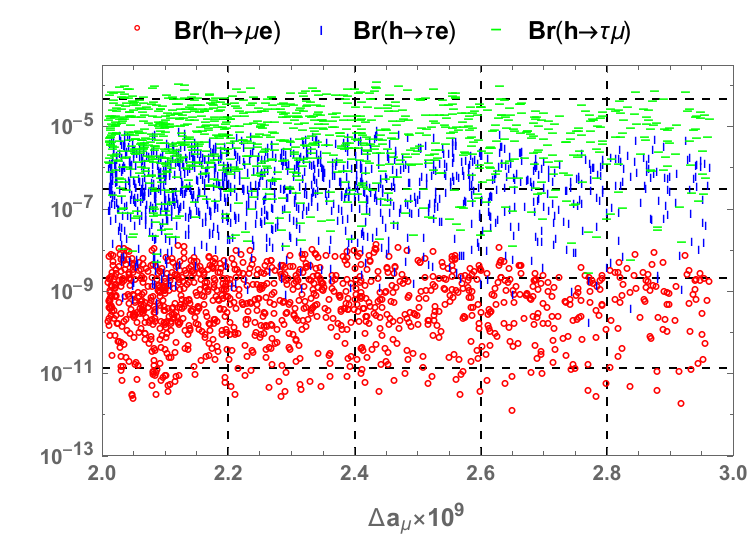}
		&
		\includegraphics[width=7.5cm]{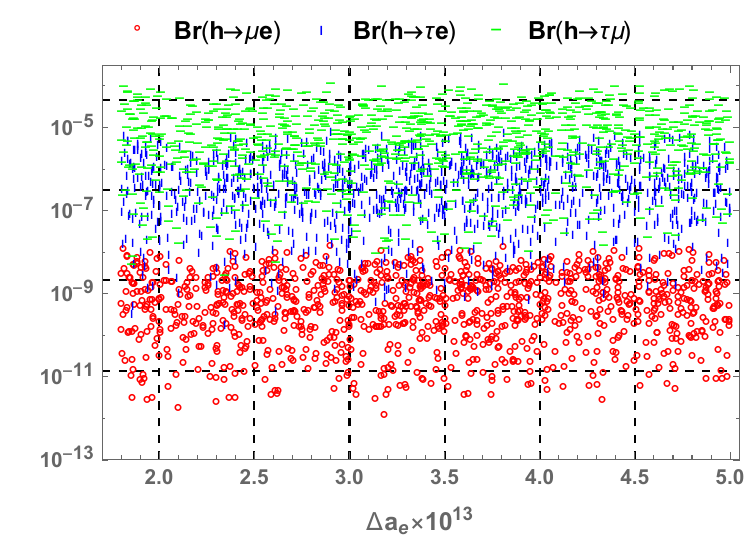} 
		\\ 
	\end{tabular}
	\caption{ The  correlations between   $\Delta a_{e,\mu}$ vs Br$(h\to e_b e_a)$ in the limits of  $\delta=0$ and  $y^d_{11},y^d_{22}\neq0$.}\label{fig_aemhba}
\end{figure}

The correlations between $\Delta a_{e,\mu}$ with LFV decays Br$(h\to e_b e_a)$ are shown in Fig. \ref{fig_aemhba}.
The upper bounds of these decays are Br$(h\to \mu^+e^-)< 1.4\times 10^{-8}$, Br$(h\to \tau^+ e^-)< 8\times 10^{-6}$, and Br$(h\to \tau^+ \mu^-)< 1.2\times 10^{-4}$. Only Br$(h\to \tau^+ \mu^-)$ can reach the near future experimental sensitivities. 

In the more general conditions of numerical investigations, the allowed regions of the parameters such as heavy neutrinos and charged Higgs boson masses, and $y^d_{11,22}$  do not change significantly. Therefore, we will not pay attention to them. The two entries $|y^d_{12}|, |y^d_{21}|<10^{-3}$  because of the strict constraint from Br$(e_b\to e_a\gamma)$. As a consequence, they  give suppressed contributions to the remaining LFV decay rates, therefore we will fix $y^d_{12}=y^d_{21}=0$ in the numerical investigation. In addition, the allowed regions prefer the small $|y_{31}|<0.05$,  we therefore consider the following constraints: 
\begin{align}
\label{eq_datall1}
y_{31}=0,\; |y_{33}|<1,\; |y_{13}|, |y_{23}|, |y_{32}|\leq 0.5.
\end{align}
The allowed ranges of $y^d_{11}, y^d_{22}$ do not change significantly with the case 1.  In contrast the upper bounds of the LFV decays enhance strongly, namely:
\begin{align}
\label{eq_LFVbr2}
\mathrm{Br}(\mu \to e\gamma) &\leq 4.2\times 10^{-13},\; \mathrm{Br}(\tau \to e\gamma)\leq 3.3\times 10^{-8},\; \mathrm{Br}(\tau \to \mu \gamma)\leq 4.4\times 10^{-8},
\crn 
\mathrm{Br}(Z \to \mu^+e^-) &\leq 7.9\times 10^{-8}, \; \mathrm{Br}(Z \to \tau^+e^-) \leq 1.9\times 10^{-6}, \; \mathrm{Br}(Z \to \tau^+\mu^-) \leq 6.5\times 10^{-6},
\crn 
\mathrm{Br}(h \to \mu e) &\leq 9.1 \times 10^{-9}, \; \mathrm{Br}(h \to \tau e) \leq 2\times 10^{-3}, \; \mathrm{Br}(h \to \tau \mu) \leq 1.4\times 10^{-3}. 
\end{align}
Therefore, five decays rates, including three cLFV decay $e_b\to e_a \gamma$, one $Z\to \tau^+\mu^-$, and two LFV$h$ decays $h\to \tau \mu,\tau e$ have upper bounds coinciding with recent experimental constraints. Only Br$(h\to \mu e)$ is much smaller than the near future experimental sensitivities.   In contrast, Br$(Z\to \mu^\pm e^\mp)$ can reach large values close to $10^{-7}$, which are different from previous discussions.  

\begin{figure}[ht]
	\centering
	\begin{tabular}{ccc}
		\includegraphics[width=5.5cm]{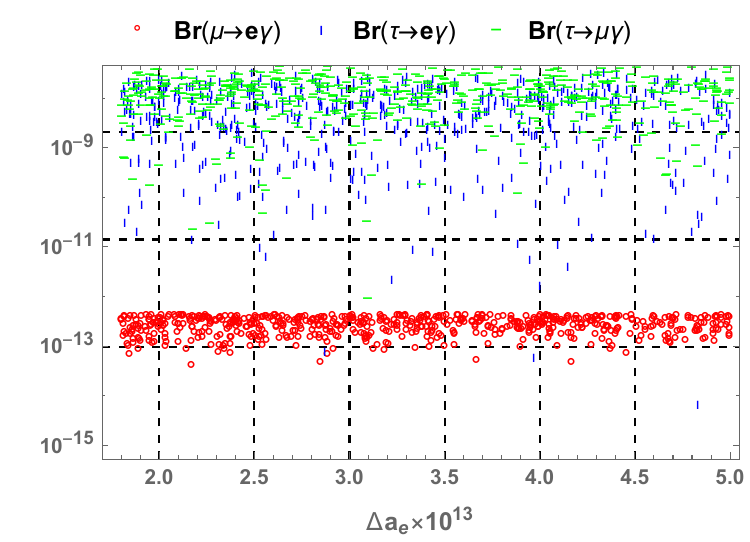}
		&
		\includegraphics[width=5.5cm]{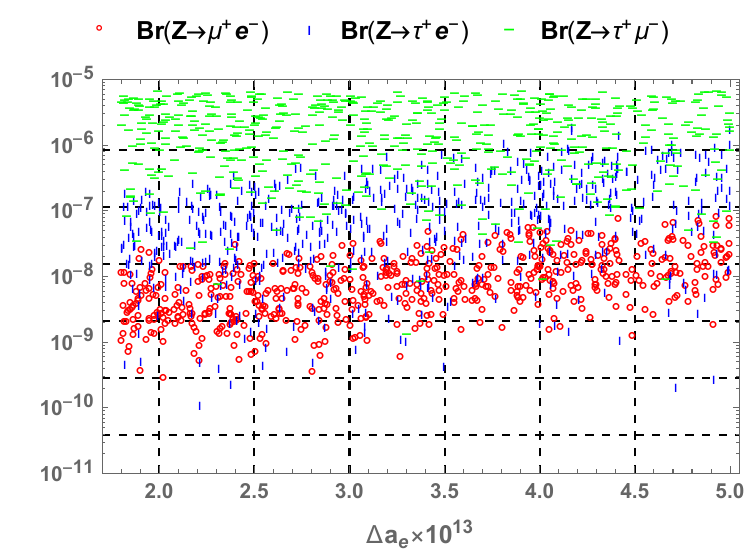} 
		& 
		\includegraphics[width=5.5cm]{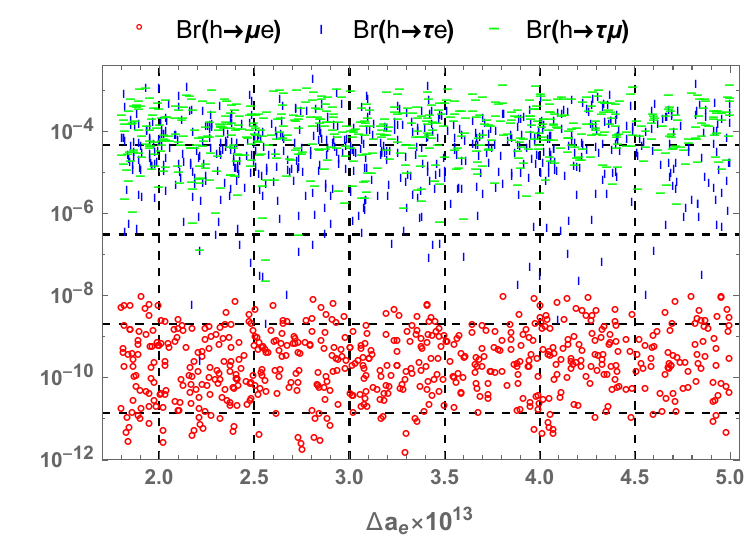} 	\\ 
	
	\end{tabular}
	\caption{ The  correlations between   $\Delta a_{e,\mu}$ vs LFV decays   in the limits given by Eq. \eqref{eq_datall1}.}\label{fig_aeLFVall}
\end{figure}

The correlations between $a_{e}$ with different LFV decays are shown in Fig. \ref{fig_aeLFVall}. 
It shows that all allowed values of $a_e$ support small LFV decays rates corresponding to the future sensitivities, hence the model will not be excluded if any of LFV decays are detected. All LFV decay rates depend weakly on  $a_{\mu}$, we therefore do not present here.  
 
\begin{figure}[ht]
	\centering
	\begin{tabular}{ccc}
		\includegraphics[width=5.5cm]{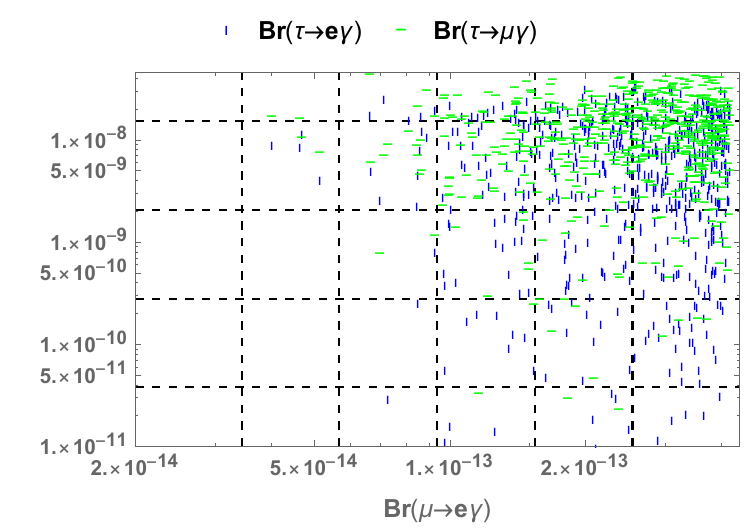}
		&
	\includegraphics[width=5.5cm]{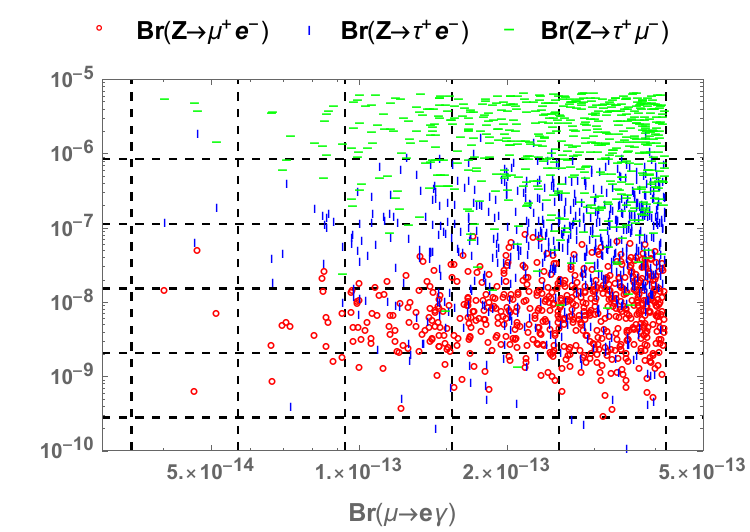} 
		& 
		\includegraphics[width=5.5cm]{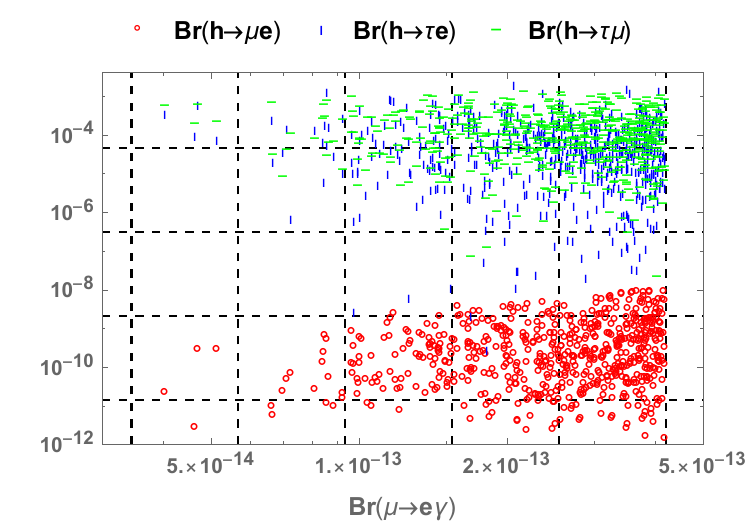} 	\\ 		
	\end{tabular}
	\caption{ The  correlations between   Br$(\mu \to e \gamma)$ vs LFV decays   in the limits given by Eq. \eqref{eq_datall1}.}\label{fig_ebaLFVall}
\end{figure}

 The dependence of the LFV decay rates vs Br$(e_b\to e_a \gamma)$ are given in Fig \ref{fig_ebaLFVall}.
Although, Br$(e_b\to e_a \gamma)$ is the most stringent from experiments, the future sensitivities of $\mathcal{O}(10^{-9})$ still support all other decays rates reaching the respective expected sensitivities, except  Br$(h\to \mu e)$, which is always invisible for future experimental searches.

\begin{figure}[ht]
	\centering
	\begin{tabular}{ccc}
		\includegraphics[width=7.5cm]{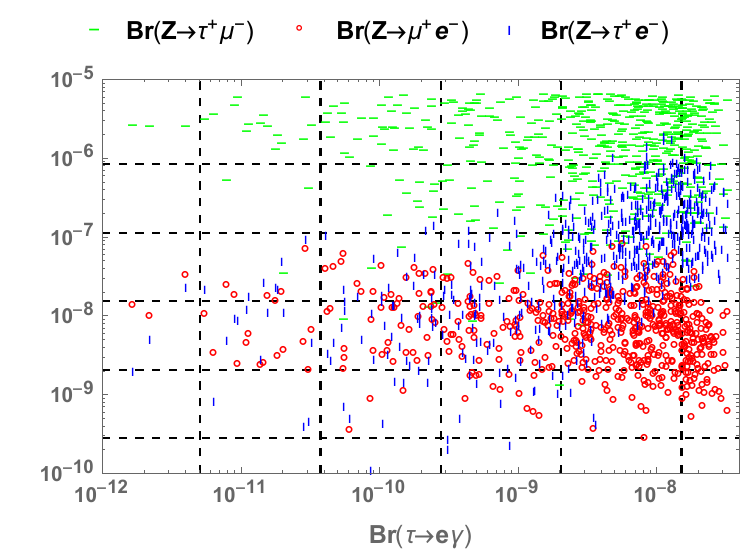}
		&
		\includegraphics[width=7.5cm]{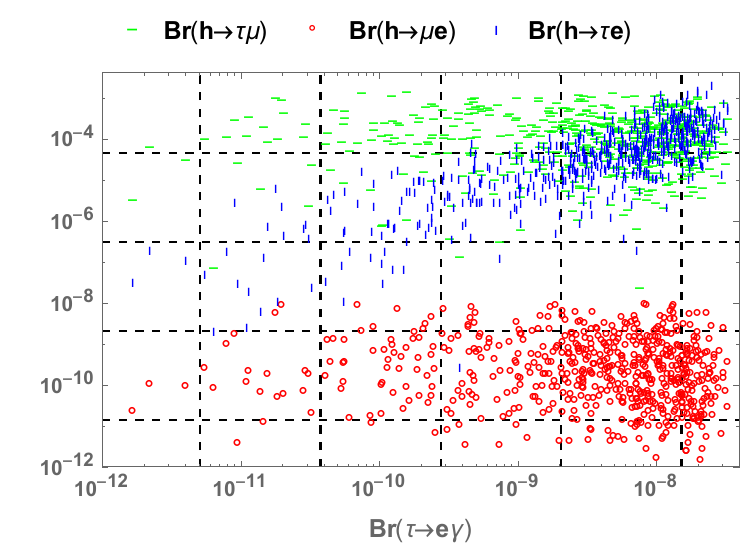} 
		\\ 		
	\end{tabular}
	\caption{ The  correlations between   Br$(\tau \to e \gamma)$ vs LFV decays   in the limits given by Eq. \eqref{eq_datall1}.}\label{fig_tegaall1}
\end{figure}

There are significant dependence between Br$(\tau \to e \gamma)$ and two decay rates Br$(h\to e_b e_a)$ and Br$(Z\to e_b^+ e_a^-)$, see  illustrations in Fig.\ref{fig_tegaall1}.
Here large Br$(\tau \to e\gamma)$ predicts large  Br$(Z \to \tau^+e^-)$ and  Br$(h \to \tau e)$. Therefore, if one of these decays are detected, there are some clues to predict the values of the two remaining ones. 

\section{Conclusions}   
We have explored the LFV decays in the allowed regions of the parameter space accommodating the $(g-2)_{\mu,e}$ in the 2HDM$N_{L,R}$ framework. We obtained some following interesting results that distinguish the 2HDM$N_{L,R}$ from other available BSM.  Firstly, there exist allowed regions predicting the large values of  Br$(e_b\to e_a \gamma)$, Br$(h\to \tau \mu,\tau e)$, and Br$(Z \to \tau^\pm \mu^\mp, \tau^\pm e^\mp)$ close to the recent experimental constraints. Furthermore, significant correlations among Br$(\tau \to e \gamma)$, Br$(Z \to \tau^\pm e^\mp)$, and Br$(h \to \tau e)$ were found, which will be interesting for testing the model if at least one of these decay channels is detected experimentally.  Secondly, the 2HDM$N_{L,R}$  model predicts large max[Br$(Z\to \mu^\pm e^\mp)]\simeq 7.9\times 10^{-8}$, but suppressed Br$(h\to \mu e)<10^{-8}$ which is invisible  for the near future experimental sensitivity. This is in contrast with the prediction from the 2HDM model discussed in Ref. \cite{Jurciukonis:2021izn}, which predicts large LFV$h$ but very small LFV$Z$ decay rates.  On the other hand, some BSM discussed in Ref. \cite{Abada:2022asx} for example, support large LFV$Z$ but small LFV$h$ decay rates. Therefore, if future experiments confirm the existence of any LFV signals, they will be used to determine the reality of many available models or constrain the model parameter space we have studied in this work.

\section*{Acknowledgments}
We are grateful Dr. Navin McGinnis for the interesting comments. This research is funded by Vietnam National University HoChiMinh City (VNU-HCM) under grant number “C2022-16-06”.

\appendix
\section{ \label{app_PVLT} PV functions for one loop contributions defined by LoopTools}
\subsection{General notations}
The PV-functions used here were listed in ref.~\cite{Hue:2017lak}, namely
\begin{align}
	&A_{0}(m)=\frac{(2\pi\mu)^{4-d}}{i\pi^2}\int \frac{d^d k}{k^2 -m^2 +i\delta},\crn
	&B_{\{0,\mu ,\mu \nu \}}(p^2_i,M^2_0,M^2_i) =\frac{(2\pi\mu)^{4-d}}{i\pi^2}\int \frac{d^d
		k \times\{1, k_{\mu}, k_{\mu}k_{\nu}\}}{D_0 D_i}, \; i=1,2, \crn
	&C_{0,\mu,\mu\nu}=\frac{(2\pi\mu)^{4-d}}{i\pi^2}\int \frac{d^d k \{ 1,k_\mu,k_\mu k_\nu\}}{D_0 D_1 D_2}, 
	\label{ABC_def}
\end{align} 
where $D_0\equiv k^2-M_0^2  +i\delta$, $D_1\equiv (k -p_1)^2-M_1^2 + i\delta$, $D_2\equiv (k +p_2)^2 -M_2^2 + i\delta$, $C_{0,\mu,\mu\nu}=C_{0,\mu,\mu\nu}(p_1^2, (p_1  +p_2)^2, p_2^2; M_0^2, M_1^2, M_2^2)$,  $\delta$ is an real positive quantity, and $\mu$ is an arbitrary mass parameter 
introduced via dimensional regularization \cite{tHooft:1972tcz}. The scalar PV-functions are defined as follows:
\begin{align}
&B_{\mu}(p^2_1,M^2_0,m^2_a) =(-p_{1\mu})B_1(p^2_1,M^2_0,M^2_1)\equiv (-p_{1\mu})B^{(1)}_1,
\crn& B_{\mu}(p^2_2,M^2_0,M^2_2) = p_{2\mu}B_1(p^2_2,M^2_0,M^2_2) \equiv p_{2\mu}B^{(2)}_1,\crn
&B_{\mu \nu}(q^2_i,M^2_0,M^2_i) =B_{00}(q^2_i,M^2_0,M^2_i) g_{\mu \nu}+B_{11}(q^2_i,M^2_0,M^2_i) p_{i\mu}p_{i\nu}
\crn &\qquad\qquad \qquad \quad   \equiv B^{(i)}_{00} g_{\mu \nu}+B^{(i)}_{11} p_{i\mu}p_{i\nu},  
\crn &C_\mu = \left( -p_{1\mu}\right)C_1 + p_{2\mu}C_2,\crn
&C_{\mu\nu} = g_{\mu\nu}C_{00} + p_{1\mu}p_{1\nu}C_{11} +p_{2\mu}p_{2\nu}C_{22}+  (p_{1\mu}p_{2\nu}+p_{2\mu}p_{1\nu})C_{12}.
\label{eq_fPVABC}
\end{align} 
The external momentum   $q^2=(p_2 +p_1)^2=m_Z^2$ and $m_h^2$ for the LFV$Z$ and LFV$h$ processes, respectively. As a result, the scalar functions $A_0,B_0,C_0,C_{00},C_{i},C_{ij}$ ($k,l=1,2$) are well-known PV functions  consistent with those in LoopTools \cite{Hahn:1998yk}. Useful well-known relations used in this work are:
\begin{align}
	\label{eq_Bi}	
	& A_0(m) p_{\mu}=\frac{(2\pi\mu)^{4-d}}{i\pi^2}\int \frac{d^d k k_{\mu}}{(k+p)^2 -m^2 +i\delta} ,
\crn & B^{(i)}_0 \equiv B_0(p_i^2; M_0^2, M_i^2)=B_0(p_i^2; M_i^2, M_0^2),
	\crn &B^{(i)}_1 \equiv B_1(p_i^2; M_0^2, M_i^2)= -\frac{1}{2p_i^2}\left[ A_0(M_i^2) -A_0(M_0^2) + f_i B^{(i)}_0\right],
\crn & B_0^{(12)} =\dfrac{(2\pi\mu)^{4-d}}{i\pi^2}\int\dfrac{d^dk}{D'_0D'_1} =  B_0(q^2;M^2_1,M^2_2),
\crn
&B_\mu^{(12)} \equiv B_\mu(q^2;M^2_1,M^2_2) =\dfrac{(2\pi\mu)^{4-d}}{i\pi^2}\int\dfrac{d^dk\times (k+p_1)_\mu}{D'_0D'_1} \equiv  B_1^{(12)}q_\mu + B_0^{(12)}p_{1\mu},
\end{align}
where $f_i=p_i^2 +M_i^2 -M_0^2$, $q=p_1+p_2$,  $D'_0 \equiv k^{2}-M^2_1 +i\delta$, $ D'_1 = (k+q)^2 -M^2_2 +i\delta$, and $B_{0,1}^{(12)}\equiv B_{0,1}(q^2;M^2_1,M^2_2)$.  
The scalar functions $A_0$, $B_0$, $C_0$ can be calculated using the techniques given in Ref.  \cite{tHooft:1978jhc}.   For simplicity, we define new notations appearing in many important formulas:
\begin{align}
	\label{eq_Xidef}
	X_0&\equiv C_0 +C_1 +C_2,
	\crn   X_1&\equiv C_{11} +C_{21} +C_1
	\crn   X_2&\equiv C_{12} +C_{22} +C_2,
	\crn   X_3&\equiv C_1 +C_2= X_0 -C_0,
	\crn   X_{012}&\equiv X_0+X_1 +X_2, \; X_{ij}= X_i+X_j.
\end{align}
For the decays $h\to e_b^\pm e_a^\mp$ and $Z\to e_b^+e_a^-$, 
we can derive all formulas of   $C_i$, and $C_{ij}$ as functions of $A_0$, $B^{(i)}_0$, and $C_0$ consistent with Ref. \cite{Hue:2017lak}, using the following  relations: 
\begin{align}
	\label{eq_PVrelation0}
	&2 m^2_a C_1 + (m^2_a+m^2_b -q^2)C_2 = - f_1 C_0 -B^{(12)}_0 +B^{(2)}_0,
	\crn &   (m^2_a +m^2_b -q^2)C_1 + 2m^2_b C_2= - f_2  C_0 -B^{(12)}_0 +B^{(1)}_0,
	\crn &  2C_{00}+  2 m^2_a C_{11} + (m^2_a +m^2_b -q^2)C_{12} =  B^{(12)}_1 +B^{(12)}_0 - f_1 C_1,
	\crn &    2 m^2_a C_{12} + (m^2_a +m^2_b -q^2)C_{22} = - B^{(12)}_1  + B^{(2)}_1 - f_1 C_2
	\crn &  2C_{00}+ (m^2_a +m^2_b -q^2)C_{12} +  2 m^2_b  C_{22} = B^{(12)}_1 - f_2 C_2,
	\crn &     (m^2_a +m^2_b -q^2)C_{11} + 2 m^2_b C_{12} = B^{(12)}_0 +B^{(12)}_1  +B^{(1)}_1 - f_2 C_1,
	\crn & 4C_{00} -\frac{1}{2}  +m^2_a C_{11} +(m^2_a +m^2_b -q^2)C_{12} +m^2_bC_{22} = B^{(12)}_0 +M_0^2 C_0, 
\end{align}
where $f_{i}= M_i^2 -M_0^2+m_{i}^2$, $q^2=m_Z^2,m_h^2$, and $C_{12}=C_{21}$ is used.

\section{\label{app_Zeab}  One-loop contributions to $Z\to e_b^+ e_a^- $}
\subsection{ \label{app_Ucomputed} Calculation  for $V^{\mu}$ exchanges in the unitary gauge}

Before going to the details of the calculation. We list here important  well-known results such as the on-shell conditions gives $p_1^2=m_a^2$, $p_2^2=m_b^2$, and $q^2=m_Z^2$, where $m_a$, $m_b$, and $m_Z$ are the masses of leptons $a,b$ ($a,b=1,2,3$), and gauge boson $Z$. The momentum conservation gives $q=p_1 +p_2$. Two  internal momenta $k_1 \equiv k -p_1$ and  $k_2 \equiv k +p_2$ with $i=1,2$  are denoted in  Fig. \ref{fig_2HDMU}.

When solving the Dirac matrices in the general dimension $d$, we will use the following identities \cite{Peskin:1995ev}:
\begin{align*}
&\gamma^{\mu}\gamma_{\mu}=d, \; 
%
\gamma^{\mu} \gamma^{\nu}\gamma_{\mu}= (2-d) \gamma^{\nu} \to \gamma^{\mu} \slashed{p}\gamma^{\mu}=(2-d) \slashed{p},
\crn & \gamma^{\mu} \gamma^{\nu}\gamma^{\rho} \gamma_{\mu}= 4 g^{\nu \rho} +(d-4) \gamma^{\nu}\gamma^{\rho} \to \gamma^{\mu} \slashed{p}_1\slashed{p}_2 \gamma^{\mu}= 4p_1.p_2 +(d-4) \slashed{p}_1\slashed{p}_2,
\crn & \gamma^{\mu} \gamma^{\nu}\gamma^{\rho}\gamma^{\sigma} \gamma_{\mu}= -2 \gamma^{\sigma} \gamma^{\rho} \gamma^{\nu}   -(d-4) \gamma^{\nu}\gamma^{\rho} \gamma^{\sigma} \to \gamma^{\mu} \slashed{p}_1\slashed{p}_2 \slashed{p}_3 \gamma_{\mu}= -2 \slashed{p}_3 \slashed{p}_2 \slashed{p}_1 - (d-4) \slashed{p}_1\slashed{p}_2 \slashed{p}_3.
\end{align*}
Then we take $d=4$ for all finite integrals.  For  all divergent integrals, after changing into the expressions in terms of the PV-functions, we take $d=4-2\epsilon$, then determining the final finite results before fixing $\epsilon=0$.  In addition, we will use the following transformation to change from integral to the notations of the PV-functions:
$$ \int{\dfrac{d^4k}{\left(2\pi\right)^4}} \to \frac{i}{16 \pi^2} \times \frac{(2\pi\mu)^{4-d}}{i\pi^2}\int d^dk= \frac{i}{16 \pi^2} \times \left( \mathrm{PV-functions} \right).$$
In practice, the overall factor $i/(16\pi^2)$ will be added in the final results.  
 
The diagram (1) in Fig. \ref{fig_2HDMU}  corresponding to the following amplitude:
\begin{align}
	i \mathcal{M}_{1}^{(U)} =& 	i \mathcal{M}_{nWW}^{(U)} 
	\crn =&\int{\dfrac{d^4k}{\left(2\pi\right)^4}} \overline{ u_a} \left[\dfrac{ie}{\sqrt{2}s_W}U_{ai}^\nu \gamma^{\beta'} P_L\right]  \dfrac{i\left(m_{n_i} +\slashed{k}\right)}{D_0} \left[\dfrac{ie}{\sqrt{2}s_W}U_{bi}^{\nu*}\gamma^{\alpha'} P_L\right]v_b 
	\crn & \times  \dfrac{-i}{D_1}\left(g_{\beta \beta'}-\dfrac{k_{1\beta}k_{1\beta'}}{m_W^2}\right) \times\left[ -ig_{ZWW} \Gamma^{\mu \beta \alpha}(q,k_1,-k_2) \varepsilon_{\mu} \right] \dfrac{-i}{D_2}\left(g_{\alpha \alpha'}-\dfrac{k_{2\alpha}k_{2\alpha'}}{m_W^2}\right) 
	\crn =& \dfrac{e^3 U_{ai}^\nu U_{bi}^{\nu*}}{2s^2_Wt_R}  \int {\dfrac{d^4k}{\left(2\pi\right)^4}} \times  \dfrac{ \overline{u_a} \left[ \gamma^{\beta'} \slashed{k}  \gamma^{\alpha'} P_L \right] v_b}{D_0 D_1D_2} \times \left(g_{\beta \beta'}-\dfrac{k_{1 \beta}k_{1\beta'}}{m_W^2}\right)\left(g_{\alpha \alpha'}-\dfrac{k_{2\alpha} k_{2\alpha'}}{m_W^2}\right)
	\crn &  \times \left[ \varepsilon^\beta \left( q - k_1\right)^{\alpha} + g^{\beta \alpha}\varepsilon \left( k_1 + k_2\right) + \varepsilon^\alpha \left( -k_2 -q\right)^{\beta}\right] \crn 
	=& \dfrac{e^3 U_{ai}^\nu U_{bi}^{\nu*}}{2s^2_Wt_R} \left[I_1+I_2+I_3\right]. 
\end{align}
For simplicity,  we omit the sum over all neutrino masses $m_{n_i}$ for $i=1,\dots,K+3$.  The final results must taken into account this sum for all diagrams, in order to simplify  many intermediate steps of calculation. The first integral $I_1$ contains only the highest order of variable $k_{\mu} k_{\nu}$ in the integrand, hence it is easily to write in terms of PV-function given in appendix \ref{app_PVLT}, namely 
\begin{align}
	I_1 =&  \int {\dfrac{d^4k}{\left(2\pi\right)^4}} \dfrac{ \overline{u_a} \left\lbrace \gamma^{\beta'} \slashed{k}  \gamma^{\alpha'} P_L \times g_{\beta\beta'}g_{\alpha\alpha'} \right\rbrace P_L v_b}{D_0 D_1D_2} 
\crn &  \times \left[ \varepsilon^\beta \left( q - k_1\right)^{\alpha} + g^{\beta \alpha}\varepsilon \left( k_1 + k_2\right) + \varepsilon^\alpha \left( -k_2 -q\right)^{\beta}\right] 
\crn 
	&= \int {\dfrac{d^4k}{\left(2\pi\right)^4}} \times  \dfrac{ \overline{u_a}}{D_0 D_1D_2}  \left\lbrace \slashed{\varepsilon} \slashed{k}\left(2\slashed{p}_1 + \slashed{p}_2 \right) - \left(D_0+m_{n_i}^2\right)\not\varepsilon + (4-2d)\varepsilon.k \slashed{k} - (4-2d)p_1.\varepsilon \slashed{k} \frac{}{}
	\right. \crn& \left. \hspace{4.3cm} -\left(D_0 + m_{n_i}^2 + \slashed{p}_1\slashed{k} + 2\slashed{p}_2\slashed{k}\right)\slashed{\varepsilon} \frac{}{}\right\rbrace P_Lv_b 
	\crn 
	=&\overline{u_a} \left\lbrace \left[-2m^2_{n_i}C_0 +(4-2d)C_{00} -m^2_aC_1 -m^2_bC_2 +2\left(m^2_Z-m^2_a-m^2_b\right)(C_1+C_2)\right]\slashed{\varepsilon} \frac{}{}
	\right.\crn	& \qquad - \left. 4(p_1.\varepsilon )C_2\slashed{p}_2 +2\slashed{p}_1\slashed{\varepsilon}\slashed{p}_2C_2 -2(p_1.\varepsilon )C_1\slashed{p}_2 +2(p_1.\varepsilon )C_2\slashed{p}_1 +(C_1+C_2)\slashed{p}_1\slashed{\varepsilon}\slashed{p}_2 
\right.\crn	& \left. \qquad -4(p_1.\varepsilon) C_1\slashed{p}_2 +2\slashed{p}_1\slashed{\varepsilon}\slashed{p}_2C_1 +(4-2d)(p_1.\varepsilon )\left[X_1 \slashed{p}_1 - X_2\slashed{p}_2\right] \frac{}{}\right\rbrace P_Lv_b\crn
	=& \overline{u_a} \left\lbrace \left[-2m^2_{n_i}C_0 -4C_{00} -m^2_aC_1 -m^2_bC_2 +2\left(m^2_Z-m^2_a-m^2_b\right)(C_1+C_2)\right]\slashed{\varepsilon} \frac{}{}
	\right.\crn	& \left. \qquad+m_a(p_1.\varepsilon )\left[4C_1 + 2C_2 -4X_1\right] \frac{}{}\right\rbrace P_Lv_b\crn
	&+ m_b \times \overline{u_a}\Big\lbrace -3m_a(C_1+C_2)\slashed{\varepsilon} +(p_1.\varepsilon) \left[4C_2 + 2C_1 -4X_2\right] \Big\rbrace P_Rv_b. 
\end{align}
In the above calculation, all terms proportional to $ \frac{1}{D_1D_2}$ without any factors $m_{n_i}$ will vanish after taking the sum $\sum_{i=1}^{K+3} U^{\nu}_{ai}U^{\nu*}_{bi}=\delta_{ab}=0$. In addition,  all PV-functions $C_{0,k,kl}$ with $k,l=1,2$ are finite, hence we will take $d=4$ for the relevant factors. Finally, all divergent parts of the PV-functions $C_{00}$ and $B^{(k)}_{0,1}$ are  independent to $m_{n_i}$, therefore vanish after the summation.  We also apply $d=4$. These comments will be applied to our calculation from now on.  The second integral $I_2$  contains dangerous terms like $\slashed{k}_2\slashed{k}\slashed{k}_2$ in the integrand, resulting  in the PV functions  originated from  higher tensor ranks such as $C_{\mu \nu\alpha},\dots$, which were not defined in appendix \ref{app_PVLT}, namely 
\begin{align*}
	I_2 =& -\dfrac{1}{m^2_W}\int {\dfrac{d^4k}{\left(2\pi\right)^4}} 
	\crn &\times  \dfrac{ \overline{u_a}}{D_0 D_1D_2} \left\lbrace \slashed{\varepsilon}\slashed{k}\slashed{k}_2\left[k_2.\left(q -k_1\right)\right] + \slashed{k}_2\slashed{k}\slashed{k}_2\left[\varepsilon.\left(k_1+k_2\right)\right]  - \left(\slashed{k}_2 +\slashed{q}\right)\slashed{k}\slashed{k}_2\left(\varepsilon k_2\right) \frac{}{}
	\right.\crn
	&\left. \hspace{2.2cm}+\slashed{k}_1\slashed{k}\left(\slashed{q}-\slashed{k}_1\right)\left(\varepsilon k_1\right) +\slashed{k}_1\slashed{k}\slashed{k}_1\left[\varepsilon\left(k_1+k_2\right)\right] 
- \slashed{k}_1\slashed{k}\slashed{\varepsilon}\left[k_1\left(k_2+q\right)\right] \frac{}{}\right\rbrace P_Lv_b \crn
	&= -\dfrac{1}{m^2_W}\int {\dfrac{d^4k}{\left(2\pi\right)^4}} \times  \dfrac{ \overline{u_a}}{D_0 D_1D_2} \times \left\lbrace \slashed{\varepsilon}\left(k^2+\slashed{k}\slashed{p}_2\right)\left(2k.p_1 +p^2_2 -k^2 +2p_1.p_2\right) \right.\crn
	&+ \left. \left(k^2+\slashed{p}_2\slashed{k}\right)\left(\slashed{k}+\slashed{p}_2\right) \varepsilon.\left(2k-2p_1\right) -\left(k^2+\slashed{p}_1\slashed{k}+2\slashed{p}_2\slashed{k}\right)\left(\slashed{k}+\slashed{p}_2\right)\varepsilon\left(k+p_2\right) \right.\crn
	&+ \left. \left(k^2-\slashed{p}_1\slashed{k}\right)\left(-\slashed{k}+2\slashed{p}_1+\slashed{p}_2\right)\varepsilon\left(k-p_1\right) +\left(k^2-\slashed{p}_1\slashed{k}\right)\left(\slashed{k}-\slashed{p}_1\right)\varepsilon\left(2k-2p_1\right) \right.\crn
	&- \left. \left(k^2-\slashed{p}_1\slashed{k}\right)\slashed{\varepsilon}\left(k^2+2kp_2-2p_1p_2-p^2_1\right)\right\rbrace P_Lv_b. 
	\end{align*}
	  Using the property of the polarization of external gauge boson $Z$ gives $ q.\varepsilon=0 \Longleftrightarrow \varepsilon.k_1=\varepsilon.k_2 \Longleftrightarrow \varepsilon.(k-p_1) = \varepsilon(k+p_2)$,  we can show that all terms with higher orders of $k^2$ in the numerators will vanish. Namely,
\begin{align}
I_2 	&= -\dfrac{1}{m^2_W}\int {\dfrac{d^4k}{\left(2\pi\right)^4}} 
\times  \dfrac{ \overline{u_a}}{D_0 D_1D_2}  \left\lbrace \slashed{\varepsilon}\left(D_0+m^2_{n_i}+\slashed{k}\slashed{p}_2\right)\left(-D_1-m^2_W+m^2_Z\right)
\right.\crn	&\left. \hspace{5.5cm} - \left(D_0+m^2_{n_i}-\slashed{p}_1\slashed{k}\right)\slashed{\varepsilon}\left(D_2 +m^2_W -m^2_Z\right) 
\right.\crn	&\left. \hspace{5.5cm} +\left[2D_0\slashed{k} +2m^2_{n_i}\slashed{k} 
+2D_0\slashed{p}_2 +2m^2_{n_i}\slashed{p}_2 
\right.\right.\crn	& \left.\left. \hspace{6cm} + 2D_0\slashed{p}_2 +2m^2_{n_i}\slashed{p}_2 +2\slashed{p}_2\slashed{k}\slashed{p}_2 -D_0\slashed{k}
\right.\right.\crn	& \left.\left.  \hspace{6cm} -m^2_{n_i}\slashed{k} -D_0\slashed{p}_1 -m^2_{n_i}\slashed{p}_1 -2D_0\slashed{p}_2 -2m^2_{n_i}\slashed{p}_2 \right.\right.\crn
	& \left.\left. \hspace{6cm}- D_0\slashed{p}_2 -m^2_{n_i}\slashed{p}_2 -\slashed{p}_1\slashed{k}\slashed{p}_2 -2\slashed{p}_2\slashed{k}\slashed{p}_2 -D_0\slashed{k} 
	\right.\right.\crn	& \hspace{6cm} \left.\left. -m^2_{n_i}\slashed{k} +2D_0\slashed{p}_1 +2m^2_{n_i}\slashed{p}_1 +D_0\slashed{p}_2 
	\right.\right.\crn	& \left.\left. \hspace{6cm}+ m^2_{n_i}\slashed{p}_2 +D_0\slashed{p}_1 +m^2_{n_i}\slashed{p}_1 -2\slashed{p}_1\slashed{k}\slashed{p}_1 -\slashed{p}_1\slashed{k}\slashed{p}_2 
	\right.\right.\crn	& \left.\left.\hspace{6cm} +2D_0\slashed{k} +2m^2_{n_i}\slashed{k} -2D_0\slashed{p}_1 -2m^2_{n_i}\slashed{p}_1
	 \right.\right.\crn	& \left.\left. \hspace{6cm}-2D_0\slashed{p}_1 -2m^2_{n_i}\slashed{p}_1 +2\slashed{p}_1\slashed{k}\slashed{p}_1\right](\varepsilon k - p_1.\varepsilon ) \right\rbrace P_Lv_b\crn
	&= -\dfrac{1}{m^2_W}\overline{u_a}  \left\lbrace  \slashed{\varepsilon} \left[-m^2_{n_i}\left(B_0^{(1)} +B_0^{(2)}\right)  -m^2_aB_1^{(1)}
	+ 2m^2_{n_i}\left(-m^2_W +m^2_Z\right)C_0 -m^2_bB_1^{(2)} \right]
	\right.\crn&\left. \hspace{2.8cm} +\left(-m^2_W +m^2_Z\right)\left(-C_1\slashed{\varepsilon}\slashed{p}_1\slashed{p}_2 +m^2_bC_2\slashed{\varepsilon} +m^2_aC_1\slashed{\varepsilon} -C_2\slashed{p}_1\slashed{p}_2\slashed{\varepsilon}\right) 
	\right.\crn	&\left.  \hspace{2.8cm} - 2\left[ m^2_{n_i}(-C_1\slashed{p}_1 +C_2\slashed{p}_2)  +m^2_{n_i}C_0\slashed{p}_2 - m^2_{n_i}C_0\slashed{p}_1 +m^2_aC_1\slashed{p}_2 -m^2_bC_2\slashed{p}_1\right](p_1.\varepsilon )
	 \right.\crn&\left.  \hspace{2.8cm}  +2\left[\dfrac{m^2_{n_i}\gamma^\mu k_\mu k_\nu}{D_0D_1D_2}  +\dfrac{m^2_{n_i}\slashed{p}_2k_\nu}{D_0D_1D_2}  -\dfrac{m^2_{n_i}\slashed{p}_1k_\nu}{D_0D_1D_2} -\dfrac{\slashed{p}_1\gamma^\mu\slashed{p}_2k_\mu k_\nu}{D_0D_1D_2}\right]\varepsilon_\nu\right\rbrace P_Lv_b\crn
	= & -\dfrac{1}{m^2_W}  \overline{u_a} \left\lbrace  \slashed{\varepsilon} \left[ \left(m^2_Z -m^2_W\right) \left( 2m^2_{n_i}C_0 +m^2_aC_1 +m^2_bC_2\right) \frac{}{}
	 \right.\right.\crn	& \left.\left. \hspace{2.4cm} - m^2_{n_i}\left(B_0^{(1)} +B_0^{(2)}\right) -m^2_aB_1^{(1)} -m^2_bB_1^{(2)}  +2m^2_{n_i}C_{00}\right] 
	 \right.\crn& \left. \hspace{1.8cm}+ 2m_a(p_1.\varepsilon )\Big[\left(m^2_Z -m^2_W +m^2_{n_i} +m^2_b\right)C_2 +m^2_{n_i}(C_0 +2C_1 +C_{11}) 
	  \right.\crn& \left.  \hspace{4 cm}+m^2_bC_{22}  +(m^2_{n_i} +m^2_b)C_{12}\Big]\right\rbrace P_Lv_b\crn
	&-\dfrac{m_b}{m^2_W}\overline{u_a}  \left\lbrace  \slashed{\varepsilon} m_a\left[2C_{00} -(m^2_Z -m^2_W)(C_1 +C_2)\right] \frac{}{}
	\right.\crn&\left. \hspace{1.8 cm} +2(p_1.\varepsilon )\left[(m^2_Z -m^2_W +m^2_{n_i} +m^2_a)C_1  \frac{}{}
	\right.\right.\crn
	& \left.\left.  \hspace{3.5 cm}+ m^2_{n_i}(C_0 +2C_2 +C_{22}) +m^2_aC_{11} +(m^2_{n_i} +m^2_a)C_{12}\right] \frac{}{}\right\rbrace P_Rv_b.
\end{align}
We can see that the higher tensor ranks of PV-functions vanish because the relevant terms in numerators were changed into the factor $D_0$, resulting in many terms of the form $1/(D_1D_2)$  independent to neutrinos masses $m_{n_i}$. They  are proportional to the sum $\sum_{i=1}^{K+3} U^{\nu}_{ai} U^{\nu*}_{bi}=0$ with $a\neq b$.   

Using above tricks to calculate $I_3$, we have
\begin{align}
	I_3 =& \dfrac{1}{m^4_W}\int {\dfrac{d^4k}{\left(2\pi\right)^4}} 
	\crn& \times  \dfrac{ \overline{u_a}}{D_0 D_1D_2} \times \slashed{k}_1\slashed{k}\slashed{k}_2  \Big\lbrace  k_1.\varepsilon\left[k_2.\left(q-k_1\right)\right] +\left(k_1.k_2\right) \left(k_1+k_2\right).\varepsilon
	-k_2.\varepsilon\left[k_1.\left(k_2+q\right)\right]  \Big\rbrace P_Lv_b
	\crn= & \dfrac{1}{m^4_W}\int {\dfrac{d^4k}{\left(2\pi\right)^4}} \times  \dfrac{ \overline{u_a}}{D_0 D_1D_2} \left(k^2-\slashed{p}_1\slashed{k}\right)\left(\slashed{k}+\slashed{p}_2\right)
\crn&\times	  \left\lbrace \varepsilon.k_1 \left(k+p_2\right).\left(-k+2p_1+p_2\right) \frac{}{}
	%
+  2\left(k_1.k_2\right) \varepsilon.k_1
	- \varepsilon.k_2 \left(k_1.\left(k+p_1+2p_2\right)\right)  \frac{}{}\right\rbrace P_Lv_b
\crn =& \dfrac{1}{m^4_W}\int {\dfrac{d^4k}{\left(2\pi\right)^4}} \times  \dfrac{ \overline{u_a}}{D_0 D_1D_2}  \left(D_0\slashed{k} +m^2_{n_i}\slashed{k}-D_0\slashed{p}_1 -m^2_{n_i}\slashed{p}_1 +D_0\slashed{p}_2 + m^2_{n_i}\slashed{p}_2 -\slashed{p}_1\slashed{k}\slashed{p}_2\right) 
\crn& \hspace{3cm}\times \left(\varepsilon.k_1\right) \left[-k.k_2 +2k_2.p_1  -k.k_1 -2k.p_2 +k_1.p_1 +2k_1.k_2 \frac{}{}\right] P_Lv_b\crn
=&  \dfrac{1}{m^4_W}\int {\dfrac{d^4k}{\left(2\pi\right)^4}} \times  \overline{u_a} \left( \dfrac{m^2_{n_i}\slashed{k}}{D_0D_1D_2}  -\dfrac{m^2_{n_i}\slashed{p}_1}{D_0D_1D_2}  +\dfrac{m^2_{n_i}\slashed{p}_2}{D_0D_1D_2} -  \dfrac{\slashed{p}_1\slashed{k}\slashed{p}_2}{D_0D_1D_2}\right) 
	\crn &\times \left[\varepsilon.(k-p_1)\right] \left[k_2.\left(q-k_1\right) -k_1.\left(q+k_2\right) +2k_1.k_2\right]P_Lv_b. 
\end{align}
Note that: $k_2\left(q-k_1\right) -k_1\left(q+k_2\right) +2k_1k_2 = \left(k_2-k_1\right).q = \left(p_2+p_1\right)q = p^2_0 = m^2_Z$. Therefore
\begin{align}
	I_3 =&  \dfrac{m^2_Z}{m^4_W}\int {\dfrac{d^4k}{\left(2\pi\right)^4}} 
	\crn& \times  \overline{u_a}  \left\lbrace\left[ \dfrac{m^2_{n_i}\gamma^\mu\varepsilon^\nu k_\mu k_\nu}{D_0D_1D_2} + \dfrac{m^2_{n_i}\varepsilon^\mu k_\mu}{D_0D_1D_2}\left(-\slashed{p}_1+\slashed{p}_2\right)  -\dfrac{2p_2^\mu\varepsilon^\nu k_\mu k_\nu}{D_0D_1D_2}\slashed{p}_1 +\slashed{p}_1\slashed{p}_2\dfrac{\gamma^\mu\varepsilon^\nu k_\mu k_\nu}{D_0D_1D_2}\right]  
	\right. \crn&\left. \qquad \quad -(p_1.\varepsilon )\left[ m^2_{n_i}\left(-C_1\slashed{p}_1+C_2\slashed{p}_2\right)  +m^2_{n_i}C_0\left(-\slashed{p}_1+\slashed{p}_2\right) +m^2_aC_1\slashed{p}_2 -m^2_bC_2\slashed{p}_1\right]\right\rbrace P_Lv_b\crn
	=&  \dfrac{m^2_Z}{m^4_W} \times \overline{u_a} \left\lbrace m^2_{n_i}C_{00} \slashed{\varepsilon} + m_a(p_1.\varepsilon )\left[m^2_{n_i} X_{01} +m^2_bX_2
\right] \frac{}{}\right\rbrace P_Lv_b\crn
&+ \dfrac{m^2_Z m_b}{m^4_W}\times \overline{u_a}\left\lbrace m_aC_{00}\slashed{\varepsilon} + (p_1.\varepsilon ) \left[m^2_{n_i}X_{02}+m^2_aX_1\right]\frac{}{} \right\rbrace P_Rv_b. 
		\end{align}
The final result of $\mathcal{M}^{(U)}_{1}$ is derived in terms of the PV-functions as follows:
	\begin{align}
	i \mathcal{M}^{(U)}_{1}=& \dfrac{e^3 U_{ai}^\nu U_{bi}^{\nu*}}{2s^2_Wt_R}  \crn
	&\times \overline{u_a} \Big\lbrace \left[-2m^2_{n_i}C_0 -4 C_{00} -m^2_aC_1 -m^2_bC_2 +2\left(m^2_Z-m^2_a-m^2_b\right)X_3 \right]\slashed{\varepsilon} 
	 \crn	& \qquad \quad +   m_a(p_1.\varepsilon )\left[4C_1 + 2C_2 -4X_1 \right]\Big\rbrace P_Lv_b 
	 \crn&+ \overline{u_a}\times m_b \left\lbrace -3m_aX_3 \slashed{\varepsilon} +(p_1.\varepsilon ) \left[4C_2 + 2C_1 -4X_2 \right] \frac{}{} \right\rbrace P_Rv_b
	 \crn 	&-\dfrac{1}{m^2_W} \overline{u_a} \left\lbrace  \slashed{\varepsilon} \left[ \left(m^2_Z -m^2_W\right)\left(2m^2_{n_i}C_0 +m^2_aC_1 +m^2_bC_2\right) \frac{}{}
	 \right.\right.\crn	& \left.\left. \hspace{2.2cm} -m^2_{n_i}\left(B_0^{(1)} +B_0^{(2)}\right) -m^2_aB_1^{(1)} -m^2_bB_1^{(2)}  +2m^2_{n_i}C_{00}\right]
	  \right.\crn	&\left. \hspace{2.2cm}  + 2m_a(p_1.\varepsilon )\Big[\left(m^2_Z -m^2_W \right)C_2 +m^2_{n_i}X_{01} +m^2_bX_2 \Big]\right\rbrace P_Lv_b
	\crn&-\dfrac{m_b}{m^2_W}  \overline{u_a} \left\lbrace  \slashed{\varepsilon} m_a\left[2C_{00} -(m^2_Z -m^2_W)X_3\right] \frac{}{}
	\right.\crn&\left.  \hspace{1.8cm} +2(p_1.\varepsilon )\left[(m^2_Z -m^2_W) C_1 + m^2_{n_i}X_{02} +m^2_aX_1 \right] \frac{}{}\right\rbrace P_Rv_b
	\crn&+\dfrac{m^2_Z}{m^4_W} \overline{u_a}  \left\lbrace \slashed{\varepsilon}m^2_{n_i}C_{00}  + m_a(p_1.\varepsilon )\left[m^2_{n_i}X_{01}   +m^2_bX_2 	\right] \frac{}{}\right\rbrace P_Lv_b
	\crn &+ \dfrac{m^2_Z m_b}{m^4_W}\overline{u_a}\left\lbrace m_aC_{00}\slashed{\varepsilon} + (p_1.\varepsilon ) \left[m^2_{n_i}X_{02} +m^2_aX_1 \right] \frac{}{}\right\rbrace P_Rv_b. 
	\label{eq_Z1} 
\end{align}
Eq. \eqref{eq_Z1} results in the form factors given in Eqs. \eqref{eq_al1u},  \eqref{eq_ar1u}, \eqref{eq_bl1u}, and \eqref{eq_br1u}. 

The amplitude originated from diagram (2) in Fig. \ref{fig_2HDMU} is 
\begin{align}
	i \mathcal{M}_{2}^{(U)} =& \int {\dfrac{d^4k}{\left(2\pi\right)^4}} \times \overline{u_a} \times \left[\dfrac{ie}{\sqrt{2}s_W}U_{ai}^\nu \gamma^{\alpha} P_L\right]  \dfrac{i\left(-\slashed{k}_1+m_{n_i}\right)}{D_1} \left[ \dfrac{ie}{2s_Wc_W} \slashed{\varepsilon}  \left( q_{ij}P_L -q_{ji}P_R\right) \right]
\crn&\times \dfrac{i\left(-\slashed{k}_2+m_{n_j} \right)}{D_2}\times \left[\dfrac{ie}{\sqrt{2}s_W}U_{bj}^{\nu*}\gamma^{\beta} P_L\right] \times v_b  \times  \frac{-i}{D_0} \left(g_{\alpha\beta} -\dfrac{k_{\alpha} k_{\beta}}{ m^2_W}\right)
	\crn=	& \dfrac{e^3U^{\nu}_{ai}U^{\nu*}_{bj}}{4s^3_Wc_W}  \int {\dfrac{d^4k}{\left(2\pi\right)^4}}\dfrac{\overline{u_a}}{D_0D_1D_2}  \left\lbrace q_{ij}\gamma^\alpha \slashed{k}_1\slashed{\varepsilon}\slashed{k}_2\gamma^\beta -q_{ji}m_{n_i}m_{n_j}\gamma^\alpha\slashed{\varepsilon}\gamma^\beta \frac{}{}\right\rbrace \left(g_{\alpha\beta} -\dfrac{k_{\alpha} k_{\beta}}{ m^2_W}\right) P_Lv_b 
	\crn= 	&  \dfrac{e^3U^{\nu}_{ai}U^{\nu*}_{bj}}{4m^2_Ws^3_Wc_W}  \int {\dfrac{d^4k}{\left(2\pi\right)^4}} \times \dfrac{\overline{u_a}}{D_0D_1D_2}
	  \Big\lbrace q_{ij}\left[\left(-2\slashed{k}_2\slashed{\varepsilon}\slashed{k}_1 +(4-d)\slashed{k}_1\slashed{\varepsilon}\slashed{k}_2\right)m^2_W - \slashed{k}\slashed{k}_1\slashed{\varepsilon}\slashed{k}_2\slashed{k} \right] 
	\crn&  \hspace{5.8cm}- q_{ji}m_{n_i}m_{n_j}\left[(2-d)\slashed{\varepsilon}m^2_W -\slashed{k}\slashed{\varepsilon}\slashed{k}\right] \Big\rbrace P_Lv_b 
\crn&= \dfrac{e^3U^{\nu}_{ai}U^{\nu*}_{bj}}{4m^2_Ws^3_Wc_W}  \int {\dfrac{d^4k}{\left(2\pi\right)^4}} \times \overline{u_a} \left\lbrace q_{ij}\left[m^2_W\left(\dfrac{-2\gamma^\mu\slashed{\varepsilon}\gamma^\nu k_\mu k_\nu}{D_0D_1D_2} -2C_1\slashed{p}_1\slashed{\varepsilon}\slashed{p}_1 +2C_2\slashed{p}_2\slashed{\varepsilon}\slashed{p}_1 \right.\right.\right.\crn
	&+ \left.\left.\left. 2C_1\slashed{p}_2\slashed{\varepsilon}\slashed{p}_1 -2C_2\slashed{p}_2\slashed{\varepsilon}\slashed{p}_2 +2C_0\Big(2p_1.\varepsilon (-\slashed{p}_1 +\slashed{p}_2) -\left(m^2_Z-m^2_a-m^2_b\right)\slashed{\varepsilon} -\slashed{p}_1\slashed{\varepsilon}\slashed{p}_2\Big) \right.\right.\right.\crn
	&+ \left.\left.\left. (4-d)\Big( \dfrac{\gamma^\mu\slashed{\varepsilon}\gamma^\nu k_\mu k_\nu}{D_0D_1D_2} -C_1\slashed{p}_1\slashed{\varepsilon}\slashed{p}_2 +C_2\slashed{p}_2\slashed{\varepsilon}\slashed{p}_2 +C_1\slashed{p}_1\slashed{\varepsilon}\slashed{p}_1 -C_2\slashed{p}_1\slashed{\varepsilon}\slashed{p}_2 -C_0\slashed{p}_1\slashed{\varepsilon}\slashed{p}_2\Big)\right) \right.\right.\crn
	&- \left.\left. \dfrac{\left(k^2-\slashed{k}\slashed{p}_1\right)\slashed{\varepsilon}\left(k^2+\slashed{p}_2\slashed{k}\right)}{D_0D_1D_2}\right] +q_{ji}m_{n_i}m_{n_j}\bigg[2C_0m^2_W\slashed{\varepsilon} +\gamma^\mu\slashed{\varepsilon}\gamma^\nu\Big(C_{00}g_{\mu\nu}+C_{11}p_{1\mu}p_{1\nu} \right.\crn
	&+ \left. C_{22}p_{2\mu}p_{2\nu} -C_{12}\left(p_{1\mu}p_{2\nu}+p_{1\nu}p_{2\mu}\right)\Big)\bigg]\right\rbrace P_Lv_b. 
\end{align}
Using a relation that 
$$ \left(k^2-\slashed{k}\slashed{p}_1\right) \slashed{\varepsilon}\left(k^2+\slashed{p}_2\slashed{k}\right) = \left(D_1+m^2_{n_i}-m^2_a +\slashed{p}_1\slashed{k}\right)\slashed{\varepsilon}\left(D_2+m^2_{n_j}-m^2_b -\slashed{k}\slashed{p}_2\right) $$  we will derive $\mathcal{M}_{2}^{(U)}$ in terms of PV-functions defined in appendix \ref{app_PVLT} as follows 
\begin{align}
	\label{eq_M2Uf}
i \mathcal{M}_{2}^{(U)} = & \dfrac{e^3U^{\nu}_{ai}U^{\nu*}_{bj}}{4m^2_Ws^3_Wc_W}
\crn&\times \overline{u_a} \left\lbrace  q_{ij} \left[m_W^2\left( (2-d)^2  C_{00} +2 m^2_aX_{01} +2m^2_b X_{02} +2m_Z^2 (C_{12}- X_0)  \frac{}{} \right)\slashed{\varepsilon}
 \right.\right.\crn	&\left.\left. \hspace{1.8cm} -(2p_1.\varepsilon)m_a \left(m_W^2\left( 4X_1 +2C_0+2C_2\right)  -m_{n_j}^2C_2  + m^2_b X_2 \right)  
\right.\right.\crn&\left.\left.  \hspace{1.8cm} -\left(  (m_{n_i}^2-m_a^2)B_0^{(1)}+(m_{n_j}^2-m^2_b)B_0^{(2)} + (m_{n_i}^2-m_a^2)(m_{n_j}^2-m^2_b)C_0
\right.\right.\right.\crn	& \left.\left.\left. \hspace{1.8cm} -m_a^2 {B_1^{(1)}}- m^2_b B_1^{(2)} -(m_{n_j}^2-m^2_b)m_a^2C_1  - (m_{n_i}^2-m_a^2)m^2_b C_2\right)\right] 
\right. \crn&\left.  \hspace{1.2cm} 	+q_{ji}m_{n_i}m_{n_j}\left[\left(4m^2_W C_0  +(2-d)C_{00} -C_{11}m^2_a -C_{22}m^2_b  \frac{}{}
 \right.\right.\right.\crn	&\left.\left.\left. \hspace{3.3 cm}  + C_{12}\left(m^2_Z-m^2_a-m^2_b\right) \frac{}{}\right)\slashed{\varepsilon} +2\varepsilon.p_1 m_a( C_{11}+ C_{12}) \right]\right\rbrace P_Lv_b 
\crn& +\dfrac{e^3U^{\nu}_{ai}U^{\nu*}_{bj} m_b}{4m^2_Ws^3_Wc_W} 
  \overline{u_a} \left\lbrace q_{ij}\left[\slashed{\varepsilon}m_a\left( -(2-d)C_{00} + 2m_W^2 X_0 +m_a^2 X_1    +m^2_b X_2   \frac{}{}
\right.\right.\right.\crn&\left.\left.\left. \hspace{5. cm} -m_{n_j}^2 C_2 - m_{n_i}^2 C_1 -m_Z^2 C_{12}\right) 
\right.\right. \crn &\left.\left. \hspace{4cm}+2(p_1.\varepsilon ) \left( -2m_W^2  X_{02}  
+ m_{n_i}^2C_1 -m_a^2X_1 \right)  \frac{}{}\right]
\right.\crn	&\left. \hspace{3.2cm} +q_{ji}m_{n_i}m_{n_j}\left[2 \varepsilon.p_1 (C_{22}+C_{12}) \right] \frac{}{}\right\rbrace P_Rv_b. 
\end{align}
From Eq. \eqref{eq_M2Uf} results in four form factors listed in Eqs. \eqref{eq_al2u},  \eqref{eq_ar2u}, \eqref{eq_bl2u}, and \eqref{eq_br2u}.

For diagram (3) in Fig. \ref{fig_2HDMU}: 
\begin{align}
	i \mathcal{M}_{3}^{(U)} =& \int {\dfrac{d^4k}{\left(2\pi\right)^4}} \times \overline{u_a}  \left[\dfrac{ie}{\sqrt{2}s_W}U_{ai}^\nu \gamma^{\alpha} P_L\right]  \dfrac{i\left(m_{n_i} +\slashed{k}\right)}{D_0} \left[\dfrac{ie}{\sqrt{2}s_W}U_{bi}^{\nu*}\gamma^{\beta} P_L\right]\crn 
	&\times \dfrac{i\left(m_b +\slashed{p}_1 \right)}{p_1^2 -m^2_b} \times \left[ ie \slashed{\varepsilon} \left(t_LP_L +t_RP_R\right)\right]  v_b  \times  \frac{-i}{D_1} \left(g_{\alpha\beta} -\dfrac{k_{1\alpha} k_{1\beta}}{ m^2_W}\right)
	\crn = & \dfrac{e^3U^{\nu}_{ai}U^{\nu*}_{bi}}{2s^2_W(m_a^2 -m^2_b)}  \int {\dfrac{d^4k}{\left(2\pi\right)^4}} \times \dfrac{\overline{u_a}\gamma^\alpha \slashed{k}\gamma^\beta P_L\left(m_b+\slashed{p}_1\right)\slashed{\varepsilon}{\left( t_LP_L +t_RP_R\right)}}{D_0D_1}  
	 \left(g_{\alpha\beta} -\dfrac{k_{1\alpha} k_{1\beta}}{ m^2_W}\right) \crn
	=& \dfrac{e^3U^{\nu}_{ai}U^{\nu*}_{bi}}{2s^2_W(m_a^2 -m^2_b)} \int {\dfrac{d^4k}{\left(2\pi\right)^4}} \times \dfrac{\overline{u_a}}{D_0D_1}  \left[(2-d)\slashed{k} -\dfrac{\slashed{k}_1\slashed{k}\slashed{k}_1}{m^2_W} \right] 
	\left( t_L\slashed{p}_1\slashed{\varepsilon}P_L + t_Rm_b\slashed{\varepsilon}P_R\right)v_b 
	\crn =& \dfrac{e^3U^{\nu}_{ai}U^{\nu*}_{bi}m^2_at_L}{2m^2_Ws^2_W(m_a^2 -m^2_b)}  \overline{u_a}  \left\lbrace 2m^2_{n_i}B_0^{(1)}- \left[(2-d)m^2_W -m^2_{n_i} -m^2_a\right]B_1^{(1)}\right\rbrace\slashed{\varepsilon}P_Lv_b
	 \crn &+ \dfrac{e^3U^{\nu}_{ai}U^{\nu*}_{bi}m_a m_bt_R}{2m^2_Ws^2_W(m_a^2 -m^2_b)} \overline{u_a} \left\lbrace 2m^2_{n_i}B_0^{(1)} -  \left[(2-d)m^2_W -m^2_{n_i} -m^2_a\right] B_1^{(1)}\right\rbrace \slashed{\varepsilon}P_Rv_b.
\end{align}
Similarly for the diagram (4) of Fig. \ref{fig_2HDMU}, we get 
\begin{align}
	i \mathcal{M}_{4}^{(U)} =  & \dfrac{e^3U^{\nu*}_{bi}U^{\nu}_{ai}}{2s^2_W(m^2_b -m_a^2)}
	\crn&\times  \int {\dfrac{d^4k}{\left(2\pi\right)^4}} \times \dfrac{\overline{u_a}}{D_0D_2} \times \slashed{\varepsilon} {\left( t_LP_L +t_RP_R\right)}\left(m_a-\slashed{p}_2\right) \times \gamma^\alpha \slashed{k}\gamma^\beta  
\left(g_{\alpha\beta} -\dfrac{k_{2\alpha} k_{2\beta}}{ m^2_W}\right)P_Lv_b 
	\crn	&= \dfrac{U^{\nu}_{ai}e^3U^{\nu*}_{bi}t_L}{2m^2_Ws^2_W(m^2_b -m_a^2)}  \overline{u_a} \left\lbrace \ 2m^2_{n_i}B_0^{(2)}- \left[(2-d)m^2_W -m^2_{n_i} -m^2_b\right]B_1^{(2)}\right\rbrace\slashed{\varepsilon} P_Lv_b 
	\crn&+\dfrac{U^{\nu}_{ai}e^3U^{\nu*}_{bi}t_R}{2m^2_Ws^2_W(m^2_b -m_a^2)} \overline{u_a} \left\lbrace  2m^2_{n_i}B_0^{(2)}- \left[(2-d)m^2_W -m^2_{n_i} -m^2_b\right]B_1^{(2)}\right\rbrace\slashed{\varepsilon} P_Rv_b.
\end{align}
Combining  two results of $ \mathcal{M}_{3}^{(U)} $  and  $ \mathcal{M}_{4}^{(U)} $, we obtain four form factors listed in Eqs. \eqref{eq_al3u},  \eqref{eq_ar3u}, and \eqref{eq_blr3u}.

\subsection{Divergent cancellation in $Z\to e^+_be^-_a$ amplitude}

The divergence cancellation in the total form factor $\bar{a}_{l,r}$ and $\bar{b}_{l,r}$ is proved  below.
\begin{align}
	\text{div}\left[\bar{a}^{nWW}_{l}\right] &= \dfrac{e^3}{2s^2_Wt_R}\left\lbrace \dfrac{1}{m^2_W}\left(2 -\dfrac{1}{2}\right) +\dfrac{m^2_Z}{4m^4_W} \right\rbrace\sum_{i=1}^6 U^\nu_{ai}U^{\nu*}_{bi}m^2_{n_i} \Delta_\epsilon\crn
	&= \dfrac{e^3}{4m^2_Ws^3_Wc_W}\left\lbrace 3c^2_W + \dfrac{1}{2} \right\rbrace\sum_{i=1}^6 U^\nu_{ai}U^{\nu*}_{bi}m^2_{n_i} \Delta_\epsilon,\crn
	\text{div}\left[\bar{a}^{Wnn}_{l}\right] &= \dfrac{e^3}{4m^2_Ws^3_Wc_W}\sum_{i,j=1}^6 U^\nu_{ai}U^{\nu*}_{bj}q_{ij}(-m^2_{n_i} -m^2_{n_j}) \Delta_\epsilon\crn
	&= \dfrac{e^3}{4m^2_Ws^3_Wc_W}(-2)\sum_{i=1}^6 U^\nu_{ai}U^{\nu*}_{bi}m^2_{n_i} \Delta_\epsilon,\crn
	\text{div}\left[\bar{a}^{nW}_{l}\right] &= \dfrac{e^3t_L}{2m^2_Ws^2_W}\left(2-\dfrac{1}{2}\right) \sum_{i=1}^6 U^\nu_{ai}U^{\nu*}_{bi}m^2_{n_i} \Delta_\epsilon\crn
	&= \dfrac{e^3}{4m^2_Ws^3_Wc_W}\left(\dfrac{3}{2}s^2_W -\dfrac{3}{2}c^2_W\right) \sum_{i=1}^6 U^\nu_{ai}U^{\nu*}_{bi}m^2_{n_i} \Delta_\epsilon,\crn
\end{align}
where $\text{div}\left[B_0^{(1)}\right] = \text{div}\left[B_0^{(2)}\right] = -2\text{div}\left[B_1^{(1)}\right] = -2\text{div}\left[B_1^{(2)}\right] =4\text{div}\left[C_{00}\right]\equiv \Delta_\epsilon$ and $t_R=\dfrac{s_W}{c_W}$; $t_L=\dfrac{s^2_W -c^2_W}{2s_Wc_W}; m_Z=\dfrac{m_W}{c_W}; q_{ij}=\left(U^\dagger U\right)_{ij}$.

Because $\left(3c^2_W + \dfrac{1}{2} -2 +\dfrac{3}{2}s^2_W -\dfrac{3}{2}c^2_W\right) = 0$,  the sum of all divergent parts of $\bar{a}_{l,r}$ and $\bar{b}_{l,r}$ is zero.  Also,  we see easily that 
$\text{div}\left[\bar{a}^{nWW}_{r}\right] = \text{div}\left[\bar{a}^{Wnn}_{r}\right] = \text{div}\left[\bar{a}^{nW}_{r}\right]
= \text{div}\left[\bar{b}^{nWW}_{l}\right] = \text{div}\left[\bar{b}^{nWW}_{r}\right] = \text{div}\left[\bar{b}^{Wnn}_{l}\right] = \text{div}\left[\bar{b}^{Wnn}_{r}\right] = 0 $.

\subsection{\label{eq_ZeabHF} The results in the 't Hoof-Feynman gauge}

The form factors calculated in the 't Hooft-Feynman corresponding to 10 diagrams including 6 diagrams in Fig. \ref{fig_2HDMHF} and 4 diagrams in Fig. \ref{fig_2HDMU} are denoted respectively as follows. Two diagrams (1) and (2) in Fig. \ref{fig_2HDMHF} give 
\begin{align}
\bar{a}'_{l,1+2} \equiv&\;\bar{a}'_{nWG,l}= \dfrac{-e^2t_R}{2s^2_W} \sum_{i=1}^{K+3} U^{\nu}_{ai} U^{\nu*}_{bi} \left(2m^2_{n_i}C_0 +m^2_a C_1 +m^2_bC_2\right) ,  
\crn  \bar{a}'_{r,1+2}\equiv&\; \bar{a}'_{nWG,r}= \dfrac{e^2t_R}{2s^2_W} \times m_a m_b \sum_{i=1}^{K+3} U^{\nu}_{ai} U^{\nu*}_{bi}X_3, 
\crn \bar{b}'_{l}\equiv&\; \bar{b}'_{nWG,l}=\dfrac{-e^2t_R}{2s^2_W} \sum_{i=1}^{K+3} U^{\nu}_{ai} U^{\nu*}_{bi} \left[2 m_a C_2\right],  
\crn \bar{b}'_{r}\equiv&\; \bar{a}'_{nWG, r}= \dfrac{-e^2t_R}{2s^2_W} \sum_{i=1}^{K+3} U^{\nu}_{ai} U^{\nu*}_{bi}  \left[ 2m_bC_1\right],  \label{eq_br1up} 
\end{align}
where $C_{0,1,2}=C_{0,1,2}(m_a^2,m_Z^2,m_b^2; m_i^2, m_W^2,m_W^2)$ and $X_{3}=X_{3}(m_a^2,m_Z^2,m_b^2; m_i^2, m_W^2,m_W^2)$. 
Diagrams (3)  in Fig. \ref{fig_2HDMHF} gives 
\begin{align}
	\label{eq_al3up} 
	\bar{a}'_{l,3} \equiv &\;  \bar{a}'_{nGG,l} = \dfrac{e^2t_L}{s_W^2m^2_W} \sum_{i=1}^{K+3} U^{\nu}_{ai} U^{\nu*}_{bi}m^2_{n_i}C_{00} , 
	\crn \bar{a}'_{r,3} \equiv &\;  \bar{a}'_{nGG,r} =   \dfrac{e^2t_L m_a m_b}{s_W^2 m^2_W} \sum_{i=1}^{K+3} U^{\nu}_{ai} U^{\nu*}_{bi}C_{00},  
	\crn \bar{b}'_{l,3} \equiv &\;  \bar{b}'_{nGG,l} =\dfrac{e^2t_L m_a}{s_W^2 m^2_W} \sum_{i=1}^{K+3} U^{\nu}_{ai} U^{\nu*}_{bi} \left[m^2_b X_2  +m^2_{n_i} X_{01}\right],
	\crn \bar{b}'_{r,3} \equiv &\;  \bar{a}'_{nGG,r} =  \dfrac{e^2t_Lm_b}{s_W^2 m^2_W}\sum_{i=1}^{K+3} U^{\nu}_{ai} U^{\nu*}_{bi} \left[m^2_aX_1 +m^2_{n_i}X_{02}\right],
\end{align}
where $X_{0,1,2}=X_{0,1,2}(m_a^2,m_Z^2,m_b^2; m_i^2, m_W^2,m_W^2)$ and $C_{00}= C_{00}(p_1^2,q^2,p_2^2; m_i^2, m_W^2,m_W^2)$.

Diagram (4) in Fig. \ref{fig_2HDMHF} gives: 
\begin{align}
	\bar{a}'_{l,4} \equiv & \bar{a}'_{Gnn,l}
	\crn =& \dfrac{e^2}{4s^3_Wc_Wm^2_W} 
	\crn&\times \sum_{i,j=1}^{K+3} U^{\nu}_{ai} U^{\nu*}_{bj} \left\lbrace  q_{ij}\left[ \left( m_a^2 m_j^2+ m^2_b m_i^2 -m_a^2 m^2_b \right) X_0 
		-m_i^2 m_j^2 C_0 - m^2_b m_i^2 C_1  - m_a^2 m_j^2 C_2 \frac{}{}\right] 
	\right.\crn&\left. \hspace{3cm} -  q_{ji}m_{n_i}m_{n_j}\left[B_0^{(12)}  -2C_{00}  +m^2_WC_0    \right]\right\rbrace ,   
	\crn  \bar{a}'_{r,4} \equiv & \bar{a}'_{Gnn,r}
	\crn= &  \dfrac{e^2 m_a m_b}{4s^3_Wc_Wm^2_W} \sum_{i=1}^{K+3} U^{\nu}_{ai} U^{\nu*}_{bj}  q_{ij}\left[B_0^{(12)} +m^2_WC_0-2C_{00}  -(m^2_{n_i}-m^2_a)C_1  -(m^2_{n_j}-m^2_b)C_2  \right], 
	\crn \bar{b}'_{l,4} \equiv & \bar{b}'_{Gnn,l} 
	\crn = & \dfrac{e^2 m_a}{4s^3_Wc_Wm^2_W} \sum_{i=1}^{K+3} U^{\nu}_{ai} U^{\nu*}_{bj} \left[  2q_{ij} \left(m_{n_j}^2C_2 -m^2_b X_2\right)  + 2q_{ji}m_{n_i}m_{n_j}(X_1 -C_1) \frac{}{}\right],
	\crn \bar{b}'_{r,4}\equiv & \bar{b}'_{Gnn,r} 
	\crn =& \dfrac{e^2m_b}{4s^3_Wc_Wm^2_W} \sum_{i=1}^{K+3} U^{\nu}_{ai} U^{\nu*}_{bj} \left\lbrace 2q_{ij}\left[  m^2_{n_i}C_1 -m^2_aX_1\right]  + 2q_{ji}m_{n_i}m_{n_j} (X_2-C_{2}) \right\rbrace,  \label{eq_ab4p} 
\end{align}
where $X_{0,1,2}=X_{0,1,2}(m_a^2,m_Z^2,m_b^2; m_i^2, m_j^2,m_W^2)$ and $C_{00}= C_{00}(m_a^2,m_Z^2,m_b^2; m_i^2, m_j^2,m_W^2)$.

Two diagrams (5) and (6) in Fig. \ref{fig_2HDMHF} give 
\begin{align}
	\bar{a}'_{l,5+6} \equiv & 	\bar{a}'_{nG,l} 
	\crn =&\dfrac{e^2 t_L}{2m^2_W s_W^2 \left(m^2_a -m^2_b\right)} 
 \sum_{i=1}^{K+3} U^{\nu}_{ai} U^{\nu*}_{bi} \left[m^2_{n_i}\left(m^2_a+m^2_b\right) \left( B_0^{(1)}- B_0^{(2)}\right) +m^2_am^2_b \left( B_1^{(1)} -B_1^{(2)}\right)  
\right.\crn& \left. \hspace{5.7cm}
+ m^2_{n_i} \left(m^2_aB_1^{(1)} - m^2_b B_1^{(2)}\right) \right],  
\crn \bar{a}'_{r,5+6}\equiv & 	\bar{a}'_{nG,r} 
	\crn =& \dfrac{e^2m_a m_b t_R}{2m^2_Ws_W^2\left(m^2_a -m^2_b\right)} \sum_{i=1}^{K+3} U^{\nu}_{ai} U^{\nu*}_{bi} \left[ 2m^2_{n_i} \left( B_0^{(1)} -B_0^{(2)} \right) +m^2_{n_i}\left(B_1^{(1)} -B_1^{(2)}\right) 
\right.\crn& \left. \hspace{5.7cm}
+  \left(m^2_a B_1^{(1)}- m^2_b B_1^{(2)}\right)\right],
	\crn \bar{b}'_{l,5+6} =& \bar{b}'_{r,5+6}=0,   \label{eq_al5up}
\end{align}
where $B^{(k)}_{0,1}=B^{(k)}_{0,1}(p_k^2; m_{n_i}^2,m_W^2)$.

Form factors corresponding to diagram (1) in Fig. \ref{fig_2HDMU}, calculated in the HF gauge: 
\begin{align}
 \label{eq_al7up}
\bar{a}'_{nWW,l} =& \dfrac{e^2}{2s^2_Wt_r} \sum_{i=1}^{K+3} U^{\nu}_{ai} U^{\nu*}_{bi} \left[-2\left(B_0^{(12)}+m^2_{n_i}C_0\right) +(4-2d)C_{00} 
\right. \crn&\left. \hspace{3.2cm}- \left(m^2_aC_1 +m^2_bC_2 \right)   +2\left(m^2_Z-m^2_a-m^2_b\right)X_3 \right],  
\crn \bar{a}'_{nWW,r} =&  \dfrac{e^2}{2s^2_Wt_r}\sum_{i=1}^{K+3} U^{\nu}_{ai} U^{\nu*}_{bi} \left( -3m_a m_bX_3\right),
\crn \bar{b}'_{nWW,l} =& \dfrac{e^2  m_a}{2s^2_Wt_r} \sum_{i=1}^{K+3} U^{\nu}_{ai} U^{\nu*}_{bi} \left[4C_1 + 2C_2 -4X_1\right], 
\crn \bar{b}'_{nWW, r}=& \dfrac{e^2  m_b}{2s^2_Wt_r} \sum_{i=1}^{K+3} U^{\nu}_{ai} U^{\nu*}_{bi} \left[4C_2 + 2C_1 -4X_2 \right],
\end{align}
where $X_{0,1,2}=X_{0,1,2}(m_a^2,m_Z^2,m_b^2; m_i^2, m_W,m_W^2)$ and $C_{00}= C_{00}(m_a^2,m_Z^2,m_b^2; m_i^2, m_W^2,m_W^2)$.
Form factors corresponding to diagram (2) in Fig. \ref{fig_2HDMU}, calculated in the HF gauge: 
\begin{align}
\bar{a}'_{Wnn,l} =&  \dfrac{e^2}{4s^3_Wc_W} \sum_{i=1}^{K+3} U^{\nu}_{ai} U^{\nu*}_{bj}  \Big\lbrace q_{ij} \left[  4C_{00} +2\left(m^2_aX_{01}  +m^2_bX_{02}  -m^2_Z X_0  - m^2_Z C_{12}\right)  \right] 
 \crn &\hspace{4.2cm} +2 q_{ji}m_{n_i}m_{n_j}C_0 \Big\rbrace ,  \label{eq_al8up} 
%
\crn \bar{a}'_{Wnn,r} =&  \dfrac{e^2m_a m_b}{4s^3_Wc_W}  \sum_{i=1}^{K+3} U^{\nu}_{ai} U^{\nu*}_{bj} \left[ 2q_{ij}X_0 \right] ,
\crn \bar{b}'_{Wnn,l} =& \dfrac{e^2 m_a}{4s^3_Wc_W}    \sum_{i=1}^{K+3} U^{\nu}_{ai} U^{\nu*}_{bj} \left[ -4q_{ij} X_{01}\right] , 
\crn \bar{b}'_{Wnn,r}=& \dfrac{e^2 m_b}{4s^3_Wc_W} \sum_{i=1}^{K+3} U^{\nu}_{ai} U^{\nu*}_{bj}\left[ -4q_{ij} X_{02}\right],
\end{align}
where $X_{0,1,2}=X_{0,1,2}(m_a^2,m_Z^2,m_b^2; m_i^2, m_j^2,m_W^2)$ and $C_{00}= C_{00}(m_a^2,m_Z^2,m_b^2; m_i^2, m_j^2,m_W^2)$. 

Form factors corresponding to diagram (3) and (4) in Fig. \ref{fig_2HDMU}, calculated in the HF gauge: 
\begin{align}
		\bar{a}'_{nW,l} =&  \dfrac{2e^2}{2s^2_W(m_a^2 -m^2_b)}  \sum_{i=1}^{K+3} U^{\nu}_{ai} U^{\nu*}_{bi} \left[ t_L \left(m^2_a B^{(1)}_1 -m^2_b B^{(2)}_1 \right)\right],  \label{eq_910up} 
	\crn \bar{a}'_{nW,r} =&  \dfrac{2e^2U^{\nu}_{ai}U^{\nu*}_{bi}}{2s^2_W(m_a^2 -m^2_b)}  \sum_{i=1}^{K+3} U^{\nu}_{ai} U^{\nu*}_{bi} \left[ t_Rm_a m_b \left( B^{(1)}_1  - B^{(2)}_1  \right) \right],
	\crn \bar{b}'_{nW,l} =& \bar{b}'_{nW,r}=0.  
\end{align}

 We note that the PV-functions $A,B$, and $C$ defined in this work are consistent with notations used   LoopTools and Ref. \cite{Jurciukonis:2021izn}. The equivalences between two final lepton states defined in our work and Ref. \cite{Jurciukonis:2021izn} are $e_b \equiv \ell_1$ and $ e_a\equiv \ell_2 $, which implies that $p_{1\mu}= p_{a \mu} \equiv p_{\ell_2\mu}=p_{2\mu}$ and $p_{2\mu}= p_{b \mu} \equiv p_{\ell_1\mu}=p_{1 \mu}$, i.e., $\{ p_{1\mu}, p_{2\mu}\} \leftrightarrow \{ p_{2\mu}, p_{1\mu}\}=\{-q_{\mu},  p_{\mu}\}$ corresponding to the definitions in appendix A of Ref. \cite{Jurciukonis:2021izn}. In addition, there is a change that $(ij)\equiv (ji)$ for diagrams containing two different lepton propagators.  As a consequence, the following identifications is needed in order  to check the consistence of the results between our work and Ref. \cite{Jurciukonis:2021izn}:
 \begin{align}
 	\label{eq_PVnotations}
 &m_a^2=m_a^2\leftrightarrow m_{\ell_2}^2, \quad m^2_b=m_b^2\leftrightarrow m_{\ell_1}^2,
\quad  \left\{ q_{ij}, m_i,m_j\right\}= \left\{ q_{ji}, m_j,m_i\right\}, 
 \crn  	& \left\{ C_{0}, C_{12},X_0,X_3,X_{12}\right\}  \leftrightarrow 	\left\{ C_{0}, C_{12},X_0,X_3,X_{12}\right\},
 \crn&  	\left\{B^{(1)}_{0,1,11,00}, B^{(2)}_{0,1,11,00},C_{1,2,11,22},X_{01,02} \right\}  \leftrightarrow 	\left\{B^{(2)}_{0,1,11,00}, B^{(1)}_{0,1,11,00},C_{2,1,22,11}, X_{02,01}\right\}. 
 \end{align}
 The above relations are realized from the properties that  the $C$-functions relating to the cLFV $Z$ decays are derived in the different orders of the propagators in the numerators $\{D_0,D_1,D_2\} \leftrightarrow \left\{ k^2 -A, (k+p)^2 -B, (k+q)^2-C \right\}$ with $-p_{1\mu}=p_{\mu} =p_{\ell_1}$ and $p_{2\mu}=q_{\mu} =-p_{\ell_2}$, for example  
 \begin{align*}
 	h_0&= C_{0}(m_{\ell_1}^2,q^2,m_{\ell_2}^2; m_{W}^2, m^2_{i}, m^2_{j}) 
 	\crn&= C_{0}(m_{b}^2,q^2,m_{a}^2; m_{W}^2, m^2_{i}, m^2_{j}) = C_{0}(m_{a}^2,q^2,m_{b}^2; m_{W}^2, m^2_{j}, m^2_{i}),
 	\crn h_1&= C_{1}(m_{\ell_1}^2,q^2,m_{\ell_2}^2; m_{W}^2, m^2_{i}, m^2_{j}) 
 	\crn&= C_{1}(m_{b}^2,q^2,m_{a}^2; m_{W}^2, m^2_{i}, m^2_{j}) = C_{2}(m_{a}^2,q^2,m_{b}^2; m_{W}^2, m^2_{j}, m^2_{i}), \dots 
 \end{align*}
 
 The transformations between our results with those given in Ref. \cite{Abada:2022asx} are done through the following  relations:
 \begin{align*}
 -	\frac{e}{16\pi^2}\bar{a}_{l} =& F^L_V -i\left( m_a F^L_T +m_bF^R_T\right),  \\
 -	\frac{e}{16\pi^2}\bar{a}_{r} =& F^R_V -i\left( m_a F^R_T +m_bF^L_T\right),  \\
-\frac{e}{16\pi^2}\bar{b}_{l,r} =& 2iF^{L,R}_T,
 \end{align*}
 where $m_a\equiv m_{\alpha}$,  $m_b\equiv m_{\beta}$, $q_{ij}\equiv C_{ij}$, and $q_{ji}\equiv C^*_{ij}$. The redundant relations given  in appendix \ref{app_PVLT} were used to check the consistence. There appears an overall sign because of the  different sign conventions in  Lagrangians. Regarding analytic formulas computed in the unitary gauge, we realize that all form factors $F^{L,R}_T$  and $F^{R}_V$, and $F^{L(c+d)}_V$ in Ref. \cite{Abada:2022asx}  are totally consistent with ours, while the remaining $F^{L(a)}_{V}$ and $F^{L(b)}_{V}$ result in two deviations as follows:
 \begin{align}
 	\label{eq_dea}
& \frac{e}{16\pi^2}\bar{a}^{Wnn}_{l} + \left[ F^{L(b)}_V -i\left( m_a F^{L(b)}_T +m_bF^{R(b)}_T\right)\right] 
\crn &= \frac{g^3 }{32c_W \pi^2} \sum_{i,j=1}^{K+3}q_{ij} \left[ -B^{(1)}_0 -B^{(2)}_0 +2 B^{(12)}_0 +m_a^2 \left( 3C_1 +C_2+C_0\right) +m_b^2 \left( C_1 + 3 C_2+C_0\right)
\right.\crn&\left. \hspace{3.2cm} +C_0(m_{n_i}^2 -m^2_{n_j}) - (C_1 +C_2)m_Z^2\right],
 \end{align}
 where $B^{(1)}_0=B_0(m_a^2; m_W^2,m_i^2)$, $B^{(2)}_0=B_0(m_b^2; m_W^2,m_j^2)$, $C_k=C_k(m_a^2,m_Z^2,m_b^2,m_W^2,m_i^2,m_j^2)$, and 
  \begin{align*}
 &\frac{e}{16\pi^2}\bar{a}^{nWW}_{l} + \left[ F^{L(b)}_V -i\left( m_a F^{L(b)}_T +m_bF^{R(b)}_T\right)\right] 
 	\crn &=. \frac{g^3c_W }{16 \pi^2} \sum_{i=1}^{K+3}\left[ B^{(1)}_0 +B^{(2)}_0 - m_a^2 \left( 3C_1 +C_2+C_0\right) -m_b^2 \left( C_1 + 3 C_2+C_0\right)
 \right.\crn&\left. \hspace{2.5cm} -2C_0(m_{n_i}^2 -m^2_{W}) +  (C_1 +C_2)m_Z^2\right]
 \end{align*}
 with $B^{(k)}_0=B_0(p_k^2;m_i^2, m_W^2)$, $C_k=C_k(m_a^2,m_Z^2,m_b^2,m_i^2,m_W^2,m_W^2)$. 
    
 \subsection{ \label{app_Sloop} Scalar one-loop contributions to the $Z\to e^+_b e^-_a$ amplitude}
 For contributions from Higgs mediations, similar to the diagrams in Fig. \ref{fig_2HDMU}, but the gauge boson propagators $W$ are replaced with scalars $c$, namely  diagram (1), (2), and (3+4)  are denoted as  $ns_1s_2$, $s_knn$, and $ns_k$. They are also 4 diagrams (3), (4), and (5+6) in Fig. \ref{fig_2HDMHF}. The Lagrangian consisting of relevant scalar couplings is 
 \begin{align}
 	\mathcal{L}_S =& \sum_{a=1}^3\sum_{i=1}^{K+3} \sum_k \overline{n_i}(L^k_{ai}P_L +R^k_{ai}P_R) e_a s_k
 	\crn&+ (-e g_{Zs_1s_2}) Z_{\mu} s^-_1s^+_2 (p_+ -p_-)^{\mu} +\mathrm{h.c.}.	\label{eq_LnSe}
 \end{align}
 The results are listed as follows. The diagram with three propagators $ns_1s_2$:
 \begin{align}
 	\label{eq_nss}
\bar{a}_{ns_{1}s_2,l} =& 2g_{Zs_{1}s_{2}}\sum_{i=1}^{K+3} L^{1*}_{ai} L^{2}_{bi} C_{00}
\crn=&2g_{Zs_{1}s_{2}} \sum_{i=1}^{K+3}  \frac{L^{1*}_{ai} L^{2}_{bi}}{2} \left[ m_{n_i}^2 C_0 -m_a^2 C_1 -m_b^2 C_2 - \left(m_a^2 +m_b^2 -m_Z^2\right) C_{12} \right],
\crn 	\bar{a}_{ns_{1}s_2,r} =& 2g_{Zs_{1}s_{2}}\sum_{i=1}^{K+3} R^{1*}_{ai} R^{2}_{bi} C_{00}
\crn=&2g_{Zs_{1}s_{2}} \sum_{i=1}^{K+3}  R^{1*}_{ai} R^{2}_{bi}  \left[ m_{n_i}^2 C_0 -m_a^2 C_1 -m_b^2 C_2 - \left(m_a^2 +m_b^2 -m_Z^2\right) C_{12} \right],
\crn \bar{b}_{ns_{1}s_2,l} =& 2g_{Zs_{1}s_{2}} \sum_{i=1}^{K+3} \left[ L^{1*}_{ai} L^{2}_{bi} m_aX_1 + R^{1*}_{ai} R^{2}_{bi} m_b X_2 - R^{1*}_{ai} L^{2}_{bi} m_{n_i} X_0\right],
\crn 	\bar{b}_{ns_{1}s_2,r} =& 2g_{Zs_{1}s_{2}} \sum_{i=1}^{K+3} \left[ R^{1*}_{ai} R^{2}_{bi} m_aX_1 + L^{1*}_{ai} L^{2}_{bi} m_b X_2 - L^{1*}_{ai} R^{2}_{bi} m_{n_i} X_0\right],
 \end{align}
  where  $C_{0,00,k,kl}=C_{0,00,k,kl}(p_1^2,m_Z^2,p_2^2; m_{n_i^2}, m_{s_1}^2, m_{s_2}^2)$.
  
 The diagram with three propagators $snn$:  
 \begin{align}
 	\label{eq_snn}
 	\bar{a}_{s_knn,l} =& - \dfrac{1}{2s_Wc_W} \sum_{i,j=1}^{K+3} \left\lbrace q_{ij}\left[\lambda^{L,k*}_{ai}\lambda^{L,k}_{bj}m_{n_i} m_{n_j}C_0 +\lambda^{R,k*}_{ai}\lambda^{L,k}_{bj}m_{n_j}m_a (C_0+C_1)  \right.\right.\crn
 	&\left.\left. \hspace{3cm}+\lambda^{L,k*}_{ai}\lambda^{R,k}_{bj} m_{n_i} m_b (C_0+C_2) +\lambda^{R,k*}_{ai}\lambda^{R,k}_{bj}m_a m_b X_0 \right]\right.\crn
 	&\left. \hspace{3cm}  -q_{ji}\left[\lambda^{L,k*}_{ai}\lambda^{L,k}_{bj}\left(2C_{00} -B_0^{(12)} -m^2_{s_k}C_0 -m^2_aC_1 -m^2_bC_2\right) \right.\right.\crn
 	&\left.\left. \hspace{3.8cm}  - \lambda^{R,k*}_{ai}\lambda^{L,k}_{bj}m_{n_i} m_a C_1 +\lambda^{L,k*}_{ai}\lambda^{R,k}_{bj}m_{nj}m_b C_2 \right]\right\rbrace ,  
 	\crn 	\bar{a}_{s_knn,r} =&  -\dfrac{1}{2s_Wc_W}    \sum_{i,j=1}^{K+3}\left\lbrace q_{ij}\left[ - \lambda^{L,k*}_{ai}\lambda^{R,k}_{bj}m_{n_i} m_a C_1 + \lambda^{R,k*}_{ai}\lambda^{L,k}_{bj}m_{n_j}m_bC_2   \right.\right.\crn
 	&\left.\left.  \hspace{3.5 cm}  +\lambda^{R,k*}_{ai}\lambda^{R,k}_{bj}\left((2-d)C_{00} -\left(m^2_Z-m^2_a-m^2_b\right)C_{12}  -m^2_aC_1 -m^2_bC_2\right) \right]\right.\crn
 	&\left.  \hspace{2.7 cm}-q_{ji}\left[  \lambda^{R,k*}_{ai}\lambda^{R,k}_{bj}m_{n_i}m_{n_j}C_0  +\lambda^{L,k*}_{ai}\lambda^{R,k}_{bj}m_{n_j} m_a (C_0+C_1) \right.\right.\crn
 	&\left.\left.  \hspace{3.5 cm}  +\lambda^{R,k*}_{ai}\lambda^{L,k}_{bj}m_{n_i}m_b (C_0+C_2) + \lambda^{L,k*}_{ai}\lambda^{L,k}_{bj}m_a m_bX_0  \right]\right\rbrace,
 	\crn \bar{b}_{s_knn,l} =&  -\dfrac{1}{s_Wc_W}    \sum_{i,j=1}^{K+3} \left\lbrace q_{ij} \lambda^{R,k*}_{ai}\lambda^{R,k}_{bj}m_b X_2  -q_{ji}\left[ \lambda^{L,k*}_{ai}\lambda^{L,k}_{bj}m_a X_1  +\lambda^{R,k*}_{ai}\lambda^{L,k}_{bj}m_{n_i} C_1  \right]\right\rbrace,  
 	\crn 	\bar{b}_{s_knn,r} =&  -\dfrac{1}{ s_Wc_W}    \sum_{i,j=1}^{K+3} \left\lbrace q_{ij} \lambda^{R,k*}_{ai}\lambda^{R,k}_{bj} m_a X_1    -q_{ji}\left[\lambda^{L,k*}_{ai}\lambda^{L,k}_{bj}m_b X_2  +  \lambda^{L,k*}_{ai}\lambda^{R,k}_{bj} m_{n_j} C_2 \right] \right\rbrace,
 \end{align}
 where  where  $C_{0,00,k,kl}=C_{0,00,k,kl}(p_1^2,m_Z^2,p_2^2;  m_{s_k}^2, m_{n_i^2},m_{n_j^2})$.
 
 The two one-loop-two-point diagrams  with two propagators $nc$:
 \begin{align}
 	\label{eq_nc}
 	\bar{a}_{ns_{k},l} =& \dfrac{-t_L}{(m^2_a -m^2_b)} \sum_{i=1}^{K+3}    \left[m_{n_i}\left({\lambda^{L,k*}_{ai}{\lambda^{R,k}_{bi}}m_b +\lambda^{R,k*}_{ai}{\lambda^{L,k}_{bi}}m_a}\right)\left(B_0^{(1)} -B_0^{(2)}\right) \right.\crn
 	&\left. \hspace{2.8cm} -\left({\lambda^{L,k*}_{ai}{\lambda^{L,k}_{bi}}m^2_b +\lambda^{R,k*}_{ai}{\lambda^{R,k}_{bi}}m_a m_b}\right)\left(B_1^{(1)} -B_1^{(2)}\right)\right],
 	\crn 	\bar{a}_{ns_{k},r} =&\dfrac{-t_R}{(m^2_a -m^2_b)} \sum_{i=1}^{K+3}  \left[m_{n_i}\left({\lambda^{L,k*}_{ai}{\lambda^{R,k}_{bi}}m_a +\lambda^{R,k*}_{ai}{\lambda^{L,k}_{bi}}m_b}\right)\left(B_0^{(1)} -B_0^{(2)}\right) \right.\crn
 	&\left.  \hspace{2.8cm} -\left({\lambda^{L,k*}_{ai}{\lambda^{L,k}_{bi}}m_a m_b +\lambda^{R,k*}_{ai}{\lambda^{R,k}_{bi}}m^2_b}\right)\left(B_1^{(1)} -B_1^{(2)}\right)\right] ,
 	\crn \bar{b}_{ns_{k},l} =&	\bar{b}_{ns_{k},r} =0.
 \end{align}

\section{\label{app_heab} Form factors for the LFV$h$ decays $h\to e_a e_b$}

Denoting that $\Delta^{(i)}_{L,R}\equiv \Delta^{(ab)(i)}_{L,R}$ for short, the private contributions of all diagrams in Fig. \ref{fig_heab2HDM} to the LFV$h$ decay amplitude are as follows \cite{Nguyen:2018rlb, Nguyen:2020ehj, Hong:2022xjg}. The first class consists of only gauge boson, namely four diagrams (1), (5), (7), and (8).  The diagram (1) $nWW$:  
\begin{align} 
	\Delta^{(1)}_{L} =& 	\Delta^{nWW}_{L} 
	\crn= & \frac{g^3 m_a}{64\pi^2 m_W^3}c_{\delta} \crn
	& \times \sum_{i=1}^9U^{\nu}_{ai}U^{\nu*}_{bi}\left\{ m_{n_i}^2\left(B^{(1)}_0+ B^{(2)}_0 + B^{(1)}_1\right) + m_b^2B^{(2)}_1- \left(2m_W^2+m_h^2\right)m_{n_i}^2C_0  \right.\crn
	& \left.\hspace{2.5cm} - \left[m_{n_i}^2 \left( 2m_W^2+m_h^2\right)+2m_W^2 \left( 2 m_W^2+m_a^2-m_b^2\right) \right]C_1\right.\crn
	&\left. \hspace{2.5cm} - \left[ 2 m_W^2\left(m_a^2-m_h^2\right) +m_b^2m_h^2\right]C_2 \frac{}{}\right\}, \label{eq_nWWL}\\
	\Delta^{(1)}_{R}=& 	\Delta^{nWW}_{R} 
	\crn= & \frac{g^3m_b}{64\pi^2 m_W^3}c_{\delta}  \crn
	& \times\sum_{i=1}^9U^{\nu}_{ai}U^{\nu*}_{bi}\left\{ m_{n_i}^2\left(B^{(1)}_0+ B^{(2)}_0+ B^{(2)}_1\right) +m_a^2B^{(1)}_1- \left(2m_W^2+m_h^2\right)m_{n_i}^2C_0  \right.\crn
	& \left. \hspace{2.5cm}  -\left[ 2 m_W^2\left(m_b^2-m_h^2\right) +m_a^2m_h^2\right] C_1\right.\crn
	&\left. \hspace{2.5cm}  -\left[m_{n_i}^2 \left( 2m_W^2+m_h^2\right)+2m_W^2 \left( 2 m_W^2-m_a^2+m_b^2\right) \right]C_2 \frac{}{}\right\}, \label{eq_nWWR}
\end{align}
where $B^{(i)}_{0,1}=B_{0,1}(p_i^2;m^2_{n_i},m_W^2)$ and $C_{0,1,2}=C_{0,1,2}(p_1^2,m_h^2,p_2^2;m^2_{n_i},m_W^2,m_W^2)$.

The diagrams (5)  with three propagators $Wnn $:
\begin{align}
	\Delta^{(5)}_{L} =&\Delta^{(Wnn)}_{L}
	\crn=& \frac{g^3 m_a}{64\pi^2   m_W^3} \times \frac{c_{\alpha}}{s_{\beta}}\sum_{i,j=1}^{9}U^{\nu}_{ai}U^{\nu*}_{bj} \left\{\lambda^{0*}_{ij}m_{n_j}\left[-B^{(12)}_0+m_W^2C_0  +\left(2 m_W^2+m_{n_i}^2-m_a^2\right)C_1\right]\right. \crn
	&\hspace{4.7cm}\left.  + \lambda^{0}_{ij}m_{n_i}\left[ B^{(1)}_1 + \left(2 m_W^2+m_{n_j}^2-m_b^2\right)C_1\right]\right\},\crn
	\Delta^{(5)W}_{R}  =&\Delta^{(Wnn)}_{R}
	\crn=& \frac{g^3 m_b}{64\pi^2 m_W^3} \times \frac{c_{\alpha}}{s_{\beta}} \sum_{i=1}^{9}U^{\nu}_{ai}U^{\nu*}_{bj} \left\{\lambda^{0}_{ij}m_{n_i}\left[ -B^{(12)}_0 +m_W^2C_0 +\left(2 m_W^2+m_{n_j}^2-m_b^2\right)C_2\right]\right. \crn
	&\hspace{5cm} +\left.\lambda^{0*}_{ij}m_{n_j}\left[ B^{(2)}_1+\left(2 m_W^2+m_{n_i}^2-m_a^2\right)C_2\right]\right\},  
\end{align}
where $\lambda^h_{ij}$ is given in Eq.~\eqref{eq_lahij},  $B^{(12)}_{0}=B_{0}(m_h^2;m^2_{n_i},m^2_{n_j})$,  $B^{(k)}_{1}=B_{1}(p_k^2;m^2_{W},m^2_{n_i})$, and $C_{0,1,2}=C_{0,1,2}(p_1^2,m_h^2,p_2^2;m^2_{W},m_{n_i}^2,m_{n_j}^2)$   with $k=1,2$. 

The diagrams (7+8) with two propagators $nW $:
\begin{align}
	\Delta^{(7+8)}_{L}=&\Delta^{nW}_{L}
	\crn=& \frac{g^3m_a}{64\pi^2 m_W^3}\times \frac{s_{\alpha}}{c_{\beta}}\times \frac{m_b^2}{m_b^2-m_a^2}\sum_{i=1}^9U^{\nu}_{ai}U^{\nu*}_{bi}\crn
	\times& \left[\left(2m_W^2 +m_{n_i}^2 \right) \left(B^{(2)}_1-B^{(1)}_1\right) +m_b^2B^{(2)}_1-m_a^2B^{(1)}_1 +2m_{n_i}^2 \left( B^{(2)}_0 -B^{(1)}_0\right)\right], \crn 
	%
	\Delta^{(7+8)}_{R} =&\Delta^{nW}_{R}
	= \frac{m_a}{m_b} \Delta^{(7+8)}_{L},\label{eq_d78R}
\end{align} 
where $\Delta^{(7+8)}_{L,R}\equiv \Delta^{(7)}_{L,R} +\Delta^{(8)}_{L,R}$, and $B^{(i)}_{0,1}\equiv B_{0,1}(p_i^2;m^2_{n_i},m^2_{W})$. 

There are 4 diagrams relating with only scalar mediation, namely  diagrams (2), (6), (9), and (10).   
The diagrams (2)  with three propagators  $n c c $:
\begin{align}
	\Delta^{(2)}_{L} =& 	\Delta^{(ncc)}_{L} 
	\crn=& 	\frac{g^2}{32 \pi^2  m_W^2} \sum_{k,l=1}^2\lambda^h_{kl} \sum_{i=1}^{9}\left[-\lambda^{R,k*}_{ai}\lambda^{L,l}_{bi}m_{n_i}C_0 +\lambda^{L,k*}_{ai}\lambda^{L,l}_{bi}m_{a}C_1 +\lambda^{R,k*}_{ai}\lambda^{R,l}_{bi}m_{b}C_2 \right],\crn
	\Delta^{(2)}_{R} =& 	\Delta^{(ncc)}_{R} 
	\crn=&\frac{g^2}{32 \pi^2 m_W^2} \sum_{k,l=1}^2\lambda^h_{kl} \sum_{i=1}^{9}\left[-\lambda^{L,k*}_{ai}\lambda^{R,l}_{bi}m_{n_i}C_0 + \lambda^{R,k*}_{ai}\lambda^{R,l}_{bi}m_{a}C_1 +\lambda^{L,k*}_{ai}\lambda^{L,l}_{bi}m_{b}C_2 \right],\label{eq_Dncc}
\end{align}
where $C_{0,1,2}=C_{0,1,2}(p_1^2,m_h^2,p_2^2;m^2_{n_i},m_{c_k}^2,m_{c_l}^2)$, and $\lambda^h_{kl}$  are given in Eq~\eqref{eq_lahkl}.

The diagrams (6)  with $c nn $:
\begin{align}
	\Delta^{(6)}_{L}=& 
	\Delta^{(cnn)}_{L}
	\crn =& \frac{g^3c_{\alpha}}{64 \pi^2  m_W^3s_{\beta}}\sum_{k=1}^{2} \sum_{i,j=1}^{9} \left\{\lambda^{0*}_{ij} \left[\lambda^{R,k*}_{ai} \lambda^{L,k}_{bj}\left(B^{(12)}_0+m_{c_k}^2C_0 +m_a^2 C_1 +m_b^2C_2\right)\right.\right.\crn
	&+\left.\left. \lambda^{R,k*}_{ai} \lambda^{R,k}_{bj} m_b m_{n_j} C_2 +\lambda^{L,k*}_{ai}\lambda^{L,k}_{bj}m_am_{n_i}C_1 \right]\right. \crn
	&+\left.  \lambda^{0}_{ij} \left[\lambda^{R,k*}_{ai}\lambda^{L,k}_{bj} m_{n_i} m_{n_j} C_0 +\lambda^{R,k*}_{ai} \lambda^{R,k}_{bj} m_{n_i}m_{b} (C_0+C_2)\right.\right.\crn
	&+\left.\left.\lambda^{L,k*}_{ai} \lambda^{L,k}_{bj} m_{a}m_{n_j}(C_0 +C_1)+ \lambda^{L,k*}_{ai}\lambda^{R,k}_{bj}m_{a}m_{b}(C_0 +C_1+C_2) \right]\frac{}{}\right\}
	, \label{eq_cnnL}\\
	\Delta^{(6)}_{R}=& \Delta^{(cnn)}_{R}
	\crn =& \frac{g^3c_{\alpha}}{64 \pi^2  m_W^3s_{\beta}} \sum_{k=1}^{2} \sum_{i,j=1}^{9}\left\{\lambda^{0}_{ij}\left[\lambda^{L,k*}_{ai} \lambda^{R,k}_{bj} \left(B^{(12)}_0+m_{c_k}^2C_0 +m_a^2 C_1+m_b^2C_2\right)\right.\right.\crn
	&+\left.\left.\lambda^{L,k*}_{ai}\lambda^{L,k}_{bj} m_bm_{n_j}C_2 +  \lambda^{R,k*}_{ai} \lambda^{R,k}_{bj} m_am_{n_i}C_1 \right]\right.\crn
	&+\left.
	\lambda^{0*}_{ij}\left[\lambda^{L,k*}_{ai}\lambda^{R,k}_{bj}m_{n_i}m_{n_j}C_0 +\lambda^{L,k*}_{ai}\lambda^{L,k}_{bj}m_{n_i}m_{b}(C_0+C_2)\right.\right.\crn
	&+\left.\left.\lambda^{R,k*}_{ai}\lambda^{R,k}_{bj}m_{a}m_{n_j} (C_0 +C_1)+ \lambda^{R,k*}_{ai}\lambda^{L,k}_{bj}m_{a}m_{b}(C_0 +C_1+C_2) \right]\frac{}{}\right\}
	, \label{eq_cnnR}
\end{align}
where $B^{(12)}_{0}=B_{0}(m_h^2;m^2_{n_i},m^2_{n_j})$,   and $C_{0,1,2}=C_{0,1,2}(p_1^2,m_h^2,p_2^2;m^2_{c_k},m_{n_i}^2,m_{n_j}^2)$,

The diagrams (9+10) with $nc^\pm_l$:
\begin{align}
	%
	\Delta^{(9 +10)}_{L} =&	\Delta^{(nc)}_{L} 
	\crn=& \frac{g^3s_\alpha}{64\pi^2c_\beta m_W^3\left(m_a^2-m_b^2\right)}  \crn
	&\times \sum_{k=1}^{2}\sum_{i=1}^{9}\left[ m_am_bm_{n_i}  \lambda^{L,k*}_{ai}\lambda^{R,k}_{bi}\left(B^{(1)}_0-B^{(2)}_0\right)+ m_{n_i}
	\lambda^{R,k*}_{ai}\lambda^{L,k}_{bi}\left(m^2_bB^{(1)}_0-m^2_aB^{(2)}_0\right)\right.\crn
	&\left.+ m_{a}m_b \left(\lambda^{L,k*}_{ai}\lambda^{L,k}_{bi}m_b + \lambda^{R,k*}_{ai}\lambda^{R,k}_{bi}m_a\right)\left(-B^{(1)}_1+ B^{(2)}_1\right)\right], 
	%
	\crn 	\Delta^{(9 +10)}_{R} =&	\Delta^{(nc)}_{R} 
	\crn=& \frac{g^3s_\alpha}{64\pi^2c_\beta m_W^3\left(m_a^2-m_b^2\right)}  \crn
	&\times\sum_{k=1}^{2} \sum_{i=1}^{9}\left[ m_am_bm_{n_i}  \lambda^{R,k*}_{ai}\lambda^{L,k}_{bi}\left(B^{(1)}_0-B^{(2)}_0\right)+ m_{n_i}
	\lambda^{L,k*}_{ai}\lambda^{R,k}_{bi}\left(m^2_bB^{(1)}_0-m^2_aB^{(2)}_0\right)\right.\crn
	&\left.+ m_{a}m_b \left(\lambda^{R,k*}_{ai}\lambda^{R,k}_{bi}m_b + \lambda^{L,k*}_{ai}\lambda^{L,k}_{bi}m_a\right)\left(- B^{(1)}_1+ B^{(2)}_1\right)\right], \label{eq_ncLR}
\end{align}
where $\Delta^{(9+10)}_{L,R}\equiv \Delta^{(9)}_{L,R} +\Delta^{(10)}_{L,R}$, and $B^{(i)}_{0,1}\equiv B_{0,1}(p_i^2;m^2_{n_i},m^2_{c_k})$.  The analytic forms of $\Delta^{(i)}_{L,R}$ given here were also cross-checked using the FORM package~\cite{Vermaseren:2000nd, Kuipers:2012rf}.

There are 2 diagrams consisting  of both scalar and vector mediation, namely  diagrams (3) and (4).  
The diagrams (3)  $nc W$:
\begin{align}
	%
	\Delta^{(3)}_{L} =&  \Delta^{(ncW)}_{L}
	\crn =& \frac{eg^2s_{\delta} f_{k}}{64\pi^2 s_Wm_W^3} 
	 \crn &\times \sum_{i=1}^{9}U^{\nu*}_{bi} \left\{  \lambda^{L,k*}_{ai}m_am_{n_i}\left[ 2m_W^2 C_0+\left(m_W^2-m_{c_k}^2+m_{h}^2\right)C_1\right]
	\right.\crn	&	\left. \qquad \qquad + \lambda^{R,k*}_{ai}\left[m_{n_i}^2 B^{(2)}_0 +m_b^2B^{(2)}_1 -m_{n_i}^2\left(m_W^2 -m_{c_k}^2 +m_{h}^2\right)C_0  - 2m_a^2 m_W^2 C_1
	\right.\right.\crn &\left.\left. 
	\qquad \qquad \qquad \quad +\left[ 2m_W^2 \left(m_{h}^2-m_a^2\right)- m_b^2\left(m_W^2 -m_{c_k}^2 +m_{h}^2\right)\right]C_2\right]\right\}, \label{eq_ncWL}\\
	\Delta^{(3)}_{R} =&  \Delta^{(ncW)}_{R}
	\crn=&  -\frac{g^3 m_b s_{\delta} f_{k}}{64\pi^2 m_W^3} \crn
	&\times \sum_{i=1}^{9}U^{\nu*}_{b i} \left\{\lambda^{L,k*}_{ai} m_{n_i}\left[ B^{(2)}_0 +B^{(2)}_1 
	%
	%
	+\left(m_W^2 +m_{c_k}^2 -m_{h}^2\right)C_0 -\left(m_W^2 -m_{c_k}^2 +m_{h}^2\right) C_2\right]
	\right.\crn	& \left. 
	\qquad \qquad -\lambda^{R,k*}_{ai}m_{a} \left[  \left(m_W^2 +m_{c_k}^2 -m_{h}^2\right)C_1 +2m_W^2 C_2\right]\right\},\label{eq_ncWR}
\end{align}
where 
$$ f_1 = c_{\phi}, \; f_2 = -s_{\phi}, $$  $B^{(2)}_{0,1}=B_{0,1}(p_2^2;m^2_{n_i},m_W^2)$,  and $C_{0,1,2}=C_{0,1,2}(p_1^2,m_h^2,p_2^2;m^2_{n_i},m_{c_k}^2,m_W^2)$. 

%
The diagrams (4) with $nWc$:
\begin{align}
	%
	\Delta^{(4)}_{L} =&  \Delta^{(nWc)}_{L}
	\crn =& -\frac{eg^2 m_{a} s_{\delta} f_{k}}{64 \pi^2 s_Wm_W^3} \sum_{i=1}^{9}U^{\nu}_{ai}\crn
	&\times\left\{\lambda^{L,k}_{bi} m_{n_i}\left[ B^{(1)}_0 +B^{(1)}_1  +\left(m_W^2 +m_{c_k}^2 -m_{h}^2\right) C_0  -\left(m_W^2 -m_{c_k}^2 +m_{h}^2\right) C_1\right]
	\right.\crn& \left. 
	\quad  - \lambda^{R,1}_{bi}m_{b}\left[ 2m_W^2C_1 +\left(m_W^2 +m_{c_k}^2 -m_{h}^2\right) C_2\right]\right\}, \label{eq_nWcL}\\
	\Delta^{(4)}_{R} =&   \Delta^{(nWc)}_{R}
	\crn=& \frac{g^3 s_{\delta} f_{k}}{64 \pi^2 m_W^3}\sum_{i=1}^{9}U^{\nu}_{ai}\crn
	&\times\left\{ \lambda^{L,k}_{bi} m_bm_{n_i} \left[ 2m_W^2C_0  +\left(m_W^2 -m_{c_k}^2 +m_{h}^2\right) C_2\right] 
	\right.\crn &\left. 
	+ \lambda^{R,k}_{bi} \left[m_{n_i}^2 B^{(1)}_0 +m_a^2B^{(1)}_1  -m_{n_i}^2 \left(m_W^2 -m_{c_k}^2 +m_{h}^2 \right) C_0\right.\right.\crn
	&\left.\left. \qquad +\left[ 2m_W^2 \left(m_{h}^2 -m_b^2\right)- m_a^2 \left(m_W^2 -m_{c_k}^2 +m_{h}^2\right)\right] C_1 -2 m_b^2m_W^2 C_2\right]\right\} ,\label{eq_nWcR}
\end{align}
where $B^{(1)}_{0,1}=B_{0,1}(p_1^2;m^2_{n_i},m_W^2)$ and $C_{0,1,2}=C_{0,1,2}(p_1^2,m_h^2,p_2^2;m^2_{n_i},m_W^2,m_{c_k}^2)$. 

The divergent cancellation of the decay amplitudes are proved  following Refs.  \cite{Hue:2015fbb, Nguyen:2018rlb,  Nguyen:2020ehj}. 

\section{ \label{app_2HDMnu} The Higgs sector of the 2HDM$N_{L,R}$}

The Higgs potential of the model 
\begin{align}
	\mathcal{V}= & \mathcal{V}_{\varphi} +\mathcal{V}_{\mathrm{2HDM}}, 	\label{eq_V2HDMnu}
	\\	\mathcal{V}_{\varphi}=& \mu^2_{\varphi} \varphi^*\varphi + \mu^2_{\chi} \chi^+\chi^-  +\lambda_{\varphi} \left(\varphi^*\varphi\right)^2  +\lambda_{\chi} \left(\chi^+\chi^- \right)^2 + \lambda_{\varphi \chi} \left(\varphi^*\varphi\right) \left(\chi^+\chi^- \right)
	\crn+ & \sum_{i=1}^2 \left[ \lambda_{i\varphi} (\varphi^*\varphi)  + \lambda_{i\chi}(\chi^+\chi^-)  \right] \left(H_i^\dagger H_i\right) + \left\{
	\lambda \left( H_1^T i\sigma_2 H_2\right) \chi^- \varphi +\mathrm{h.c.}\right\} , \label{eq_Vnew}
	\\  \mathcal{V}_{\mathrm{2HDM}}=&  \mu^2_{11} H^\dagger_1H_1 + \mu^2_{22} H^\dagger_2H_2  - \left( \mu^2_{12} H_1^{\dagger} H_2 +\mathrm{h.c.}\right) + \frac{1}{2} \lambda_1 \left(H_1^{\dagger} H_1\right)^2 + \frac{1}{2}  \lambda_2 \left(H_2^{\dagger} H_2\right)^2 
	\crn &+  \lambda_3 \left(H_1^{\dagger} H_1\right)  \left(H_2^{\dagger} H_2\right)  +  \lambda_4 \left(H_1^{\dagger} H_2\right)  \left(H_2^{\dagger} H_1\right) 
	\crn &+ \frac{\lambda_5}{2}  \left[ \left(H^{\dagger}_1H_2\right)^2 + \left(H^{\dagger}_2H_1\right)^2\right]. \label{eq_2HDM} 
\end{align}
The minima conditions of the Higgs potential:
\begin{align}
	r_1: \quad \mu_{11}^2 &=-\frac{1}{2} c_{\beta}^2 \lambda_1 v_H^2-\frac{1}{2} \lambda_{345} s_{\beta}^2 v_H^2+t_{\beta} \mu_{12}^2-\frac{\lambda_{1\varphi} v_{\varphi}^2}{2}, 
	\crn	r_2: \quad  \mu_{22}^2& = -\frac{1}{2} c_{\beta}^2 \lambda_{345} v_H^2-\frac{1}{2} \lambda_2 s_{\beta}^2 v_H^2+\frac{\mu_{12}^2}{t_{\beta}}-\frac{\lambda_{2\varphi} v_{\varphi}^2}{2},
	\crn	r': \quad  \mu_{\varphi}^2 & = -\frac{1}{2} c_{\beta}^2 \lambda_{1\varphi} v_H^2-\frac{1}{2} \lambda_{2\varphi} s_{\beta}^2 v_H^2-\lambda_{\varphi} v_{\varphi}^2,
\end{align}
where $\lambda_{345}= \lambda_3 +\lambda_4 +\lambda_5$. 

Neutral CP-odd Higgs bosons. We find that $z'$ is one massless eigeinstate corresponding the   Goldstone  of  $B'_{\mu}$. The squared mass matrix in the basis $(z_1,z_2)$ is:
\begin{align}
	\label{eq_m2z}
	\mathcal{M}^2_A &=  \left(
	\begin{array}{cc}
		t_{\beta} \mu_{12}^2-s_{\beta}^2 v_H^2 \lambda_5 & c_{\beta} s_{\beta} v_H^2 \lambda_5-\mu_{12}^2 \\
		c_{\beta} s_{\beta} v_H^2 \lambda_5-\mu_{12}^2 & \frac{\mu_{12}^2}{t_{\beta}}-c_{\beta}^2 v_H^2 \lambda_5 \\
	\end{array}
	\right). 
\end{align}
The mixing matrix $R(\beta) $ diagonalisizing  $\mathcal{M}^2_A$:
\begin{align}
	\label{eq_OA}
	R(\beta)=& \left(
	\begin{array}{cc}
		c_{\beta} & s_{\beta} \\
		-s_{\beta} & c_{\beta} \\
	\end{array}
	\right)  \Rightarrow 
R(\beta) \mathcal{M}^2_A R^T(\beta)= \mathrm{diag}\left( 0, \; \frac{\mu_{12}^2}{c_{\beta} s_{\beta}} -\lambda_5 v_H^2\right) ,
\end{align}
where $R(x)$ is a real $2\times2$ rotation matrix of angle $x$ satisfying $R^T(x)R(x)=R(x)R^T(x)=I_2$.
This gives
\begin{align}
	\label{eq_mA}
	m_{G_Z} &=0,\; m_A^2 = \frac{\mu_{12}^2}{c_{\beta} s_{\beta}} -\lambda_5 v_H^2,
	\crn\begin{pmatrix}
		z_1\\
		z_2
	\end{pmatrix}&= R^T(\beta) \begin{pmatrix}
		G_Z\\
		A
	\end{pmatrix}= \left(
	\begin{array}{c}
		c_{\beta} G_Z-s_{\beta} A \\
		G_Z s_{\beta}+c_{\beta} A \\
	\end{array}
	\right). 
\end{align}

Neutral CP-even Higgs bosons. In the original basis $ (r_1,r_2,r')$, the squared mass matrix is:
\begin{align}
	\label{eq_M2r}
	\mathcal{M}^2_h =& \left(
	\begin{array}{ccc}
		c_{\beta}^2 \lambda_1 v_H^2+t_{\beta} \mu_{12}^2 & c_{\beta} s_{\beta} v_H^2 \lambda_{345}-\mu_{12}^2 & c_{\beta} v_H v_{\varphi} \lambda_{1\varphi} \\
		c_{\beta} s_{\beta} v_H^2 \lambda_{345}-\mu_{12}^2 & s_{\beta}^2 \lambda_2 v_H^2+\frac{\mu_{12}^2}{t_{\beta}} & s_{\beta} v_H v_{\varphi} \lambda_{2\varphi} \\
		c_{\beta} v_H v_{\varphi} \lambda_{1\varphi} & s_{\beta} v_H v_{\varphi} \lambda_{2\varphi} & 2 v_{\varphi}^2 \lambda_{\varphi} \\
	\end{array}
	\right).
\end{align}
Because det$\left.\mathcal{M}^2_h\right|_{v_H=0}=0 $, there are at least one Higgs bosson having  mass at the electroweak scale, which will be identified with the real one found experimentally. 
In this work we limit  $\lambda_{1\varphi}=\lambda_{2\varphi}=0$, leading to the consequence that  $\mathcal{M}^2_h$  has one physical state  $r'$ with very heavy mass  $m^2_{r'}=2 v_{\varphi}^2 \lambda_{\varphi} \varpropto v^2_{\varphi}$. The remaining $2\times 2$ matrix, $\mathcal{M}'^{2}_h$, is diagonalized  by  a transformation $	O(\alpha) $\footnote{This definition give the popular forms of  the SM-like Higgs couplings}:
\begin{align}
	\label{eq_OHp}
	O(\alpha) =& \left(
	\begin{array}{cc}
		-s_{\alpha} & c_{\alpha} \\
		c_{\alpha} & s_{\alpha} \\
	\end{array}
	\right)  \Rightarrow 
	O(\alpha)  \mathcal{M}'^{2}_h 		O^T(\alpha) = \mathrm{diag}\left(  m_h^2, m_H^2\right),  \\
	\begin{pmatrix}
		r_1\\
		r_2
	\end{pmatrix}&= 		O^T(\alpha)  \begin{pmatrix}
		h\\
		H
	\end{pmatrix}= \left(
	\begin{array}{c}
		-s_{\alpha} h +c_{\alpha} H \\
		c_{\alpha} h+s_{\alpha}H \\
	\end{array}
	\right),
\end{align}
where $h$ and  $H$ are  SM-like Higgs and new heavy Higg boson.  
Their masses and mixing parameter $\alpha$ are determined as follows
\begin{align}
	\label{eq_mh0}
	m^2_h =& M^2_{11}s^2_{\beta -\alpha} +M^2_{22} c^2_{\beta -\alpha} +M^2_{12}s_{2(\beta -\alpha)}, \\
	m^2_H =&  M^2_{11}c^2_{\beta -\alpha} +M^2_{22} s^2_{\beta -\alpha} -M^2_{12}s_{2(\beta -\alpha)},
\crn \tan{2(\beta- \alpha)}&=\frac{2 c_{\beta}^2 s_{\beta}^2 v_H^2 \left[ c_{\beta}^2 \lambda_1 -s_{\beta}^2 \lambda_2 + \lambda_{345}(s^2_{\beta} -c^2_{\beta})\right]}{c_{\beta}
	s_{\beta} v_H^2 \left(c_{\beta}^4 \lambda_1-c_{\beta}^2 s_{\beta}^2 (\lambda_1+\lambda_2-4 \lambda_{345})+\lambda_2 s_{\beta}^4\right)-\mu_{12}^2} 
\end{align}
where the entries  $M^2_{ij}$ are:
\begin{align}
	\label{eq_M2ij}
	\crn	M^2 &= \begin{pmatrix}
		v_H^2 \left(c_{\beta}^4 \lambda_1 + \frac{1}{2} s_{2\beta}^2 \lambda_{345}  +\lambda_2 s_{\beta}^4\right) & \frac{1}{2}s_{2\beta}  v_H^2
		\left( c_{2\beta}\lambda_{345} -c_{\beta}^2 \lambda_1 +s_{\beta}^2  \lambda_2\right) \\
		\star  & c_{\beta}^2 s_{\beta}^2 v_H^2  (\lambda_1 +\lambda_2 -2 \lambda_{345}) +\frac{\mu_{12}^2}{c_{\beta} s_{\beta}} \\
	\end{pmatrix}.
\end{align}
It is noted that $O(\alpha -\beta)M^2O^T(\alpha -\beta)= \mathrm{diag}\left(  m_h^2, m_H^2\right) $.
The final results are 
\begin{align}
	\label{eq_H0i}
	h &= \frac{1}{\sqrt{2}} \left[v_H c_{\beta} +c_{\alpha} H- s_{\alpha} h +ic_{\beta} G_Z -i s_{\beta} A\right],
	\crn H &= \frac{1}{\sqrt{2}} \left[v_H s_{\beta} +s_{\alpha} H + c_{\alpha} h +is_{\beta} G_Z +i c_{\beta} A\right].
\end{align}

Singly charged Higgs bossons. The basis $(H^-_1,H^-_2, \chi^-)$ gives the following squared mass matrix: 
\begin{align}
	\label{eq_mc2}
	\mathcal{M}^2_C &=\frac{1}{2} \begin{pmatrix}
		2 t_{\beta} \mu_{12}^2- \lambda_{45} s_{\beta}^2 v^2_H 	&c_{\beta} s_{\beta} \lambda_{45} v_H^2  -2 \mu_{12}^2& \lambda  s_{\beta} v_H v_{\varphi} \\
		\star 	& \frac{2 \mu_{12}^2}{t_{\beta}} -c_{\beta}^2  \lambda_{45} v_H^2 &  -\lambda  c_{\beta} v_H v_{\varphi} \\
		\star 	&  	\star  & 2 \mathcal{M}^2_{+33} 
	\end{pmatrix},
\end{align}
where 
\begin{align}
	\label{eq_Mc33}
	\lambda_{45}& =\lambda_4 + \lambda_5,  \\
	\mathcal{M}^2_{+33} &=(c_{\beta}^2 \lambda_{1\chi} +\lambda_{2\chi} s_{\beta}^2) v_H^2   +\lambda_{\varphi \chi} v_{\varphi}^2 +2 \mu_{\chi}^2. 
\end{align} 
The mixing matrix  $O_C$ is defined as :
\begin{align}
	\label{eq_OC}
	O_C 	\mathcal{M}^2_CO_C^T = \mathrm{diag} \left(0,\; m^2_{c_1},\; m^2_{c_2}\right), \; O_CO_C^T= O_C^T O_C=I_3.
\end{align}
where $(c^\pm_0, c^\pm_1, c^\pm_2)$ are mass eigeinstates with $c^\pm_0$  being the Goldstone bosons of $W^\pm$:
\begin{align}
	\label{eq_Vc}
	\begin{pmatrix} 
		H^\pm_1\\
		H^\pm_2\\
		\chi^\pm
	\end{pmatrix} =O^T_C \begin{pmatrix} 
		c^\pm_0\\
		c^\pm_1\\
		c^\pm_2
	\end{pmatrix} =  \left(
	\begin{array}{ccc}
	c_{\beta} & -c_{\phi} s_{\beta} & s_{\beta} s_{\phi} \\
	s_{\beta} & c_{\beta} c_{\phi} & -c_{\beta} s_{\phi} \\
	0 & s_{\phi} & c_{\phi} \\
	\end{array}
	\right) \begin{pmatrix} 
	c^\pm_0\\
	c^\pm_1\\
	c^\pm_2
	\end{pmatrix}, 
\end{align}
where 
\begin{align}
	\label{eq_t2phi}
	t_{2\phi}\equiv \tan(2\phi) &= \frac{ 2\lambda  s_{2\beta} v_H v_{\varphi}}{s_{2\beta} \lambda_{45}  v_H^2+2  s_{2\beta} M_{+33}^2- 4 \mu_{12}^2}. 
\end{align}
This gives 
\begin{align}
	m^2_{c_1} &= c_{\phi}^2 \left[ \frac{2\mu_{12}^2}{ s_{2\beta}} -\frac{\lambda_{45} v_H^2}{2} \right]  + s_{\phi}^2 \left[\mathcal{M}^2_{+33} \right] +s_{2\phi} \left[ -\frac{1}{2} \lambda v_H v_{\varphi}\right],
	\crn 	m^2_{c_2} &=s_{\phi}^2 \left[ \frac{2\mu_{12}^2}{ s_{2\beta}} -\frac{\lambda_{45} v_H^2}{2} \right]  + c_{\phi}^2 \left[\mathcal{M}^2_{+33} \right] -s_{2\phi} \left[ -\frac{1}{2} \lambda v_H v_{\varphi}\right].  
\end{align}
In the numerical calculation, the following quantities relating to the Higgs potential are independent inputs:
\begin{align}
	\label{eq_Higginput}
\lambda_1,\; \lambda_4,\; \lambda_5,  \; m_h,\; m_H, \;m_{c_{1}}, \;m_{c_{2}},\; \delta,\; \phi,
\end{align}
where $\delta$ was defined in Eq. \eqref{eq_deltadef} for accommodating the SM results.  
The dependent parameters are:
\begin{align}
	\label{eq_fparameter}
\crn m_A^2&=c_{\phi}^2 m_{c_1}^2+s_{\phi}^2 m_{c_2}^2+\frac{1}{2} \left(\lambda _4 -\lambda _5\right) v_H^2,
\crn \lambda_2&= \frac{c_{\beta}^2 \lambda _1 s_{\alpha}^2}{c_{\alpha}^2 s_{\beta}^2}+\frac{c_{2\alpha} m_h^2 +\left(\lambda _5 v_H^2+m_A^2\right)s_{\delta} \left(s_{\alpha} s_{\beta}- c_{\alpha} c_{\beta}\right)}{c_{\alpha}^2 s_{\beta}^2 v_H^2},
\crn \lambda_{3}=&\frac{c_{\beta} \lambda _1 s_{\alpha}}{c_{\alpha} s_{\beta}} +\frac{s_{\delta} s_{\beta} \left(\lambda _5 v_H^2+m_A^2\right) -s_{\alpha} m_h^2}{c_{\alpha} c_{\beta}s_{\beta}
	v_H^2} -\lambda_4 -\lambda_5,
\crn m_H^2 =& \frac{v_H^2 \left(c_{\beta}^2 \lambda _1+\lambda _5 s_{\beta}^2\right) -s_{\alpha}^2 m_h^2 +s_{\beta}^2 m_A^2}{c_{\alpha}^2}. 
\end{align}
We note that  the following intermediate parameters are absorbed in the Higgs potential: 
\begin{align*}
	\mu_{12}^2 =& c_{\beta} s_{\beta} \left(\lambda_5 v_H^2+m_A^2\right),
\crn \lambda =&\frac{s_{2\phi} (m_{c_2}^2-m_{c_1}^2)}{v_H v_{\varphi}},
\crn \mathcal{M}^2_{+33} = & s_{\phi}^2 m_{c_1}^2 +c_{\phi}^2 m_{c_2}^2,
\end{align*}
In the numerical discussions, all Higgs couplings were checked to satisfy the two conditions of bounded from below  and the unitarity limits \cite{Branco:2011iw}, namely 
\begin{align}
&4\pi\geq	\lambda_1,\lambda_2 \geq0,
\crn &\lambda_3+\sqrt{\lambda_1 \lambda_2}\geq 0,\;   \lambda_3+ \lambda_4 +\sqrt{\lambda_1 \lambda_2} -|\lambda_5| \geq 0,\;  
\crn & \left| \frac{3}{2} \left( \lambda_1+\lambda_2\right) +\sqrt{\frac{9}{4} (\lambda_1- \lambda_2)^2 +(2\lambda_3 +\lambda_4)^2}\right|< 8\pi,
\crn &\left| \frac{1}{2} \left( \lambda_1+\lambda_2\right) +\frac{1}{2}\sqrt{ (\lambda_1- \lambda_2)^2 +4\lambda_4^2}\right|< 8\pi,
\crn &\left| \frac{1}{2} \left( \lambda_1+\lambda_2\right) + \frac{1}{2}\sqrt{ (\lambda_1- \lambda_2)^2 +4\lambda_5^2}\right|< 8\pi,
\crn& \left| \lambda_3+2\lambda_4 -3\lambda_5 \right|<8\pi, \; \left| \lambda_3- \lambda_5 \right|<8\pi, \; \left| \lambda_3 + 2\lambda_4 + 3\lambda_5 \right|<8\pi,
\crn & \left| \lambda_3+ \lambda_5 \right|<8\pi, \; \left| \lambda_3+ \lambda_4 \right|<8\pi, \; \left| \lambda_3- \lambda_4 \right|<8\pi. 
\end{align}
It can be seen in Eq. \eqref{eq_fparameter} that, large $t_{\beta}$ will result in  large  $|\lambda_3|$ unless $s_{\delta} \left(\lambda _5 v_H^2+m_A^2\right)/(c_{\beta}v_H^2)\varpropto \mathcal{O}(1)$, therefore $s_{\delta}$ must be small enough.

\end{document}